\documentclass[12pt]{article}
\pdfoutput=1
\usepackage{a4wide}
\usepackage{amsmath,amssymb,amsfonts}
\usepackage{latexsym}
\usepackage[paper=letterpaper,margin=1.0in]{geometry}
\usepackage{graphicx}
\usepackage{color}
\usepackage{comment}

\def\mP{\mathbb{P}}
\def\s{\sigma}
\newcommand{\be}{\begin{equation}}
\newcommand{\ee}{\end{equation}}
\newcommand{\bea}{\begin{eqnarray}\displaystyle}
\newcommand{\eea}{\end{eqnarray}}

\newcommand{\pd}{\partial}

\newcommand{\ba}{\begin{array}}
\newcommand{\ea}{\end{array}}
\newcommand{\ben}{\begin{enumerate}}
\newcommand{\een}{\end{enumerate}}
\newcommand{\bi}{\begin{itemize}}
\newcommand{\ei}{\end{itemize}}
\newcommand{\bc}{\begin{center}}
\newcommand{\ec}{\end{center}}
\newcommand{\bfig}{\begin{figure}}
\newcommand{\efig}{\end{figure}}
\newcommand{\bq}{\begin{quotation}}
\newcommand{\eq}{\end{quotation}}
\newcommand{\bt}{\begin{table}}
\newcommand{\et}{\end{table}}
\newcommand{\btab}{\begin{tabular}}
\newcommand{\etab}{\end{tabular}}
\newcommand{\bmi}{\begin{minipage}}
\newcommand{\emi}{\end{minipage}}
\newcommand{\bs}{\begin{slide}}
\newcommand{\es}{\end{slide}}
\newcommand{\nn}{\nonumber}
\newcommand{\eref}[1]{(\ref{#1})}
\newcommand{\cN}{{\cal N}}

\newcommand{\sgb}{\sigma_{B}}
\newcommand{\sgw}{\sigma_{W}}
\newcommand{\sgi}{\sigma_{\infty}}
\newcommand{\tsgb}{\widetilde \sigma_{B}}
\newcommand{\tsgw}{\widetilde \sigma_{W}}

\newcommand{\bmQ}{\overline{\mathbb{Q}}}

\newcommand{\mQ}{ \mathbb{Q} }

\newcommand{\IC}{{\mathbb C}}

\newcommand{\IZ}{{\mathbb Z}}
\newcommand{\la}{\langle}
\newcommand{\ra}{\rangle}

\newcommand{\mT}{ \mathbb{T} }
\newcommand{\Pp}{ P^{\prime} }
\newcommand{\tr}{ {\rm Tr}  }

\newcommand{\cZ}{\mathcal{Z}}
\def\g{ {\gamma} }

\newcommand{\AdS}[1]{{\rm AdS}_{#1}}
\newcommand{\pa}{\partial}
\begin{document}

{}~
{}~
\hbox{QMUL-PH-10-16}
\break

\vskip .6cm

\centerline{{\LARGE \bf Toric CFTs, Permutation Triples, and Belyi Pairs}}

\medskip

\vspace*{4.0ex}

\centerline{
{\large \bf Vishnu Jejjala}\footnote{v.jejjala@qmul.ac.uk},
{\large \bf Sanjaye Ramgoolam}\footnote{s.ramgoolam@qmul.ac.uk},
{\large \bf Diego Rodriguez-Gomez}\footnote{drodrigu@physics.technion.ac.il}}

\vspace*{4.0ex}

\begin{center}
{\large ${}^{1,2}$ Department of Physics\\
Queen Mary, University of London\\
Mile End Road\\
London E1 4NS, UK\\
}
\vspace*{4.0ex}
{\large ${}^{3}$ Department of Physics\\
Technion, Haifa, 3200, Israel\\}
\vspace*{3.0ex}
{\large ${}^{3}$ Department of Mathematics and Physics\\
University of Haifa at Oranim, Tivon, 36006, Israel\\
}
\end{center}

\vspace*{5.0ex}

\centerline{\bf Abstract} \bigskip

Four-dimensional CFTs dual to branes transverse to toric Calabi--Yau threefolds have been described by bipartite graphs on a torus (dimer models).
We use the theory of dessins d'enfants to describe these in terms of triples of permutations which multiply to one.
These permutations yield an elegant description of zig-zag paths, which have appeared in characterizing the toroidal dimers that lead to consistent SCFTs.
The dessins are also related to Belyi pairs, consisting of a curve equipped with a map to $\mP^1$, branched over three points on the $\mP^1$.
We construct explicit examples of Belyi pairs associated to some CFTs, including $\IC^3$ and the conifold.
Permutation symmetries of the superpotential are related to the geometry of the Belyi pair.
The Artin braid group action and a variation thereof play an interesting role.
We make a conjecture relating the complex structure of the Belyi curve to $R$-charges in the conformal field theory.

\newpage

\setcounter{footnote}{0}

\tableofcontents

\section{Introduction}

The AdS/CFT correspondence~\cite{Maldacena:1997re, Gubser:1998bc, Witten:1998qj} provides a bridge between gravitational physics and gauge theory.
First formulated as the statement that the maximally supersymmetric four-dimensional Yang--Mills theory and type IIB string theory on the $\AdS{5}\times S^5$ are dual, it was later understood that more generic versions, in particular with less supersymmetry, can be constructed if the internal space is replaced by some other positive curvature manifold.
To be precise, one can consider $N$ D$3$-branes spanning the Minkowski part of $\mathbb{R}^{1,3}\times \mathcal{M}$, where $\mathcal{M}$ is a non-compact conical toric Calabi--Yau threefold.
At the singular point of the CY$_3$ moduli space, when the CY$_3$ is $\mathbb{R}^+\times \mathcal{B}$ ($\mathcal{B}$ being the base of the cone), the low-energy theory on the D$3$-branes flows in the infrared (IR) to a
four-dimensional superconformal field theory (SCFT).
On the other hand, through the AdS/CFT correspondence, this IR SCFT must be holographically dual to type IIB supergravity on $\AdS{5}\times \mathcal{B}$.
The first such example was the celebrated Klebanov--Witten theory~\cite{Klebanov:1998hh}, which arises in the near-horizon limit of a stack of D$3$-branes probing a conifold singularity.
In~\cite{Benvenuti:2004dy} the explicit field theory duals for D$3$-branes probing generic cones over $Y^{p,\, q}$ manifolds were proposed, while in~\cite{Benvenuti:2005ja, Butti:2005sw, FHMSVW05} the examples of the duality were further extended to the $L^{p,\, q,\, r}$ spaces.

Understanding the details of the IR SCFT dual to the gravity background represents {\em a priori} an enormous challenge.
However, in the case of toric CY$_3$, due to tremendous progress in the last five years ({\em e.g.},~\cite{FHMSVW05, Hanany:2005ve, FHKVW05, hananyvegh, Franco:2006gc, Feng:2005gw}), it has been shown that these theories can be encoded in bipartite tilings of a torus, also frequently called a {\em dimers} in this context (for detailed reviews see~\cite{Kennaway:2007tq, yam08}).
In a few words, a dimer on the torus is a graph drawn on ${\mathbb T}^2$ consisting of two types of vertices (distinguished, for example, by black or white coloring) and edges that each connect a black vertex to a white vertex.
Similar bipartite constructions have appeared in statistical physics and mathematics (see, {\em e.g.},~\cite{Kenyon2}).
The dimer can be thought of as the dual graph to a periodic quiver.
The faces of the dimer represent the {\em equal rank} $SU(N)$ gauge groups of the SCFT.\footnote{
The addition of fractional branes and the extension of our discussion to non-conformal theories, while very interesting, is beyond the scope of this paper.
Thus, from now on equal $SU(N)$ gauge factors will be assumed.}
The edges represent the fields of the theory, which are in the bifundamental representation of the two adjacent faces (gauge groups) they separate.
The edges meet at the nodes, which then naturally encode the interactions as superpotential terms.
Encircling the black nodes in a clockwise manner supplies an orientation for the edges,
 which is kept fixed and  induces the
anti-clockwise  encircling  for the white nodes.
The black nodes correspond to positive terms $W_+$ in the superpotential $W$ while the white ones correspond to negative terms $W_-$.
In this way, the construction ensures that each field appears precisely two times in $W$, once with each sign.
Thus the dimer captures the toric character of the theory.

The origin of dimers is understood from several perspectives.
As described in~\cite{FHKVW05}, two T-dualities along the appropriate $U(1)^2$ subgroup of the $\mathbb{T}^3$ fiber of the Calabi--Yau when regarded as a toric variety yield a certain five-brane system  (see, {\em e.g.}, the review~\cite{yam08} and references therein).
The distribution of the five-branes on the $\mathbb{T}^2$ is then intrinsically related to the dimer.
An alternative perspective, developed in~\cite{Feng:2005gw}, emerges when one considers the type IIA mirror of the original system of D$3$-branes probing the CY$_3$.
The D$3$-branes are mapped into D$6$-branes wrapping three-cycles.
These three-cycles intersect in a given manner, and a direct connection to the dimer appears.

In this paper we will explore a fascinating interplay between dimer models and the theory of {\em dessins d'enfants} and {\em Belyi pairs}.\footnote{
For earlier work along these lines, see \cite{Stienstra:2007dy}.
Dessins and Belyi pairs have also appeared in string theory in the context of Seiberg--Witten curves for $\mathcal{N}=2$ theories~\cite{Ashok:2006du, ACD06} and Matrix Models~\cite{BI,Looijenga,dMR}.}
A dimer, being a bipartite graph on a torus, is a {\em dessin d'enfant}~\cite{grotesquisse}.
(See the compilation in~\cite{schnepsbook}.)
A well-known combinatoric description characterizes any dessin in terms of three permutations which multiply to one.
These are permutations of natural numbers $1, \ldots, d $, where $d$ is the number of edges of the bipartite graph, {\em i.e.}, we have three permutations in $S_d$.
The condition of three group elements multiplying to one also defines the fundamental group of a sphere with three punctures.
This shows, using standard facts from covering space theory, that the combinatoric data of a dimer determine a unique (up to equivalences defined later) holomorphic map from the torus ${\mathbb T}^2$ to a projective space $\mathbb{P}^1$  with three marked points, which are branch points of the map.
This means that any point on the torus where the derivative of the holomorphic map vanishes is in the inverse image of one of these three points.
These points can be chosen to be fixed at $0,1,\infty$.
From the holomorphic map, the dessin d'enfant or dimer can be reconstructed.
The black vertices are inverse images of $0$, the white vertices are inverse images of $1$, and the edges are the inverse images of the $[0,1]$ interval.
The graph divides the torus into faces, each of which contains one inverse image of $\infty$.

Holomorphic maps from a Riemann surface to $\mP^1$, branched over three points, play a distinguished role due to a result of Belyi~\cite{Belyi} which provides a deep connection between these holomorphic maps and Riemann surfaces which can be defined by means of algebraic equations with coefficients living in $\bmQ$, the field of algebraic numbers.
These are numbers that arise as solutions to polynomial equations with rational coefficients.
This has generated a substantial interest, in the mathematics literature, in explicit constructions of these holomorphic maps, in terms of {\em Belyi pairs}.
A Belyi pair consists of a curve, in our case of toroidal topology, defined by some algebraic equations and a map, also given by algebraic equations;
so that all the numbers appearing in these equations are algebraic numbers (see, for example,~\cite{schnepsbook}).
Unfortunately, given a dessin, there is no general algorithm for writing the equations for the curve and the map.
Belyi's theorem only assures the existence of such a map.

In this paper we set up the study of the implications of the Belyi construction for AdS/CFT arising from conical toric CY$_3$.
We find that the tools developed in the analysis of dimers, such as zig-zag paths~\cite{FHMSVW05, Hanany:2005ve, FHKVW05, hananyvegh, Franco:2006gc, Feng:2005gw}, are naturally incorporated in an efficient way in the combinatorial framework of permutation triples.
The Belyi pair provides a geometric realization of many aspects of the SCFT, in particular discrete permutation symmetries of the superpotential which are realized as automorphisms of algebraic curves.

It is natural to ask about the physical meaning of the Belyi construction.
Our strategy to approach this question is to focus, in the first instance, on the complex structure of the Belyi curve.
Indeed, the Belyi pair determines a unique complex structure, which we denote as $\tau_B$, on the torus supporting the Belyi map.
On the other hand the dimer can be drawn on an arbitrary torus.
It is particularly interesting to consider the dimer drawn in an isoradial embedding, where one considers marking the center of each face and drawing the dimer such that all nodes surrounding a center lie upon a circle of unit radius~\cite{hananyvegh}.
Isoradial embeddings have also previously appeared in the mathematical literature (see, {\em e.g.},~\cite{Kenyon1}) related to statistical mechanics.
While it is not understood how this prescription might arise --- if, indeed, it does --- from the available brane constructions, it turns out that, remarkably, one can encode further details of the IR superconformal fixed point by drawing the edges of the dimer as straight lines and fixing the angles between the edges meeting at a node in terms of the $R$-charges~\cite{hananyvegh}.
At the superconformal fixed point, there is a unique $U(1)_R$ which sits in the same supermultiplet as the stress-energy tensor.
Indeed, the $R$-charges of fields (under this particular $U(1)_R$) are directly related to the scaling dimension of the chiral operators in the theory.
As is well-known, the corresponding $R$-charges can be determined by $a$-maximization~\cite{Intriligator:2003jj}, which, from this point of view, selects from the moduli space of isoradial dimers the $R$-charges in the SCFT.
In particular, the $R$-charges fixed by the $a$-maximization procedure arise from a system of quadratic algebraic equations, and therefore are {\em algebraic numbers}.
In turn, through the construction in~\cite{hananyvegh}, the dimer drawn according to this prescription determines a particular complex structure $\tau_R$ for the torus.

The combinatoric data of a toroidal dimer determine {\em two} complex structures:
on one hand the Belyi complex structure $\tau_B$ of the curve supporting the Belyi map, and on the other hand the $\tau_R$ determined by the $R$-charges.
We investigate the relation between these two complex structures and find two infinite families (including orbifolds) where $\tau_B = \tau_R$.
{\em This motivates the proposal that $\tau_B =\tau_R $ for all dimers associated with toric SCFTs.}
While at the moment a proof of this proposal remains elusive, it is supported by explicit computation in the available examples and it has interesting implications for algebraic values of the Klein $j$-function at transcendental arguments.
This is a fascinating avenue for future research.

The structure of the paper is as follows.
In Section~2, we introduce the combinatorial description of the dimer in terms of a triple of permutations.
By the Riemann existence theorem, this leads to a holomorphic map from the torus ${\mathbb T}^2$ to a sphere $\mathbb{P}^1$ branched over three points.
We briefly review the connection to algebraic numbers by Belyi's theorem.
In Section~3, we explore the (discrete)  symmetries of the superpotential arising from permutations of
the chiral  fields, which are related to symmetries of the dimer, in terms of the associated permutation triple and Belyi pair.
We also  consider the operation of twisting which keeps fixed the permutation around the black vertices and inverts the permutation around the white vertices.
This provides the combinatoric data for another Belyi pair, which we will denote as as the untwisted Belyi pair.
The source of the holomorphic map here is a curve $\Sigma $ in the mirror geometry of the CY$_3$ which has been called the ``untwisted'' curve~\cite{Stienstra:2007dy,Feng:2005gw}.
In Section~4, we illustrate these ideas with simple examples:
$\cN=4$ SYM, the conifold, the del Pezzo zero ($dP_0$), and the suspended pinch point (SPP).
In Section~5, we discuss orbifold theories.
In Section~6, we introduce the zig-zag paths, which have played an important role in AdS/CFT literature in constructing the data of the toric diagram from the dimer.
They have a particularly simple combinatorial description in terms of permutations.
We give a simple algorithmic formulation of the condition that zig-zags do not self-intersect in terms of strings of permutations.
This is one of the consistency conditions on dimers, which mean that not all toroidal dimers lead to consistent SCFTs.
In Section~7, we develop the proposal of the equality of the complex structure parameter of the Belyi and the $R$-charge tori.
In Section~8, we summarize our results and discuss future directions.
The Appendices collect background information and additionally present the orbifold constructions in greater detail.

\section{Toric CFTs, dimer models, and permutations}\label{sec:dimerperms}

Since toric CFTs have been described by dimers,\footnote{
We give a lightning review of the use of dimers in characterizing SCFTs in Appendix~\ref{nutshell}, and refer to, {\em e.g.},~\cite{Kennaway:2007tq, yam08} for a thorough introduction.}
{\em i.e.}, bipartite graphs on a torus, we will start with describing general bipartite graphs on Riemann surfaces $\Sigma_h$, which are also known as dessins d'enfants.
We will explain their description in terms of triples of permutations and to algebraic numbers via Belyi theory.

\subsection{Dessins and permutations}\label{DesPerm}

The bipartite condition simply means that the vertices (or nodes) in the graph can be divided into two disjoint sets such that lines connect vertices in one set to vertices in the other, and that no lines connect vertices within the same set.
We color the vertices in the first set black and the vertices in the second set white. The lines, which are the edges of the graph, separate the faces, which are topological disks. This graph is called a {\em dessin d'enfant}, French for ``children's drawing.''

Given a bipartite graph with $d$ edges on a Riemann surface, we can label the edges $1, \ldots, d$.
Choosing an orientation on the Riemann surface, we traverse the edges that are incident on the black vertices in a direction compatible with the orientation.
This enables us to construct a permutation in the symmetric group of $d$ elements, $\sgb \in S_d$. Traversing the edges incident on the white vertices in the same way, we construct a second permutation $\sgw \in S_d$.
It is useful to define $\sgi = (\sigma_B\, \sigma_W)^{-1}$, so that we have
\bea
\sgb\, \sgw\, \sgi = 1 ~.
\eea
This can be recognized as a sequence of three permutations which obey the same relation as the generating elements of the fundamental group of a sphere with three punctures.
This information, by covering space theory (see a physics review with mathematical references in~\cite{CMR}), is exactly what is required to describe a unique holomorphic branched cover of the sphere of degree $d$, with three fixed branch points.
These can be chosen to lie at $0, 1, \infty$.
Each cycle of $ \s_i $ ($ i \in \{ B , W , \infty \} $ ) corresponds to a point in the inverse image, respectively, of $ \{ 0 , 1, \infty \} $.
A cycle of length $n$ corresponds to a point where the map locally looks like $ w = z^n $, with $w$ being a local coordinate on the target $\mP^1$ and $z$ a local coordinate on the covering Riemann surface $\Sigma_h$.
If $n >1$, the point on $\Sigma_h$ is said to be a {\em ramification point} and $n$ is called the {\em ramification index}. The derivative of 
the map, calculated using the local coordinates, vanishes at a ramification 
point.  
In the current terminology, which is frequent but not universal in the literature on branched covers, the points $ \{ 0 , 1, \infty \} $ on $\mP^1$, which are images of any point where the derivative of the map vanishes, are called branch points.
The points on $ \Sigma_h$ where the derivative vanishes are ramification points.
The branch points are also often called critical values of the map while the ramification points are called critical points.

Note that the permutation $\sigma_{\infty}$ contains information about the faces of the dimer.
Indeed, the number of cycles in $\sigma_{\infty}$ counts the number of faces, which is in turn the number of gauge groups of the SCFT.
We will elaborate further on the description of the faces in terms of permutations in Section~\ref{consdimzig}.
It turns out that a permutation in $ S_{2d} $, related to $ \s_{\infty}$, contains a precise description of the faces.

\subsection{Riemann existence theorem, automorphisms, and Belyi pairs}
\label{RetAutBel}

The one-to-one equivalence between permutation triples and holomorphic maps holds with the understanding of equivalences on both sides.
This is the content of a deep and powerful theorem called the {\bf Riemann Existence Theorem}, discussed at length, for example, in~\cite{lando-zvonkin}.
Two permutation triples $\{\sigma_B,\,\sigma_W,\,\sigma_{\infty}\}$ and $\{\sigma'_B,\,\sigma'_W,\,\sigma'_{\infty}\}$ define equivalent holomorphic maps $\beta$, $\beta'$ if
\be
\s_B^{\prime} = \gamma\, \s_B\, \gamma^{-1} ~, \qquad \s_W^{\prime} = \gamma\, \s_W\, \gamma^{-1} ~,
\ee
for some $\gamma\in S_d$. Two such holomorphic maps $\beta, \beta^{\prime}$ are equivalent if there is a holomorphic one-to-one map $\phi$ from the source to itself such that $\beta^{\prime} = \beta \circ \phi$, or, in other words, if the following diagram commutes.
\begin{equation}
\begin{array}{c c c}
& \phi & \\
\Sigma_h & \longrightarrow & \Sigma_h\\ & & \\
\beta'\,\searrow & &\swarrow \, \beta\\ & &\\
& \mP^1 &
\end{array}
\end{equation}

It is appropriate, at this point, to consider permutations $\gamma$ that leave the pair $(\sgb, \sgw)$ fixed, {\em i.e.},
\be\label{autcomb}
\gamma\, \sgb \, \gamma^{-1} = \sgb ~, \qquad
\gamma \, \sgw\, \gamma^{-1} = \sgw ~.
\ee
These define a group that we denote ${\rm Aut}(\sgb, \sgw)$.
This can be identified with the {\em automorphisms of the pair} $(\Sigma_h, \beta)$ denoted as ${\rm Aut}(\Sigma_h, \beta)$, which are holomorphic one-to-one maps $\phi: \Sigma_h \rightarrow \Sigma_h$ such that $ \beta \circ \phi = \beta $.
The group ${\rm Aut}(\Sigma_h, \beta)$ is a subgroup of the group  ${\rm Aut}(\Sigma_h)$ of all one-to-one holomorphic maps from $\Sigma_h$ to $\Sigma_h$.
For the case $h=1$ case, $\Sigma_h$ is an elliptic curve, and much is known about this group, as we describe in Appendix~\ref{ellipticcurves}.
When $h=0$, and $\Sigma_h$ is a sphere, the automorphism group is $PSL(2, \IC)$, which is the group
of M\"obius transformations
\bea
x \rightarrow {( ax + b ) \over ( cx + d ) }
\eea
with $ ad - bc = \pm 1 $. For $\Sigma_h$ of higher genus, the order of the group is bounded by $84(h-1)$~\cite{griffhar}.
The direct connection between the combinatorially defined automorphism group ${\rm Aut}(\sgb, \sgw)$ and the analytically defined ${\rm Aut}(\Sigma_h, \beta)$ will lead to a geometrical formulation of symmetries of the superpotential in Section~\ref{PermW}.

For a covering space of genus $h$, we have
\bea\label{riemhur}
2h-2 = d - C_{\sgb} - C_{\sgw} - C_{\sgi} ~,
\eea
where $C_\sigma$ is the number of cycles in $\sigma$ and $d$ is the degree of the map.
This is a special case of the Riemann--Hurwitz relation, which says in general, for a map
\bea
\beta: \Sigma_h \rightarrow \Sigma_G
\eea
from a Riemann surface of genus $h$ to one of genus $G$ and branching number $B$, the following equation is satisfied
\bea
2h-2 = d\, (2G-2) + B ~.
\eea
Let us specialize to the case $h=1$, or $\Sigma_h = \mathbb{T}^2$, and $G=0$, corresponding to the target $\mP^1$.
The torus condition is expressed as
\bea\label{genonecond}
2h-2 = 0 = -2d + (d - C_{\s_B}) + (d - C_{\s_W}) + (d - C_{\s_{\infty}}) = d - C_{\sgb} - C_{\sgw} - C_{\sgi} ~.
\eea
The branching number $B$ is given as
\be
B = \sum_{i\in \{ B,W,\infty\} } (d - C_{\s_i}) ~.
\ee

The above discussion connecting triples of permutations to holomorphic covering maps branched over three points generalizes to arbitrary tuples of $L$ permutations corresponding to holomorphic covers branched over $L$ points.
The case of three permutations has particular interest because of Belyi's theorem~\cite{Belyi}.
It states that if a Riemann surface admits a map to $\mP^1$ branched over three points on the $\mP^1$, then it can be described by algebraic equations involving only algebraic number coefficients.
Algebraic numbers live in the field $\bmQ$ and are all possible solutions of polynomial equations with coefficients in $\mQ$.\footnote{
We recall that almost all real numbers are not algebraic, {\em i.e.}, they are transcendental.}
Conversely, if a Riemann surface can be defined over $\bmQ$, then it admits a map to $\mP^1$ branched over three points.
A Belyi pair is an explicit realization of the holomorphic curve $\Sigma_h$ and map $\beta$ in terms of algebraic equations involving algebraic numbers.
The absolute Galois group ${\rm Gal}( \bmQ/ \mQ ) $ acts as a symmetry of the field of algebraic numbers, which preserves the rational numbers.
Grothendieck~\cite{grotesquisse} observed that the 
action of ${\rm Gal}( \bmQ/ \mQ ) $ on the Belyi pair induces a faithful action on dessins, which were thus a simple combinatoric structure containing information about a mysterious group of deep number theoretic importance.

To summarize, the data of a bipartite graph on the Riemann surface $\Sigma_h$ specify a meromorphic function from $\Sigma_h$ to $\mP^1$ with three branch points, chosen to be $0,1,\infty$.
This is the {\em Belyi map}, which we label as $\beta$.
The source Riemann surface together with the map $(\Sigma_h, \beta)$ 
is called the {\em Belyi pair}.
From the surface $\Sigma_h$ and the map $\beta$, 
we can construct the dessin d'enfant as follows.
We associate to the points $\beta^{-1}(0)$ the black nodes, 
and to the points $\beta^{-1}(1)$ the white nodes.
The edges are the inverse images of the interval $[0,1]$.
Points which are the inverse image of infinity correspond to 
the faces of the dessin.

In this way, all the information about the bipartite graph is 
captured by the {\em Belyi pair} $(\Sigma_h, \beta)$.
Unfortunately, there is no known general algorithm which starts 
with  $\Sigma_h$ and the bipartite graph and yields the algebraic 
equations for $ \Sigma_h $ and the map $\beta$.
The Belyi pair $ ( \Sigma_h , \beta ) $ has generally to be 
constructed on a case-by-case basis, although some general 
procedures are known for genus zero covers.

\subsection{The basic example: $\cN=4$ SYM  and $ \IC^3$ }

\begin{figure}
\begin{center}
\includegraphics{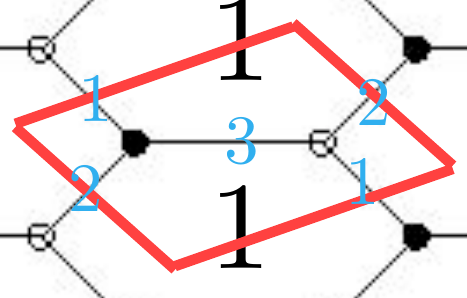}
\end{center}
\caption{Dimer for the $\mathcal{N}=4$ SYM theory.}
\label{C3dimer}
\end{figure}

In order to make our discussion more concrete, let us consider the simplest example, namely D$3$-branes transverse to
$ \mathbb{C}^3$.
The worldvolume theory on the branes is $\cN=4$ SYM, which is encoded in the dimer shown in Figure~\ref{C3dimer}.

We see that there are three fields corresponding to the edges labelled as $1$, $2$, $3$.
As there is one face corresponding to the $SU(N)$ gauge group, all these fields $\Phi_i$ are in the adjoint representation ({\em viz.}\ $(N, \overline{N})$).
Circling the black node anticlockwise gives  the ${\rm Tr} (X_1  X_2 X_3)$ interaction.
Circling the white node clockwise, we find the ${\rm Tr} (X_1 X_3 X_2) $ interaction.
Thus, the superpotential is
\be\label{spotC3}
W = {\rm Tr}\ ( X_1 X_2  X_3 -  X_1 X_3 X_2 ) ~.
\ee

Following the procedure explained at the beginning of Section~\ref{DesPerm}, from Figure~\ref{C3dimer}, we can read off the permutation structure
\begin{equation}
\label{sigmaC3}
\sigma_B=(123) ~, \qquad
\sigma_W=(123) ~, \qquad
\sigma_{\infty}=(\sigma_B\sigma_W)^{-1} = (123) ~.
\end{equation}
Each of these permutations consists of a single cycle.
The fact that the dessin lives on a genus one surface follows immediately from the cycle structures of the three permutations using the Riemann--Hurwitz formula (\ref{genonecond})
\be
2h-2 = 0 = d - C_{ \s_B }  - C_{\s_W }  -  C_{ \s_{\infty} } = 3 -1 -1 -1  ~.
\ee
Given the connection between dessins and branched covers explained in Sections~\ref{DesPerm} and \ref{RetAutBel}, we infer that this dessin corresponds to a map of degree three, with three ramification points of degree three each.

In the Weierstrass parametrization, consider a torus given by the defining equation
\begin{equation}
y^2=x^3+1 ~.
\label{eq:elcurve}
\end{equation}
As we review in Appendix B.1, the $j$-function classifies elliptic curves up to isomorphism.
This torus has $j$-invariant $j(\tau)=0$.
Inverting this, $j^{-1}(0) = e^{\pi i/3} = \frac12 + i \frac{\sqrt{3}}{2}$ gives the modular parameter of the torus.

In the case of the $\IC^3$ theory, let us consider 
\begin{equation}
\beta=\frac{y+1}{2} ~.
\label{eq:beln4}
\end{equation}
 The derivative of the map, defined with respect to local coordinates, 
 vanishes at the ramification points. The images of these ramification 
points are $ \{ 0 , 1, \infty \} $. 
Thus, the $\beta$ defined in~\eref{eq:beln4} is a Belyi map.
The Belyi pair consisiting of~\eref{eq:elcurve} and~\eref{eq:beln4} has previously been given in~\cite{khadscha}.

To demonstrate explicitly that this is indeed a Belyi pair, let us examine the preimages of the relevant points.
\bi
\item $\beta^{-1}(0)$:
the preimage of $0$ is $y=-1$ corresponding to the point $(x,y) = (0,-1)$ on the elliptic curve.
Close to this point, we write $(x,y)=(\epsilon,-1+z)$ so that
\be
(-1 + z)^2 = \epsilon^3 + 1 ~,
\ee
or $-2z=\epsilon^3$.
A good local coordinate is then $x=\epsilon$, $z=-\frac12\epsilon^3$.
Thus, locally the map $\beta$ is
\be
\beta(-1+z) = -\frac{\epsilon^3}{4} ~.
\label{eq:17}
\ee
On the target $\mP^1$, $w=\beta(-1+z)-\beta(-1) = -\frac{\epsilon^3}{4}$ is a good local coordinate.
The derivative $\pa_\epsilon w$ vanishes at $\epsilon=0$.
The cubic power in~\eref{eq:17} indicates a ramification point with ramification index $3$ and corresponds to a black node with three incident edges in the associated dessin.
\item $\beta^{-1}(1)$:
the preimage of $1$ is $y=1$ corresponding to the point $(x,y) = (0,1)$ on the elliptic curve.
Close to this point, we write $(x,y)=(\epsilon,1+z)$ so that
\be
(1 + z)^2 = \epsilon^3 + 1 ~,
\ee
or $2z = \epsilon^3$.
A good local coordinate is then $\epsilon$ where 
 $x=\epsilon$, $z=\frac12 \epsilon^3$.
Using the map $\beta$
\be
\beta(1+z) = 1+\frac{\epsilon^3}{4}
\label{eq:19}
\ee
we find the map in terms of the  local coordinate 
$w = \beta(1+z) - \beta(1) $ on $\mP^1$ as  
$ w =  \frac{\epsilon^3}{4}$.
Again, the derivative $\pa_\epsilon w$ vanishes at $\epsilon=0$.
Since the exponent in~\eref{eq:19} is $3$, we have another ramification point with ramification index $3$, this one corresponding to a white node with three incident edges in the dessin.
\item $\beta^{-1}(\infty)$:
the preimage of infinity is the point infinity on the elliptic curve.
There, locally, the torus looks like $y^2=x^3$, so the appropriate local coordinate is $y=\epsilon^{-3}$, $x=\epsilon^{-2}$.
Thus, near $\beta \sim \infty$, 
\be
\beta  \simeq \frac{1}{2\epsilon^3} ~.
\ee
On the target $\mP^1$, the local coordinate $w$ around this point is
\be
w = \frac{1}{\beta} \simeq 2\epsilon^3 ~.
\label{eq:20}
\ee
The derivative of the local coordinate $w$ with respect to $\epsilon$ vanishes at $\epsilon=0$.
We observe from~\eref{eq:20} that the ramification index of the point at infinity is $3$.
\item $\beta^{-1}(\frac12)$:
Let us consider a point in $\mP^1$, which is not in the set $\{ 0, 1 , \infty \}$.
For simplicity, we choose $\beta=\frac12$ corresponding to $(\sqrt[3]{-1}, 0)$, though any other point works as well.
Around $(-1,0)$, we can write the elliptic curve as
\be
z^2 = (-1 + \epsilon)^3 + 1 ~,
\ee
or $z^2 = 3\epsilon$.
Close to this point, a good local coordinate is $\epsilon = \varepsilon^2$, $z=\sqrt{3}\,\varepsilon$.
Thus,
\be
\beta(z) = \frac{1+\sqrt{3}\,\varepsilon}{2}
\ee
describes a local coordinate $w = \beta(z) - \beta(0) = \frac{\sqrt{3}}{2}\,\varepsilon$.
The derivative $\pa_\varepsilon w$ is non-vanishing.
This point is unramified.
\ei

Summarizing, we have a Belyi map from $\mathbb{T}^2$ into the marked $\mathbb{P}^1$ such that the preimages of $0$ and $1$ are a single point each, both with ramification $3$, and the preimage of infinity is again a single point with ramification $3$.
With these ramification data,  the branching number is $B=2+2+2=6$.
The Riemann--Hurwitz formula tells us that
\be
2h-2 = 0 = d \,(2G-2) + B = 3\times -2 + 6 ~.
\ee

\section{Permutation symmetries of superpotential and Automorphisms of Belyi curves}\label{PermW}

The dimer encodes the superpotential terms (see the brief discussion in Appendix~\ref{nutshell}).
We will write $ W = W_+ - W_-$ where $ W_+, W_- $ are both sums of positive monomials in the fields.
Fields are associated with labelled edges.
Labelling the edges $1, 2, \ldots, d$, we have fields $ X_1, X_2, \ldots, X_d$.
A black vertex with $k$ edges labelled $ (i_1 i_2 \ldots i_k )$ read according to the orientation of the torus gives a monomial ${\rm Tr} X_{i_1} X_{i_2} \ldots X_{i_k } $ in $W_+ $ and a cycle $( i_1 i_2 \ldots i_k ) $ in $\sgb$.
A white vertex with $l$ edges labelled $ ( j_1 j_2 \ldots j_l ) $ read according to the orientation of surface leads, on the contrary, to a cycle $ (j_1 \ldots j_{l-1} j_l ) $ in $\sgw$ and a monomial ${\rm Tr} ( X_{j_l} X_{j_{l-1}} \ldots X_{j_1} )$ in $W_-$.
In other words the monomials in $W_+ $ are read off from the cycles of $ \sgb $ whereas the monomials in $W_-$ are read off from the cycles of $ \sgw^{-1} $.
These remarks have been illustrated above using the dimer for the $\mathcal{N}=4$ theory.

The same data of the SCFT gauge groups, matter content, and superpotential can be encoded alternatively in the permutation triple
\bea
\{\widetilde{\sigma}_B\, , \widetilde{\sigma}_W\, , \widetilde \s_{ \infty} =
 ( \widetilde{\sigma}_B\,  \widetilde{\sigma}_W )^{-1}  \}
= \{ \s_B , \s_{W}^{-1} ,   \s_W\, \s_B^{-1}  \}
\eea
as discussed previously in~\cite{Stienstra:2007dy}.
This in turn leads to a branched cover $ \widetilde \beta: \Sigma \rightarrow \mP^1 $ with three branch points at $ \{ 0 , 1, \infty \} $.
Following the literature~\cite{Feng:2005gw} we will call $ ( \Sigma , \widetilde \beta )$ the untwisted Belyi pair.
So we start with the standard dimer, dual of the periodic quiver, which lives on a torus $\mathbb{T}^2$, to which we have associated a dessin and the permutation triple $ ( \s_B , \s_W , \s_{ \infty} =  ( \s_B\,  \s_W)^{-1} ) $.
The permutation triple $  \{\widetilde{\sigma}_B\, \widetilde{\sigma}_W  , \widetilde \s_{\infty} \} $ defines, by the usual association of permutation triples to dessins described in Section~\ref{DesPerm}, a dessin on the untwisted curve $ \Sigma$, which needn't be a torus.
Indeed, the genus of the untwisted curve is given by the number of four-cycles in the Calabi--Yau geometry (in other words, by the number of internal points in the toric diagram) transverse to the D$3$-branes.
The genus can also be computed by using the Riemann--Hurwitz relation (\ref{riemhur}) applied to the untwisted triple $( \widetilde \s_B , \widetilde  \s_W , \widetilde \s_{\infty} )$.

\subsection{Permutation symmetries fixing $W_+ $ and $W_-$}
\label{permfix}

In the following, we will consider symmetries of the superpotential which act by permuting the $d$ fields $X_1 , \ldots , X_d$.
Among such permutation operations, those which leave fixed the sum of terms  $W_+$ as well as  the sum of terms  $ W_- $ form a group we call ${\rm Sym}  (W_+ , W_-) $.
Note that these symmetries do not necessarily fix the monomials in $W_+$ (or $W_- $) individually.
We will find various constraints on ${\rm Sym}  (W_+ , W_-) $ using the geometry of Riemann surfaces.

We will derive the following result
\bea\label{firstres}
{\rm Sym} (W_+ , W_-) = {\rm Aut} (\mathbb{T}^2 , \beta) = {\rm Aut} (\Sigma , \widetilde \beta) ~,
\eea
which implies that
\bea
&& {\rm Sym}  (W_+ , W_-) \subset {\rm Aut} (\mathbb{T}^2) ~, \cr
&& {\rm Sym}  (W_+ , W_-) \subset {\rm Aut} ( \Sigma) ~.
\eea
This will rely on the Riemann existence theorem, which, as explained in Section~\ref{RetAutBel} gives the combinatoric formulation of the geometrical symmetries in (\ref{firstres}):
\bea
&& {\rm Aut} (\sigma_B , \sgw) = {\rm Aut} (\tsgb , \tsgw) ~, \cr
&& {\rm Aut} (\mathbb{T}^2 , \beta) = {\rm Aut} (\Sigma , \widetilde \beta) ~.
\eea

The group  $ {\rm Aut} (\sgb , \sgw) $ consists of permutations $ \gamma$ leaving fixed the pair  $ \s_B , \s_W $ under the action of conjugation.
The permutation $ \gamma $ acting by conjugation on a cycle $ ( i_1\, i_2\, \ldots\, i_k ) $
\bea
\gamma (i_1 \ldots i_k) \gamma^{-1} \rightarrow
(\gamma ( i_1 ) \ldots \gamma (i_k)) ~.
\eea
Using the correspondence between cycles of $\sgb$ and monomials in $W_+$, we see that the monomial $ {\rm Tr} ( X_{i_1}  X_{i_2} \ldots X_{i_k}) $ goes to $ {\rm Tr} (X_{\gamma (i_1)} \ldots X_{\gamma (i_k)}) $.
Now, the condition that $ \gamma \in {\rm Aut} (\sgb , \sgw) $ means that the set of cycles of $ \sgb $ goes back to the same set of cycles, and the set of cycles of $\sgw $ goes back to the same set of cycles.
Note that this does not necessarily mean that the individual cycles are fixed.
To see the connection to $ {\rm Aut} (\Sigma , \widetilde \beta) $, we use the fact that if $ \gamma\, \sgw\, \gamma^{-1} = \sgw $, then $ \gamma\, \sgw^{-1} \,\gamma^{-1} = \sgw^{-1} $.
This completes the proof of (\ref{firstres}).

For  the $\mathcal{N}=4$ theory, the group is generated by $(123)$ and its powers, thus forming a $\mathbb{Z}_3$.
In terms of the fields, this permutation translates into the action
\bea
X_1 \rightarrow X_2 ~, \qquad
X_2 \rightarrow X_3 ~, \qquad
X_3 \rightarrow X_1 ~.
\eea
Making use of the explicit form of the $\mathcal{N}=4$ SYM superpotential (\ref{spotC3}), it is clear that these transformations do indeed leave invariant $W_{\pm}$.

In our discussion, we have assumed that the couplings of the 
terms in the superpotential are all equal to plus or minus $1$. 
This guarantees that the moduli space is a toric Calabi--Yau. 
These couplings can be adjusted while preserving supersymmetry, 
but typically global symmetries may be broken and there are
also constraints due to conformal invariance. 
Our considerations of automorphisms of the Belyi pair 
in connection with symmetries of the superpotential 
apply, in simplest form, to the case where all the couplings 
are $\pm 1 $, but should admit extension to the more general case. 
We leave the elaboration of this for the future.

\subsection{Symmetries of exchange of $ W_+$ and $ W_-$}
\label{secartin}

Having related the superpotential terms $ W_+ , W_- $ to permutations $ \sigma_B , \,\sigma_W ,\, \sigma_{\infty } $, we will turn to a discussion of symmetries which exchange $ W_+ $ and $W_-$.
This will require, in some sense, exchanging $ \s_B $ and $ \s_{W}^{-1} $.
The exchange has to be formulated in a way that allows consistent action on the equation $\sigma_B \, \sigma_W \, \sigma_{ \infty } = 1 $.
We wish to develop, in analogy to Section~\ref{permfix}, an elegant geometrical picture for the symmetries which include this exchange.
We will not provide a definitive  answer in the form a simple geometric characterization to match the result (\ref{firstres})  of Section~\ref{permfix}.
But we will develop some arguments in this direction, supported by examples in later sections, which will hopefully  lead to  such a definitive answer in the near future.

 First, we note that, on the geometrical side the exchange of
 $ \s_B $ with an element in the conjugacy class of $ \s_W $ (which
 is the same as that of $ \s_W^{-1} $) would correspond to
 exchanging the branch points at $0$ and $1$.  On this geometrical side,
 there is no obvious need to  distinguish
 between exchanging the branch points $0$ and $1$
 with the operation of exchanging the branch points   $1$ and $\infty$.
 This naturally leads us to consider, on the same footing,
 all permutations of the three branch points.  The na\"{\i}ve $S_3$  action of
 permuting $ ( \s_B,\, \s_W ,\, \s_{ \infty} ) $ does not preserve
 the relation $ \s_B \,\s_W\, \s_{\infty} = 1 $. Under this na\"{\i}ve action,
 only the $\mathbb{Z}_3$ subgroup preserves the relation.

  Rather the correct framework, which in fact works for arbitrary
   sequences of permutations $ \s_1\, \s_2\, \ldots\, \s_k = 1 $
  is to consider the Braid group action \cite{lando-zvonkin}.
 The standard braid group relations are
\bea\label{basicbraid}
&& B_i \,B_{i+1}\, B_i = B_{i+1} \,B_i\, B_{i+1} ~, \cr
&& B_{i }\, B_j = B_j\, B_i ~~~ \hbox{for} ~~  |i-j| > 1 ~,
\eea
for $ i = 1, \ldots, {k-1} $. There is an
 action of $B_i$ (called the {\em Artin action}) on the permutations
which takes
\bea
&& \s_i \rightarrow \s_{i+1} ~, \cr
&& \s_{i+1} \rightarrow \s_{i+1}^{-1} \,\s_i \,\s_{i+1} ~,
\eea
and leaves the other permutations fixed. This action obeys the braid relations
and  preserves the product
 $ \s_1 \s_2 \ldots \s_k = 1 $. Defined as an action on conjugacy classes,
{\em i.e.}, where $ \gamma ( \s_1 ,\, \ldots ,\, \s_k ) \gamma^{-1 } $  is regarded as
equivalent to $ ( \s_1 , \,\ldots ,\, \s_k )$, one can show that there
is an extra relation
\bea\label{addrel}
B_1\, B_2\, \ldots\, B_k \,B_k\, B_{k-1}\, \ldots\, B_1 = 1 ~.
\eea
This, along with (\ref{basicbraid}), defines the spherical braid group
 $\mathcal{B}_k$.  For the case $k=3$, it is in fact true
the two generators also obey $ B_1^2  = B_2^2 = 1 $,
when acting on conjugacy classes.  In other words, using
\bea\label{artinact}
&& B_1 ( \s_1 , \s_2 , \s_3 ) = ( \s_2 , \s_2^{-1} \s_1  \s_2  , \s_3 ) ~, \cr
&& B_2 (  \s_1 , \s_2 , \s_3 ) = ( \s_1 , \s_3 , \s_3^{-1} \s_2 \s_3 ) ~,
\eea
we can show that
\bea\label{Artinsquare1}
B_1^2 ( \s_1 , \s_2 , \s_3 ) & = &  \s_3 ( \s_1 , \s_2 , \s_3 ) \s_3^{-1} ~, \cr
B_2^2 ( \s_1 , \s_2 ,  \s_3 ) & = &  \s_1 ( \s_1 , \s_2 , \s_3 ) \s_1^{-1} ~.
\eea
The braid relations along with $B_1^2 = B_2^2 = 1 $ in fact
define the symmetric group $S_3$. So, the permutation group
does act on the triples $ \s_1 \s_2 \s_3 =1 $ where $\s_i \in S_d $,
not in the obvious way, but as a degeneration of the spherical braid
group action for general $k$.

Given the Artin braid group action, specialized to $k=3$ (which
as we explained above is in fact also an $S_3$ action), we may consider
this action on the triple $ (\s_B , \s_W , \s_{ \infty} ) $
describing a toroidal dimer. It will in fact be
more fruitful, in terms of finding evidence for the geometric
realization $ W_+ , W_- $ exchange, to look at the Artin braid action on
the untwisted triple
\bea
\widetilde B_1  ( \widetilde \s_B  , \,\widetilde \s_W , \, \widetilde \s_{\infty} )
 =     ( \widetilde \s_W ,  \,\widetilde \s_W^{ -1}\,  \widetilde \s_B \,\widetilde \s_W , \,
          \widetilde \s_{\infty} ) ~, \cr
 \widetilde B_2 (  \widetilde \s_B  , \,\widetilde \s_W , \,\widetilde \s_{\infty} )
 =  ( \widetilde \s_B ,  \,\widetilde \s_{ \infty} ,  \, \widetilde \s_{ \infty}^{-1}\,
   \widetilde \s_W \, \widetilde \s_{\infty} ) ~.
\eea
We will find, in  the  few examples of Belyi pairs we have been able to
explicitly construct, that this Artin action on the untwisted
curve will have a geometric realization in terms of the
untwisted Belyi pair $ \widetilde \beta: \Sigma_h \rightarrow \mP^1$.

Before we get to geometric realizations, let us describe an additional
braid group action (which also degenerates to $S_3$) on the permutation triple.
To distinguish this action from the Artin, we will give the
$B_i$ a superscript, $I$, since it involves inversions
\bea\label{BiI}
B_1^{ I  } ( \s_B ,\, \s_W ,\, \s_{ \infty} ) =
( \s_W^{-1} ,\, \s_B^{-1} ,\, \s_{\infty}^{-1} ) ~, \cr
B_2^{I } ( \s_B ,\, \s_W ,\, \s_{ \infty} ) =  ( \s_B^{-1}  ,\, \s_{\infty}^{-1}  ,\,
\s_{W }^{-1} ) ~.
\eea
One checks that these actions preserve the relation $ \s_B \s_W \s_{\infty} =1 $
and that the braid group relation is satisfied
\bea\label{braidrels2}
B_1^I\,  B_2^I\,  B_1^I  = B_2^I\,  B_1^I\,  B_2^I ~,
\eea
as well as
\bea\label{Isquare1}
(B_1^I)^2 ( \s_B ,\, \s_W ,\,\s_{ \infty} ) = ( \s_B ,\, \s_W ,\,\s_{ \infty} ) ~, \cr
(B_2^I)^2 ( \s_B , \,\s_W ,\,\s_{ \infty} ) = ( \s_B , \,\s_W ,\,\s_{ \infty} ) ~,
\eea
 which implies the spherical braid group
relation, and in fact reduces the group to $S_3$.
This additional action will prove useful in our experimental
investigations of the geometric counterpart to $ W_+ , W_- $
exchange using the few toroidal Belyi pairs we have been able to
construct.

In  Section~\ref{braidgeom}, we will explain the geometrical meaning
of these actions, which we have found to hold in
all the explicit examples of Belyi pairs that we constructed.

\subsubsection{Braid group actions and geometry}
\label{braidgeom}

The exchange of $0 $ and $1$ on the target $\mP^1$
is performed by a holomorphic invertible map
in $ PSL(2 , \IC)$, which
 $\beta\rightarrow 1-\beta$. Let us suppose that the
corresponding action on $ \Sigma_h $ is a holomorphic
invertible map $B_1$ in $ {\rm Aut} ( \Sigma_h ) $. In later applications
$ \Sigma_h $ can be either $ \mT^2 $ supporting the toroidal dimer
or the untwisted curve $ \Sigma $.  $B_1$ maps a point $P$  on $ \Sigma_h $
to  $B_1( P )$
\begin{equation}
\label{1and0}
\beta( P )=1-\beta( B_1( P ) ) ~.
\end{equation}
We might now write $ P =B_1\circ B_1^{-1} ( P ) $.
To simplify the  notation, let us now call  $B_1^{-1}( P )=
P^{\prime} $. Then, we have
\begin{equation}
\beta(B_1 ( P^{\prime} ) )=1-\beta(B_1\circ B_1 (\Pp )) ~.
\end{equation}
We could now drop the primes  and use (\ref{1and0}), so that
\begin{equation}\label{B1square}
1-\beta( P )=1-\beta(B_1\circ B_1 (P )) \ \Longrightarrow \ \beta(P )=
\beta( B_1\circ B_1 (P )) ~.
\end{equation}
That is, $B_1\circ B_1=1$ up to automorphisms of the pair.
Comparing with \eref{Artinsquare1}, \eref{Isquare1}
this shows promise that there can be a  matching of geometric transformations
of $\Sigma_h $ implementing $ \beta \rightarrow ( 1- \beta ) $
with a corresponding braid group  action on permutation triples.

Turning to the exchange of  $1$ and $\infty $  in the target,
we consider  an element $B_2$  in $ {\rm Aut} ( \Sigma_h ) $,  mapping
$P $ to $ B_2 (P)$, and  implementing
  $\beta\rightarrow \frac{ - \beta }{1-\beta}$, so
\begin{equation}
\label{0andinfinity}
\beta( P )=\frac{ - \beta ( B_2 ( P )  ) }{1-\beta(B_2( P ))} ~.
\end{equation}
 We can then run the same operations as above to have that again $B_2\circ B_2=1$ up to automorphisms of the pair.

Since we have two operations, it is natural to consider
 $P =B_1\circ B_1^{-1}( P )=B_1( P^{\prime } )$,
 and substitute this into (\ref{0andinfinity}):
\begin{equation}
\beta(B_1( \Pp ))=\frac{- \beta (B_2\circ B_1 ( \Pp )    }
{1-\beta(B_2\circ B_1(  \Pp  ) )} ~.
\end{equation}
Using (\ref{1and0}) and dropping primes  we have
\begin{equation}
1-\beta( P )=\frac{ - \beta (B_2\circ B_1 ( P  )    }
{1-\beta(B_2\circ B_1( P ))} ~.
\end{equation}
 Setting  $P = B_2(\Pp  )$, we then obtain
\begin{equation}
1-\beta(B_2( \Pp ))=\frac{ - \beta (B_2\circ B_1\circ B_2 ( \Pp )  ) }{1-\beta(B_2\circ B_1\circ B_2(
 \Pp ))} ~,
\end{equation}
which, upon using (\ref{0andinfinity}) and dropping again the primes, becomes
\begin{equation}
1 + \frac{ \beta ( P  ) }{1-\beta( P )}=\frac{ - \beta (B_2\circ B_1\circ B_2(  P ) ) }{1- \beta(B_2\circ B_1\circ B_2(  P ))} ~.
\end{equation}
Repeating the above steps with the roles of $B_2 $ and $B_1$ exchanged,
 one can show that
\begin{equation}
\beta(B_1\circ B_2\circ B_1( P ))=\beta(B_2\circ B_1\circ B_2( P )) ~,
\end{equation}
which then implies
\begin{equation}
\label{braid}
B_1\circ B_2\circ B_1=B_2\circ B_1\circ B_2
\end{equation}
up to automorphisms of the pair. This, together with the conditions $B_1\circ B_1=B_2\circ B_2=1$ reproduce the expected relations (again, up to automorphisms of the pair) for  $S_3$.

To summarize, there is an $S_3$ of permutations of the set of
branch points $\{0,\,1,\,\infty\}$ on the target $\mathbb{P}^1$,
which are realized by holomorphic automorphisms of  $\mathbb{P}^1$
living in $PSL(2, \IC)$. For each  element $ \psi $ in this set of
transformations, we can look for automorphisms $ \phi$ in $ {\rm Aut} ( \Sigma_h )$
such that the following diagram commutes
\begin{center}
\begin{equation}
\begin{array}{c c c}
 & \phi & \\
\Sigma_h & \longrightarrow & \Sigma_h\\ & & \\
\beta\,\, \downarrow & &\downarrow\,\,\beta \\ & \psi & \\ \mathbb{P}^1 & \longrightarrow & \mathbb{P}^1
\end{array}
\end{equation}
\end{center}

 In terms of the combinatorial data, we expect that these transformations
 are implemented by appropriate
 braid group actions.
The above diagram translates into
\begin{equation}
\gamma_{B_i}\, \sigma_W\,\gamma_{B_1}^{-1} = B_i(\sigma_W) ~, \qquad \gamma_{B_i}\, \sigma_B\,\gamma_{B_i}^{-1}=B_i(\sigma_B) ~, \quad i=1,\,2 ~.
\end{equation}
A first guess is that the $ \gamma_{ B_i} $ implementing the Artin action on
$ ( \sigma_B , \,\sigma_W ,\, \sigma_{ \infty} ) $ matches the
geometrical transformations for the pair $ ( \mT^2 ,\, \beta ) $
and the Artin action on $  ( \widetilde \sigma_B , \,\widetilde \sigma_W ,\, \widetilde \sigma_{ \infty} ) $ matches the geometrical transformations for the pair
$ ( \Sigma  , \,\widetilde \beta ) $. While this expectation is borne out
by the available examples of  $ ( \Sigma  ,\, \widetilde \beta ) $,
it does not work for our examples of  $ ( \mT^2,\, \beta ) $.
We do have partial success matching the geometrical braiding transformations
 for $ ( \mT^2 ,\, \beta ) $ with the second braid group action (\ref{BiI}).

\section{Basic examples of  SCFT--Belyi pair correspondence}
\label{basexamp}

We now describe the dimer description for basic examples of $\cN=1$ SCFTs
along with the permutation triples and Belyi pairs.
 We demonstrate the geometrical
realization, in terms of the toroidal Belyi pair,  of the permutation
 symmetries fixing $W_+ , W_- $ separately, illustrating the general result
in Section \ref{permfix}. We also illustrate  the points raised in
Section \ref{secartin} with regard to the geometric realization of
symmetries exchanging $ W_+, W_- $.

\subsection{Branes transverse to  $\mathbb{C}^3$ and Belyi pair }

 Let us collect our results for $N$ D$3$-branes transverse to
 $\mathbb{C}^3$. The worldvolume theory on the branes
 is maximal  super-Yang--Mills theory,
 where the $\cN=1$ supersymmetry is enhanced to $ \cN=4$. The
gauge group, matter content, and superpotential is   encoded in the
 dimer in Figure~\ref{C3dimer}. The combinatorial data that define the map are the permutations
\begin{equation}
 \sigma_B=(123) ~, \qquad  \sigma_W= (123) ~, \qquad \sigma_{\infty}=(123) ~.
\end{equation}
The analytic expression for the Belyi pair is
\begin{equation}\label{recallc3}
\beta=\frac{y+1}{2} ~,\qquad y^2=x^3+1 ~.
\end{equation}

\subsubsection{Automorphisms  }
We can see the connection between the combinatoric and geometric
realizations of automorphisms at work in our $\mathbb{C}^3$ example.
 The set of permutations leaving the set
$\{\sigma_B,\, \sigma_W , \sigma_{ \infty} \}$ invariant under conjugation
according to (\ref{autcomb}) is given by
\begin{equation}
{\rm Aut}(\mT^2 ,\,\beta)=\{1,\,(123),\, (132)\} ~.
\end{equation}
According to our discussion in Section \ref{RetAutBel}
, this $\mathbb{Z}_3$ is the automorphism group which leaves both the elliptic curve and the Belyi map invariant. We can check this from an analytic perspective. The set of transformations leaving invariant the Belyi function and the curve is  of the form
\begin{equation}
\label{N=4aut}
(x,\, y)\rightarrow (\omega\, x,\, y) ~,\quad \omega^3=1 ~.
\end{equation}
These transformations generate a $\mathbb{Z}_3$, in agreement with
 the combinatorial calculation.
Another obvious symmetry of the curve, which lies
in ${\rm Aut} ( \mathbb{T}^2 )  $ is the operation
\begin{equation}
(x,\, y)\rightarrow (x,\, -y) ~,
\end{equation}
which is a symmetry of the curve but crucially not an
automorphism  of the Belyi pair.

\subsubsection{Braid group action }

We are firstly interested in exchanging  black with white nodes of the dessin,
namely the inverse images under the Belyi map of $\{ 0 ,1 \} $.
Applying (\ref{1and0}) to this case,  where the curve is described by
an equation involving $(x,y)$ as in (\ref{recallc3}), we have
\begin{equation}
\beta(x,\,y)=1-\beta(B_1(x,\,y)) ~.
\end{equation}
This is implemented by $(x,\, y)\rightarrow (\omega\, x,\, -y)$. Up to  the automorphism group of the pair
(\ref{N=4aut}),
 we can choose
\begin{equation}
B_1(x,\,y)=(x,\, -y) ~.
\end{equation}
 It is clear that $B_1^2=1$, which  is an illustration
 of (\ref{B1square}) from the general discussion.

The braid group action
also contains an  exchange of $ \{ 1 , \infty \} $ on the target
$\mP^1$, leaving $ 0 $ fixed. Following (\ref{0andinfinity}), this
 is implemented by an element $B_2$ in ${\rm Aut} ( \mT ) $, obeying the relation
\begin{equation}\label{needB2}
\beta(x,\, y)=\frac{ - \beta ( B_2(x, y) ) }{1-\beta(B_2(x,\,y))} ~.
\end{equation}
One  checks that  $B_2$ is given by
\begin{equation}
B_2 (x,\,y)=(  \frac{2\,x}{y - 1 },\, \frac{y+3}{y-1}) ~.
\end{equation}
This leaves the curve invariant, obeys (\ref{needB2}),
 and satisfies $B_2^2=1$. Together with $B_1$, it
satisfies  the relations of $S_3$, which is a quotient of the spherical
braid group $ \mathcal{B}_3$.

In terms of the permutations, the action (\ref{BiI}) leads to
an answer which matches the above geometrical discussion.
Indeed, to implement the action of $B_1$ according to   (\ref{BiI}),
we look for $ \gamma_{B_1}$ such that
\begin{equation}
\gamma_{B_1}\, \sigma_W\,\gamma_{B_1}^{-1}=\sigma_B^{-1} ~, \qquad \gamma_{B_1}\, \sigma_B\,\gamma_{B_1}^{-1}=\sigma_W^{-1} ~.
\end{equation}
In this case, since $\sigma_B =\sigma_W $, the equation
amounts to a   single condition for $\sigma=(123)$, $\sigma^{-1}=(132)$. The solution is $\gamma_{B_1}=(1)\,(23)$. (Other solutions $ (13)\,(2) $
and $ (12)\,(3)$ are related to $(1)\,(23)$ by conjugation with
 $ (123) $, which is part of the symmetry ${\rm Aut} ( \s_B , \s_W ) = {\rm Aut} ( \mT^2 , \beta ) $.)
 So we have, up to the obvious equivalence, one solution $ \gamma_{B_1}$ obeying  $\gamma_{B_1}^2=1$,  defining an expected  $\mathbb{Z}_2$.
If we try this exercise with the Artin action (\ref{artinact}),
 we do not find a  non-trivial
$\IZ_2$ realized as conjugations by elements of $S_{d=3}$.

For the exchange of $ \{ 1 , \infty \} $ we  solve
\begin{equation}
\gamma_{B_2}\, \sigma_W\,\gamma_{B_2}^{-1}=\sigma_{\infty}^{-1} ~, \qquad \gamma_{B_2}\, \sigma_B\,\gamma_{B_2}^{-1}=\sigma_B^{-1} ~.
\end{equation}
In this particular case $\sigma_{\infty}=(123)=\sigma_B=\sigma_W$,
so these equations are satisfied again by $\gamma_{B_2}=(1)\,(23)$.

It is somewhat puzzling that the geometrical side produces two
distinct elements of $ {\rm Aut} ( \Sigma_h )$, unrelated by
$ {\rm Aut} ( \Sigma_h ,\, \beta ) $, whereas the combinatoric
side produces only one conjugacy class which effects the
braiding for both $B_1$ and $B_2$. This should be taken
into account in a correct general formulation of the relation between
braid group actions for geometry and combinatorics.

\subsection{Branes at  conifold and Belyi pair }

\begin{figure}[h!]
\begin{center}
\includegraphics{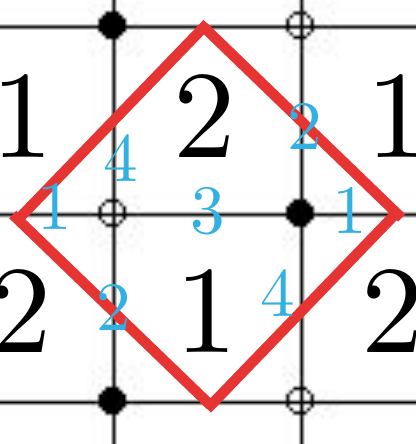}
\end{center}
\caption{Dimer for the conifold theory.}
\label{Conifoldfig}
\end{figure}

We are interested in the Klebanov--Witten field theory, which describes D$3$-branes probing a conifold singularity.
The dimer is shown in Figure~\ref{Conifoldfig}.

A unit cell is drawn in red. The four edges labelled $1, \ldots, 4$ (in blue)
correspond to four chiral multiplets $ X_1, \ldots, X_4$. Following
the description at the beginning of Section \ref{PermW},
we have a superpotential
\bea
W = \tr ( X_1 X_2 X_3 X_4 - X_1 X_4 X_3 X_2 ) ~.
\eea
There are two faces per unit cell labelled $1,2$.
This indicates a gauge group $ SU(N) \times SU(N)$.
If we choose an orientation on the surface, the  edges
can be assigned to representations   $(N,\,\overline{N}) $
or $ (\overline{N},\,N ) $.  We read off that the theory contains two
 pairs of chiral superfields in each of the
$(N,\,\overline{N})$ and $(\overline{N},\,N)$ representations. A common convention is
to label the fields with lower indices describing the gauge group,
and an additional upper multiplicity.  Thus $X_1$ is written as
$X^1_{ 12} $ since it has faces $1,2$ incident on it, and an orientation
on the torus fixes the representation $ ( N , \overline{N} ) $ for
$SU(N) \times SU(N)$ where the first $SU(N)$ is associated to the
face labelled $1$ and  the second $SU(N)$ to the face labelled $2$.
Extending this to all the edges
\begin{equation}\label{transconv}
1\sim X_{12}^1 ~, \quad 2\sim X_{21}^1 ~, \quad 3\sim X_{12}^2 ~, \quad 4\sim X_{21}^2 ~.
\end{equation}
With this notation, the same superpotential can be written in the
 conventional form  as
\begin{equation}
\label{KWW}
W={\rm Tr}\, X_{12}^i\, X_{21}^m\, X_{12}^j\, X_{21}^n\, \epsilon_{ij}\, \epsilon_{mn} ~.
\end{equation}

Viewing the dimer as a  dessin d'enfant, the permutations $ \s_B, \s_W $
below describe the edges around the black and white node,
traversed anti-clockwise.
\begin{equation}
\label{sigmaT11}
\sigma_B=(1234) ~,\qquad \sigma_W =(1234) ~, \qquad \sigma_{\infty}=(13)\,(24) ~.
\end{equation}
The permutation $\sigma_{\infty} $ is constructed as $ (\s_B \s_W)^{-1} $
following the general description in Section \ref{DesPerm}.  The
fact that there are two faces corresponds to there being two cycles in $ \s_{\infty} $.

These data define a Belyi map   $\beta$, which is   a degree four  map, {\em i.e.},
 $d=4$,  from $\mathbb{T}^2$ to $\mP^1$, with a single ramification point of
order four over each of $0$ and $1$,  and two ramification
points of order two over $\infty$.   The Belyi pair is specified by giving
an algebraic equation for  $\mathbb{T}^2$, which is
\begin{equation}
y^2=x\,(x-1)\,(x-\frac{1}{2}) ~.
\end{equation}
This $\mathbb{T}^2$ has $j(\tau)=1728$.
Inverting this, $j^{-1}(1728) = \tau = e^{\pi i/2} = i$, which gives the modular parameter of the elliptic curve.
(For more on the $j$-invariant please refer to Appendix~\ref{appsec:jfunction}.)
The relevant Belyi map to a $\mP^1$ is
\begin{equation}
\label{belyiT11v1}
\beta(x)=\frac{x^2}{2x-1} ~.
\end{equation}
By analyzing $\beta^{-1} (0 ) = \{ ( 0 , 0 )  \}$, $\beta^{-1} (1  ) = \{ ( 1, 0 ) \}$,
 $\beta^{-1} (\infty ) = \{  ( { 1\over 2 } , 0 )  , ( \infty , \infty )  \} $,
the above equations are shown to be a correct description of the
Belyi pair with the desired properties.

\subsubsection{Automorphisms }

Let us consider a rational transformation preserving the set of points $x=\{0,\, \frac{1}{2},\, 1,\, \infty\}$ where the map ramifies
\begin{equation}
\phi_{ \pm }: (x,\, y)\,\rightarrow \,(\frac{x}{2x-1},\, \pm \frac{i\,y}{(2x-1)^2}) ~.
\end{equation}
It is easy to check that under this transformation both the map and the curve are left invariant. Therefore, this is an automorphism. Some properties of
$ \phi_{\pm } $ are
\bea
\phi_+^4 & = & 1 ~, \cr
\phi_+^3  & = &  \phi_+^{-1} =  \phi_- ~, \cr
\phi_+^2  & = &  \phi_-^2: (x,\, y)\rightarrow (x,\, -y) ~.
\eea
We conclude that these geometrical automorphisms form a
$\IZ_4 $ generated by $\phi_+ $ (or $\phi_-$)
\begin{equation}
 {\rm Aut}(\mathbb{T}^2,\,\beta)=\mathbb{Z}_4 \subset {\rm Aut} ( \mT^2 ) ~.
\end{equation}

From the combinatorial perspective, the
element $\gamma_A=(1234)$ leaves the pair  $ ( \s_B , \s_W ) $
invariant under conjugation and generates
the $\mathbb{Z}_4$ automorphism group ${\rm Aut} ( \s_B , \s_W ) $, which
by the Riemann existence theorem, should be equal to ${\rm Aut} ( \mT^2 , \beta )$.

\subsubsection{Braid group action  }

In order to study the exchange of white and black vertices it is
better to switch to the Weierstrass form of the elliptic curve.
For $j=1728$, we can consider
\begin{equation}
y^2=x^3-x ~.
\end{equation}
In these coordinates, the map is
\begin{equation}
\label{belyiT11v2}
\beta=\frac{(x+1)^2}{4x} ~.
\end{equation}
One can see that the change of coordinates is just $x\mapsto 2x-1$, which transforms (\ref{belyiT11v2}) into  (\ref{belyiT11v1}). In these coordinates, the automorphisms of the pair are generated by
\begin{equation}
\label{autT11}
\qquad (x,\, y)\rightarrow (\frac{1}{x},\, -i\, \frac{y}{x^2}) ~,
\end{equation}
where we have chosen the positive sign with no loss of generality.

In the new coordinates, the operation of exchanging black and white nodes can be represented analytically as
\begin{equation}
\phi_1(x,\,y)=(-x,\, i\,y) ~,\qquad \phi_2(x,\,y)=(-\frac{1}{x},\, \frac{y}{x^2}) ~,
\end{equation}
under which
\begin{equation}
\beta(x,\, y)=1-\beta(\phi_i(x,\, y)) ~, \quad i=1,\,2 ~.
\end{equation}
The two maps $\phi_i$ above can be connected by an automorphism of the pair. Explicitly, upon considering the transformation in (\ref{autT11}) one can go from $\phi_1$ to $\phi_2$. Note that this element of the automorphism group exchanges the point at infinity with $(0,\, 0)$. These points are indeed the preimages of infinity under $\beta$, so the only fixed set of this automorphism are the branching points corresponding to the puncture at infinity. Since the two elements above are connected by an automorphism of the pair, we can just keep $\phi_1=B_1$, so that we have, as expected, that up to automorphisms of the pair $B_1^2=1$.
This is as expected from the $\mathbb{Z}_2$-operation of switching
black and white nodes.

From the combinatorial perspective, following the discussion in Section \ref{braidgeom}, we look for solutions
$ \gamma_{B_1}$ to
\begin{equation}
\gamma_{B_1}\, \sigma_W\,\gamma_{B_1}^{-1}=\sigma_B^{-1} ~, \qquad \gamma_{B_1}\, \sigma_B\,\gamma_{B_1}^{-1}=\sigma_W^{-1} ~.
\label{eq:temp74}
\end{equation}
Since in the conifold $\sigma_W=\sigma_B=(1234)$, we have to solve
\begin{equation}
\gamma_{B_1}\, (1234)\,\gamma_{B_1}^{-1}=(1432) ~.
\end{equation}
The solutions to this equation are
\bea
 \gamma_{B_1}^{(1)} & = & (13) (2) (4) ~, \cr
 \gamma_{B_1}^{(2 )} & = &  (1) (3)  (24) ~, \cr
 \gamma_{B_1}^{(3 )}  & = &  (14) ( 23) ~, \cr
  \gamma_{B_1}^{(4 )}  & = &  (12) (34) ~.
\eea
Now observe the first two are conjugate to each other
by $ (1234) $ which is in ${\rm Aut} ( \s_B , \s_W ) $.
Likewise the second pair are conjugate to each other
 by $(1234) $. Further note that
\bea
  \gamma_{B_1}^{(4 )} = (1234)  \gamma_{B_1}^{(1)} (1234)^2 ~.
\eea
Since any solution $ \gamma_{B_1} $ for the braiding
generates other solutions  $ \g_1 \gamma_{B_1} \g_2$ by multiplication
with  $ ( \g_1 , \g_2 ) $ which are any pair in ${\rm Aut} ( \s_B , \s_W ) $,
we should count solutions modulo left and right multiplication
by the automorphism of the pair.
This is because the permutations $\sigma_B$ and $\sigma_W$ are only defined up to conjugation equivalence.
Hence, we have only one equivalence class of combinatoric solution.
This matches the analytic discussion.

\subsection{Branes at $ dP_0 $ and Belyi pair }

\begin{figure}
\begin{center}
\includegraphics{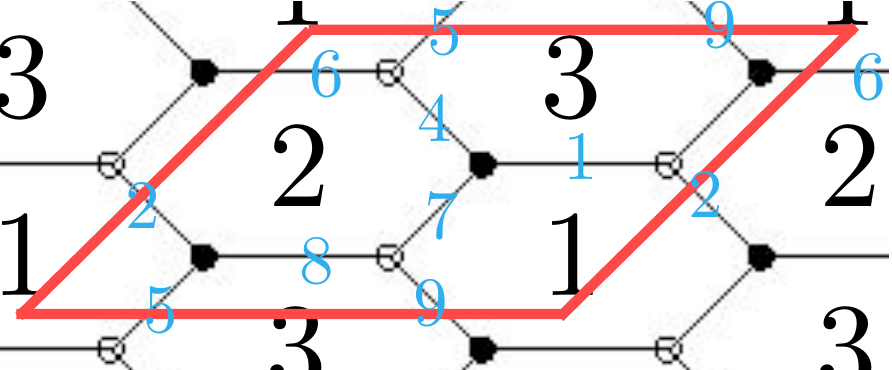}
\end{center}
\caption{Dimer for the $dP_0$ theory.}
\label{dP0dimer}
\end{figure}

Consider the $\mathbb{Z}_3$ orbifold of $\mathbb{C}^3$ which acts like
\begin{equation}
\mathbb{C}^3/\mathbb{Z}_3=\{\,(x,\,y,\,z)\in \mathbb{C}^3\,/\,(x,\,y,\,z)\sim(\omega\,x,\,\omega\,y,\,\omega\, z),\quad \omega^3=1\,\} ~.
\end{equation}
This orbifold leaves invariant the natural holomorphic three-form in $\mathbb{C}^3$, and thus preserves $\mathcal{N}=1$ supersymmetry. Moreover, this geometry is an $\mathcal{O}(-3)$ bundle over $\mathbb{P}^2$, and thus corresponds to $dP_0$, the cone over the zeroth del Pezzo surface.\footnote{
The $n$-th del Pezzo surface is a blowup into $\mathbb{P}^1$ of $n$ points on $\mathbb{P}^2$, where $n=0,\ldots,8$.
These are manifolds of complex dimension two with positive first Chern class.}
All the information is encoded in the dimer in Figure~\ref{dP0dimer}.
From the dimer one can see that the dual SCFT has three gauge nodes. The matter content involves three triplets of chiral superfields respectively in the $(N,\,\overline{N},\,\mathbf{1});\, (\mathbf{1},\,N,\,\overline{N});\, (\overline{N},\,N,\,\mathbf{1})$ representation. The superpotential is
\begin{equation}
W={\rm Tr}\, X_{12}^i\, X_{23}^j\, X_{31}^k\, \epsilon_{ijk} ~.
\end{equation}

From the dimer, we have a dessin with three black and three white nodes, both with ramification three, and three faces, also with ramification three. The associated permutations are
\begin{equation}
\label{sigmadP0}
 \sigma_B=(147)\,(258)\,(369) ~, \qquad \sigma_W=(123)\, (456)\,(789) ~, \qquad \sigma_{\infty}=(195)\,(276)\,(384) ~.
\end{equation}
Note that the first permutation is obtained from the canonically ordered $\sigma_W$ by taking the first element of each cycle, the second element of each cycle, and the third element of each cycle to write the cycles in $\sigma_B$.
The explicit permutations for this and a class of related Belyi pairs was studied in \cite{dMR}.

One can check that the isomorphism group of these permutations is generated by
\begin{equation}
\gamma_{A_1}=(1\,4\,7)\,(2\,5\,8)\,(3\,6\,9) ~,\qquad \gamma_{A_2}=(1\,5\,9)\,(2\,6\,7)\,(3\,4\,8) ~.
\end{equation}
These operators satisfy
\begin{equation}
\gamma_{A_1}^3=\gamma_{A_2}^3=1 ~.
\end{equation}
Moreover, one can verify that they commute, so that the automorphism group is $\mathbb{Z}_3\times \mathbb{Z}_3$.

The relevant Belyi map in this case can be obtained as a special case
of examples discussed in \cite{jsmongal}.
Consider the Fermat cubic in $\mathbb{P}^2$,
\begin{equation}
F=\{(x,\, y,\, z)\in \mathbb{P}^2/\, x^3+y^3=z^3\} ~.
\end{equation}
This is the projective closure of the affine variety
\begin{equation}
F^{\rm aff}=\{(x,\, y)\in\mathbb{C}^2/\,x^3+y^3=1\} ~.
\end{equation}
We can now consider the function
\begin{equation}
f:\, F^{\rm aff}\rightarrow \mathbb{C} ~,
\end{equation}
which takes $(x,\, y)$ into $x$. On the other hand, $x^3=1-y^3$ generically has three solutions. However, when $x$ is one of the three cubic roots of unity, $y$ can only take one value $y=0$. Thus, any of the cube roots of one is a critical point of $f$. However, at these critical points $f$ is clearly not $1$. But   composing  $f$ with $g=x^3$ we have  $\beta=g\circ f=x^3$, which
is  a Belyi map
\begin{equation}
\beta(x,\,y)=x^3 ~.
\end{equation}
One can check that it satisfies the ramifications expected from the dessin specified by (\ref{sigmadP0}).

The automorphisms of the Belyi pair are
\begin{equation}
\gamma_{A_1}:\,(x,\,y)\rightarrow (\omega_1\, x,\,y) ~, \qquad \gamma_{A_2}:\,(x,\,y)\rightarrow (x,\,\omega_2 \, y) ~, \quad \omega_i^3=1 ~,
\end{equation}
and these commute. We thus generate the expected $\mathbb{Z}_3\times \mathbb{Z}_3$.

Note that the fixed points under $\gamma_1$ are of the form $(0,\,y)$. The only such point in the curve is $(0,1)$, which corresponds to a critical point of the Belyi map. The fixed points under $\gamma_2$ are of the form $(x,\, 0)$. This time, only $(1,\,0)$ lies on the curve, and actually coincides with a critical point of the Belyi map. Thus the automorphisms do not have (regular) fixed points : the only fixed points being at critical points of the map. This is as expected from covering space theory.

\subsubsection{Braid group action }

As discussed above, in order to exchange the black and white nodes we must find solutions to
\begin{equation}
\gamma_{B_1}\, \sigma_W\,\gamma_{B_1}^{-1}=\sigma_B^{-1} ~, \qquad \gamma_{B_1}\, \sigma_B\,\gamma_{B_1}^{-1}=\sigma_W^{-1} ~.
\end{equation}
In terms of the combinatorial data (\ref{sigmadP0}), a solution is
\begin{equation}
\gamma_{B_1}=(3)\,(5)\,(7)\,(19)\,(26)\,(48) ~.
\end{equation}

One can check that $\gamma_{B_1}^2=1$, so the action exchanging black and white nodes is a $\mathbb{Z}_2$ action. On the other hand, from the point of view of the $W$ the action exchanging black and white nodes (thus contributing an overall minus sign to $W$) is
\begin{eqnarray}
\label{BWdP0}
(X_{12}^1,\, X_{12}^2,\, X_{12}^3)&\rightarrow& (X_{12}^1,\, X_{12}^3,\, X_{12}^2) ~, \nonumber \\ (X_{23}^1,\, X_{23}^2,\, X_{23}^3)&\rightarrow& (X_{23}^1,\, X_{23}^3,\, X_{23}^2) ~, \\ \nonumber (X_{31}^1,\, X_{31}^2,\, X_{31}^3)&\rightarrow& (X_{31}^1,\, X_{31}^3,\, X_{31}^2) ~.
\end{eqnarray}
Note how this action mimics the $\gamma_{B_1}$ above: in each triplet a member is kept fixed and the other two are exchanged. Furthermore, the square of this transformation is the identity.

The transformation (\ref{BWdP0}) is not unique. For example, we could have chosen to keep fixed not the first element of each triplet but say the second. Conversely, from the more abstract point of view starting from
 $\gamma_{B_1}$, any pair  $\gamma, \gamma^{ \prime}  \in {\rm Aut}(\mathbb{T}^2,\, \beta)$,
 generates additional solutions  $\gamma^{ \prime} \gamma_{B_1} \, \gamma$
 which effect the exchange of  black and white vertices. We can check that all the solutions
 are indeed obtained from a single one by these left-right multiplications, so that modulo
 these multiplications, there is a unique equivalence class of braiding transformations corresponding to $B_1$.
Any product of two elements among the possible $ \gamma_{B_1} $ yields
an element  in $ {\rm Aut} ( \Sigma_h , \beta ) $, reflecting the  relation $B_1^2 =1 $
in the braiding action the 3-punctures sphere (\ref{Artinsquare1}).

From an analytic point of view, the $B_1$ arises in this case from
\begin{equation}
1-B_1(x,\, y)^3=x^3 ~.
\end{equation}
One can check that
\begin{equation}
B_1(x,\, y)=(y,\, x) ~.
\end{equation}
As it is obvious, $B_1^2=1$, we have, as expected, a $\mathbb{Z}_2$ symmetry. Thus, we have again
\begin{equation}
{\rm  Aut}_s(\mathbb{T}^2,\, \beta)=\mathbb{Z}_2 ~.
\end{equation}

Furthermore, we can consider the exchange of the puncture corresponding to $\infty$ with that corresponding to $1$, that is, $\beta\rightarrow -\beta/(1-\beta)$. The transformation implementing this is
\begin{equation}
B_2(x,\, y)=(-\frac{x}{y},\, \frac{1}{y}) ~.
\end{equation}
It is clear that $B_2^2=1$. Furthermore, one may verify that the two transformations $B_1$ and $B_2$ again satisfy the spherical braid group relations. From a combinatorial perspective, the corresponding permutation implementing this action should satisfy
\begin{equation}
\gamma_{B_2}\, \sigma_W\, \gamma_{B_2}^{-1}=\sigma_{\infty}^{-1} ~,\qquad \gamma_{B_2}\, \sigma_{\infty}\, \gamma_{B_2}^{-1}=\sigma_{W}^{-1} ~.
\end{equation}
One solution to this equation is
\begin{equation}
\gamma_{B_2}=(25)\,(39)\,(47)\,(6)\,(1)\,(8) ~.
\end{equation}
Other solutions are related to this by multiplication on left and right by elements in
$ {\rm Aut} ( \s_B , \s_W ) $.\footnote{
These statements are checked explicitly using computational number theory software such as {\tt SAGE}~\cite{sage}.}   So we have a unique equivalence class for $B_2$ on the
combinatoric side, in agreement with the geometry.

\subsection{Branes at SPP and Belyi pair }

\begin{figure}[!h]
\centering
\includegraphics[scale=.7]{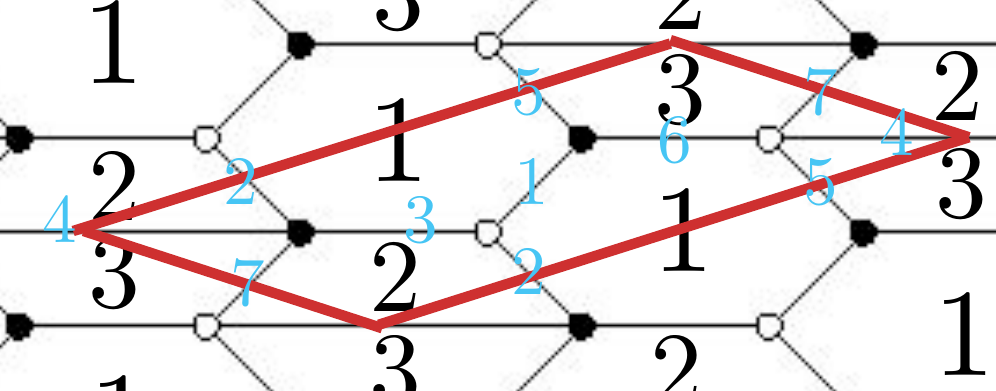}
\caption{Dimer for the SPP theory.}
\label{SPPdimer}
\end{figure}

So far we have considered orbifolds of theories with just one superpotential term. Let us now consider a more generic theory.
 We will now consider the suspended pinch point (SPP), which is defined in terms of $\{(x,\,y,\, u,\, v)\in \mathbb{C}\}$ such that
\begin{equation}
{\rm SPP}=\{(x,\,y,\, u,\, v)\,/\,x\,y=u\,v^2\} ~.
\end{equation}
The gauge theory associated to $N$ D$3$-branes at the SPP singularity is studied in~\cite{mopl, uranga}.
The dual theory is encoded in the dimer in Figure~\ref{SPPdimer}.
From the dimer we read off that the theory has three gauge groups and seven matter fields, one of which is in the adjoint under the first gauge group. The superpotential is
\begin{equation}
W={\rm Tr}\Big(\, X_{11}\, X_{12}\, X_{21}-X_{11}\, X_{13}\, X_{31}+X_{31}\,X_{13}\,X_{32}\,X_{23}-X_{21}\, X_{12}\,X_{23}\,X_{32}\,\Big) ~.
\end{equation}

Note that now we have superpotential terms with three and four fields in each, that is, cycles in white (and black) nodes of different length. The corresponding permutations are
\begin{equation}
\sigma_W=(123)\, (4567) ~,\qquad \sigma_B=(156)\,(3742) ~,\qquad \sigma_{\infty}=(153)\,(27)\,(46) ~.
\end{equation}
The ramification is $2+3+2+3+2+1+1=14$, while the degree is seven, thus corresponding as it should to a dimer drawn on a genus one curve. We have not been able to construct the Belyi pair for this case. This is a very interesting open problem, especially in the light of the discussion in Section \ref{tauBtauR}.

\subsection{Untwisting of the toroidal Belyi pair}
\label{Untwisting}

In Section \ref{PermW} we briefly described the procedure of untwisting. As described in \cite{Stienstra:2007dy}, this procedure has a very natural translation into the combinatorial language, since it  just
 amounts to encircling  the white nodes in the opposite direction as the black ones, that is, we should consider $\{\widetilde{\sigma}_B,\, \widetilde{\sigma}_W\}=\{\sigma_B,\,\sigma_W^{-1}\}$. In view of our preceding discussion, it is natural to define yet a third permutation $\widetilde{\sigma}_{\infty}$ such that $\widetilde{\sigma}_B\, \widetilde{\sigma}_W\,\widetilde{\sigma}_{\infty}=1$. This way we obtain another holomorphic map
 to a marked $\mathbb{P}^1$. This time the source of the map is not a fixed
  genus one $\mathbb{T}^2$, but rather a Riemann surface $\Sigma_g$ whose genus is given
  by the number of internal points in the toric diagram of the original CY$_3$, that is, the number
   of four-cycles. This $\Sigma_g$ corresponds to the untwisted curve introduced in \cite{Feng:2005gw},
   over which the (untwisted) dimer lives.  Using the theory of dessins, we have here a specific complex structure on
   $ \Sigma_g$ along with a holomorphic map to $\mP^1$. The  geometric and combinatoric
realizations of the  exchange of black and white nodes
described in Section~\ref{secartin}, in connection with symmetries that exchanges $W_+$ and $W_-$
 apply equally well here. We will  now give the explicit form of this untwisted Belyi pair
 and its properties  in some examples.

\subsubsection{Untwisted Belyi map for branes at  $\mathbb{C}^3$}

The first example we consider is the $\mathcal{N}=4$ SYM case for 3-branes transverse to $\IC^3$.
 Untwisting the dimer translates to considering the orientation of the black vertices to be reversed with respect to that of the white vertices. That is to say, we circle one vertex clockwise and the other anticlockwise. The combinatorial data are now
\begin{equation}
\label{sigmaC3U}
\widetilde{\sigma}_B=(132),\qquad \widetilde{\sigma}_W=(123),\qquad \widetilde{\sigma}_{\infty}=(1)\, (2)\, (3) ~.
\end{equation}
Now the branching number is $B=2+2=4$. With degree  $d=3$, the Riemann--Hurwitz formula (\ref{riemhur})
shows that this dimer now lives on a $\mathbb{P}^1$. This is to be expected, since, as anticipated, the genus of the Riemann surface coincides with the number of four-cycles in the geometry; for $\mathbb{C}^3$ there are none. We can write the Belyi map from the source $\mP^1$ to the target $\mP^1$
\begin{equation}
\widetilde{\beta}=\frac{x^3}{x^3+(x-1)^3} ~.
\end{equation}
The automorphisms of the map
are given by
\begin{equation}
\widetilde{\phi}(x)= \frac{2x}{-1+i\,\sqrt{3}\,(x-1)+3\,x}   =     \frac{ x ( i + \sqrt{3} ) /2 }  { x \sqrt{3} + ( i -  \sqrt{3} )/2 } ~,
\end{equation}
which is obtained by solving $ \beta = \beta ( \phi ) $ and, as expected is in $ PSL(2, \IC)$.
Because $\widetilde\phi^3=1$
\begin{equation}
{\rm \widetilde{Aut}}(\mP^1,\,\mP^1)=\mathbb{Z}_3 ~,
\end{equation}
 where the tilde stresses that we are considering the untwisted dimer.

We now turn to the map which exchanges black and white dots. In this case, modulo automorphisms, it reads
\begin{equation}
B_1(x) =-x+1 ~.
\end{equation}
This map squares to one, thus yielding the expected $\mathbb{Z}_2$.

From the combinatorial side,  the  braiding action in the untwisted curve is given by the Artin action
\begin{equation}
\gamma_{B_1}\,\,\widetilde{\sigma}_W\,\,\gamma_{B_1}^{-1}=\widetilde{\sigma}_W ~, \qquad \gamma_{B_1}\,\, \widetilde{\sigma}_B\,\,\gamma_{B_1}^{-1}=\widetilde{\sigma}_W^{-1}\,\widetilde{\sigma}_B\,\widetilde{\sigma}_W ~.
\end{equation}
The solution is $\gamma_{B_1}=(1)\,(23)$, which reproduces the $\mathbb{Z}_2$.

The Artin exchange $B_2$ is not performed by any $\gamma_{B_2}$ since
$ \widetilde \s_W $ and $ \widetilde \s_{ \infty} $  are in different
conjugacy classes.

\subsubsection{Untwisted Belyi map for branes at the conifold}

Our next example will be the conifold. The untwisted permutations are
\begin{equation}
\label{sigmaT11U}
\widetilde{\sigma}_B=(1432) ~,\qquad \widetilde{\sigma}_B=(1234) ~,\qquad \widetilde{\sigma}_{\infty}=(1)\,(2)\,(3)\,(4) ~.
\end{equation}
By using the Riemann--Hurwitz formula (\ref{riemhur}), one can see that these permutations encode maps from $\mP^1$ to $\mP^1$, again to be expected given that in the conifold $b_4=0$. The appropriate Belyi map in this case is
\begin{equation}
\widetilde{\beta}=\frac{x^4}{x^4+(x-1)^4} ~.
\end{equation}
In this case, the automorphisms are generated by
\begin{equation}
\widetilde{\phi}(x)=\frac{i\, x}{-1+(1+i)\,x} ~.
\end{equation}
It follows that $\widetilde{\phi}^4=1$, and so
\begin{equation}
{\rm \widetilde{Aut}}(\mathbb{P}^1,\,\mP^1)=\mathbb{Z}_4 ~.
\end{equation}
Turning to the action exchanging black and white nodes.
In this case, there are two transformations in $ P SL(2, \IC )$, the
group of holomorphic automorphisms of the sphere, namely
\bea
&& \phi_1 ( x ) = 1 - x ~, \cr
&& \phi_2 ( x ) = { 1 - x \over 1 - 2x } ~,
\eea
which both satisfy $ 1 - \beta ( x ) = \beta ( \phi_i ( x ) ) $.
We can further verify that $ \phi_i ( \phi_i ( x ) ) = x $
for $ i = 1 $ or $i=2$.

On the combinatoric side, the Artin action
\bea
\gamma_{B_1} \widetilde \s_B \gamma_{ B_1 }^{-1} & = &  \widetilde \s_W ~, \cr
\gamma_{B_1} \widetilde \s_W \gamma_{ B_1 }^{-1}  & = &  \widetilde \s_W^{-1} \widetilde \s_B
\widetilde \s_W
\eea
becomes
\bea
 \gamma_{ B_1} ( 1234) \gamma_{B_1}^{-1} & = &  ( 1432) ~, \cr
  \gamma_{ B_1} ( 1234) \gamma_{B_1 }^{-1} & = &  ( 1432) ~,
\eea
which has solutions $ \gamma_{ B_1 } = (13) (2) (4) ; (24)(1)(3) ; ( 14) (32) $.
The condition coming from conjugation of $ \widetilde \sigma_{\infty} $
is also satisfied, since $ \gamma_{B_1} \gamma_{B_1}^{_1} = 1 $.
 The first two are
 related by conjugation. So there are two independent
solutions, each of which squares to $1$. This gives a geometrical
counterpart for each of the two permutation operations on the fields of
the conifold theory, which have the effect of exchanging $ W_+ $ and $W_- $.

For the braid group element $B_2$, we look for
\bea
\gamma_{B_2} \widetilde \s_W \gamma_{ B_2 }^{-1} & = &  \widetilde \s_{ \infty} ~, \cr
\gamma_{B_2} \widetilde \s_{ \infty}  \gamma_{ B_2 }^{-1} & = &
 \widetilde \s_{\infty}^{-1} \widetilde \s_W \widetilde \s_{ \infty} ~.
\eea
There is no solution since $  \widetilde \s_W $ and $  \widetilde \s_{ \infty} $
are in different conjugacy classes. Correspondingly, on the analytic side,  solving $ { 1 \over \beta ( x )}
 = \beta ( \phi ( x ) $ does not yield any $ \phi $ in $ PSL(2 , \IC ) $

\section{Orbifolds, automorphisms, and complex structures}
\label{sec:orbs}

Starting from the periodic tiling corresponding to a given dimer describing
the SCFT for a Calabi--Yau, the operation
of orbifolding  the Calabi--Yau amounts to an enlargement of the unit cell of the tiling \cite{hananyvegh}.
This corresponds to going to an unbranched  cover of the $\mT^2$.

This has in turn has implications for the associated Belyi pairs.
We recall that the Belyi map is a branched cover  from $\mathbb{T}^2$ to $\mathbb{P}^1$:
\begin{equation}
\beta:\, \mathbb{T}^2\,\rightarrow \, \mathbb{P}^1 ~.
\end{equation}
with degree $d$ equal to the number of edges in the dimer
and ramifications over  $ \{ 0 , 1 , \infty \} $ related to the structure
of the dimer. Consider now an unbranched cover $\widehat{\mathbb{T}}^2$ of the torus $\mathbb{T}^2$:
\begin{equation}
f   :\, \widehat{\mathbb{T}}^2\, \rightarrow \mathbb{T}^2 ~.
\end{equation}
of degree $N$. Such a map $f$ has $N$ inverse images for every point on $ \mathbb{T}^2$
and its derivative is never vanishing. These properties are clear from the picture of enlarging
the unit cell.

We consider the composition of $\beta$ and $f $
\begin{equation}
\beta_{f }:\, \widehat{\mathbb{T}}^2\rightarrow \mathbb{P}^1 ~,
\end{equation}
where
\begin{equation}
\beta_{f }=\beta\circ f  ~.
\end{equation}
With a local coordinate $ z$ on $  \widehat{\mathbb{T}}^2$ we have that
\bea
 \pd_z \beta_f = \pd_z \beta ( f ( z )) =  \pd_f \beta  \pd_z f
\eea
The only zeros of $  \pd_z \beta_f  $ occur when  $\pd_f \beta ( f ) ) = 0 $
where $ \beta ( f ) \in \{ 0 , 1, \infty \} $. We deduce that $ \beta_f $ is also a
Belyi map.  Each ramification point of $ \beta$ lifts to $N$ ramification
points of $ \beta_f$ with the same ramification number at each.
The number of faces of the new dimer is $N$ times that of the
original dimer. This translates into multiplying the number of factors
in the gauge group by $N$ as expected  \cite{Douglas:1996sw}.
Thus, the pair  $ ( \ \widehat{\mT}^2 ,    \beta_{ f } ) $ is the Belyi  pair
associated to the orbifolded SCFT.

\noindent
{\bf Automorphisms of the cover as automorphisms of the orbifold}

There is a permutation description of the covers $ f  $ in terms of pairs $ s_1 , s_2 \in S_N $ such that
\bea
s_1\, s_2\, s_1^{-1}\, s_2^{- 1} = 1 ~,
\eea
where equivalent pairs are related by conjugation.
Automorphisms correspond to
$ \gamma s_i \gamma^{-1} =s_i $.
These correspond to  holomorphic  automorphisms
$ \phi : \widehat{\mT}^2  \rightarrow \widehat{\mT}^2 $
of the covering map $f$  which obey
\bea
f \circ \phi = f  ~.
\eea
A simple manipulation with the Belyi maps
shows that these lead to automorphisms of the Belyi pair corresponding
to the orbifolded theory
\bea
\beta_{f } \circ \phi = \beta \circ f  \circ \phi = \beta \circ f  = \beta_{f} ~.
\eea
which in turn, following the discussion in Section \ref{PermW},   correspond to
permutation symmetries preserving $W_+ , W_- $
in the SCFT for the orbifold

In  Appendix \ref{orbexamps} we consider a class of orbifolds with $\IZ_N$ automorphism. In particular, the permutation triples
$ (\Sigma_B , \Sigma_W , \Sigma_{ \infty} =  ( \Sigma_B  \Sigma_W )^{-1} )$
 for the orbifolds will be explicitly described. The  expected $\IZ_N$ conjugation symmetries demonstrated.
While we will use the picture of enlarging unit cells to write these down
in some examples, they are  determined from the permutation data $ ( \s_B , \s_W ) $
for the original dimer along with  $ (s_1 , s_2 ) $ describing the unbranched cover.
We expect that it should be possible to construct a general formula
 expressing $\Sigma_B, \Sigma_W$ in terms of  $ (\s_B , \s_W, s_1, s_2)$.
It would be a useful technical tool in studying orbifolds of toric Calabi--Yaus to find this construction,
especially in connection with  the investigations  in \cite{Hanany:2010cx, Davey:2010px, Hanany:2010ne}.
We leave this  for the  future.

\noindent
{ \bf Complex structure of the cover}

\begin{figure}
\centering
\includegraphics[scale=1]{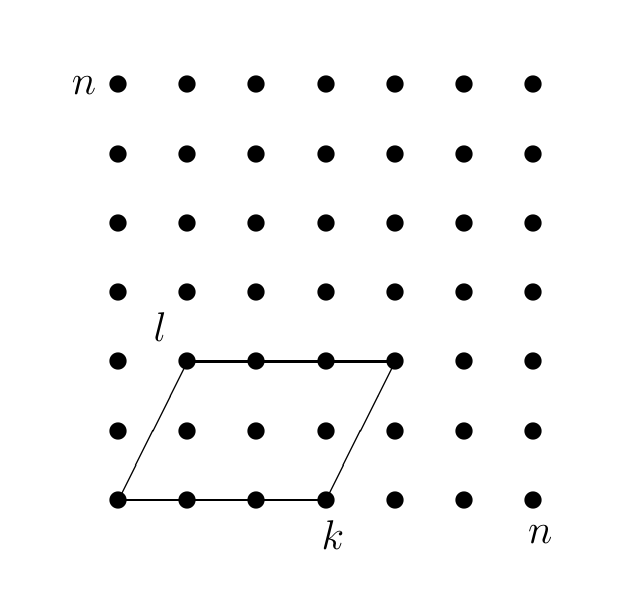}
\caption{Unit cell for the torus.}
\label{toruscover}
\end{figure}

The complex structure of the cover
$ \widehat{\mathbb{T}}^2 $ can be described as a function
of the complex structure of $ \mathbb{T}^2 $.
The covers of degree $N$ are known to be in one-to-one correspondence
with positive integers $ k , p , l $, such that  $ k \, p = N $ and $ l \le N $ (we refer to, {\em e.g.}, \cite{GT}). (See Figure~\ref{toruscover}.)

For each such $(p,k,l)$, the complex structure of the cover
$\tau_{\rm cover}$ is given in terms of the complex structure of
 the target torus $\tau$ by
\bea\label{tauofcover}
\tau_{ \rm cover} (p,k,l) = { p  + l\  \tau \over k } ~.
\eea

Note that while covers of the toroidal Belyi curve play an important role
in understanding  the orbifolded theory, there is no such role for  covers
of the untwisted curve that is immediately obvious.

\section{Consistent dimers and zig -zag paths}
\label{consdimzig}

While consistent four-dimensional SCFTs are associated with bipartite graphs on a torus, it is not true that all such graphs determine a four-dimensional SCFT.
The known consistency conditions are expressed in terms of {\em zig-zag paths}.\footnote{
Historically, these were called {\em train tracks} or {\em rhombus loops} in the mathematics literature~\cite{Kenyon1}.}
Zig-zag paths are also used to construct the $(p,q)$-web which gives rise to the SCFT.
The $(p, q)$ charges can be easily extracted from the toric diagram.
This yields a string construction in terms of NS$5$-branes via the so-called ``fast-inverse algorithm''~\cite{hananyvegh}.

It is important to understand the zig-zag paths and their consistency conditions in terms of the permutation triples.
This will be the topic of this section.

\subsection{Zig-zag paths from permutations }
\label{zzint}

Zig-zags will be described in terms of permutations of $2d$ objects $ \{1^- , 2^- , \ldots , d^- , 1^+ , \ldots , d^{+} \} $.
A related combinatoric discussion  has appeared in the literature in~\cite{GarciaEtxebarria:2006aq, Franco:2007ii}.
The contribution here is to develop that into a simple formulation in terms of operations on symmetric group
elements.

Let us recall how the zig-zag paths are defined in terms of the dimer.
Pick any edge. Start a zig-zag path by approaching a black vertex, while staying close to the edge and choosing to be on the side of the edge such that the black vertex is on the left.
The corresponding string is $(i^- \ldots)$, where $i$ is the label of the edge on which we start.
We turn right at the black vertex, cross the next edge, and approach a white vertex on the right.
We append the label $j$ of that edge to the string with a plus sign to get $ (i^- j ^+ \ldots) $.
Turning left at the white vertex, we find ourselves along the edge labelled $k$.
We cross that edge and extend the string to $ (i^- j^+ k^- \ldots) $, continuing in this manner until the starting point is again reached.
We then repeat the procedure at another edge to construct another zig-zag path and an associated cyclic string of numbers labelled alternately by $ - $ and $ + $ until all the edge labels are exhausted with both signs.

The corresponding construction in terms of permutations is as follows.
Define a permutation $ \cZ (\s_B , \s_W) $ of the $2d$ objects
$ \{1^- , 2^- , \ldots , d^- , 1^+ , \ldots , d^{+} \} $ by
\bea\label{constructS2dperm}
\cZ(k^-) = \s_B (k)^+ ~, \cr
\cZ(k^+) = \s_W^{-1} (k)^- ~,
\eea
where $ 1 \le k \le d $.
Equivalently we can describe this as a permutation of $ \{1, \ldots, 2d \}$.
Given $ \s_B , \s_W $ which are permutations in $S_d $ define a permutation $ \cZ (\s_B , \s_W) $ in $ S_{2d} $ as follows:
\bea
\cZ (k) = \s_B (k) + d && 1 \le k \le d ~, \\
\cZ (k) = \s_W^{-1} (k - d) && d+1 \le k \le 2 d ~. \nn
\eea
It is easy to put these ideas to work in our $\IC^3$ example.
Inspection of $\sigma_B$ and $\sigma_W$ tells us that the zig-zag paths are
\begin{equation}
z_1=(1^-\, 2^+) ~, \qquad z_2=(2^-\, 3^+) ~, \qquad z_3=(3^-\, 1^+) ~.
\end{equation}

Since $ \s_B $ and $ \s_W $ are by definition (see Section \ref{sec:dimerperms})
giving the permutations of edges around the black and white nodes,
and given the use of  $ j = \sigma_B (i) $, and $ k = \sigma_W^{-1} (j) $ in (\ref{constructS2dperm}),
it follows that the standard zig-zag construction ~\cite{Kenyon1, hananyvegh, GarciaEtxebarria:2006aq, Franco:2007ii}
 agrees with what we have given.

We will now make a few observations about $\cZ$.

\begin{itemize}

\item
The number of cycles in $ \cZ $ is the number of zig-zag paths in the dimer
and it is equal to the number of cycles of $ \s_B \s_{W}^{-1}$.
This is the number of  faces in the untwisting of the dimer.

\item $ \cZ^2(k) = \s_B \s_W^{-1} (k) $ maps $\{1, \ldots, d \} $ to $ \{1, \ldots , d \} $.
Similarly, $\cZ^2$ maps $ \{d+1 , \ldots , 2d \} $ to itself, where the permutation $ \s_W^{-1} \s_B$ of $S_{d} $ is embedded in $S_{2d} $ by shifting $\{1 , \ldots , d \} $ to $\{d+ 1 , \ldots , 2 d \} $.

\item
Given a cycle in $S_{2d}$, we may reduce to a string with possibly repeated indices
by taking the integers modulo $d$. A cycle with repeated indices is a self-intersecting zig-zag path.
Any dimer defined by a pair $ \s_B , \s_W $, which leads to such a self-intersecting zig-zags is inconsistent.

\item
There is another condition on consistency invoking pairs of zig-zag paths.
This says that a pair can intersect only once when lifted to the universal cover.
We have yet to find an elegant description of this condition in terms of the permutation group.

\item
Zig-zags can as well be defined for the pair $ (\s_B , \s_W^{-1}) $.
$ \cZ (\s_B , \s_W^{-1})$ gives the zig-zag paths on the untwisted curve.
Here we  get a permutation in $S_{2d}$ with the same number of
cycles as $ \s_B \s_W $ or $ \s_W \s_B $. For the case of $ \cN=4$,
the $ \cZ ( \s_B , \s_W^{-1} ) =  ( 1^- 2^+ 3^- 1^+ 2^- 3^+ ) $.
This permutation gives a precise description of the face in the toroidal
Belyi curve of $\cN=4$. Indeed this procedure gives the edges around each   face,
hence the fields charged under each  gauge group, for the SCFT. This
gives the set of fields appearing in each of the equations for $a$-maximization,
which is reviewed in Appendix \ref{nutshell}.  In  the IIA brane picture
the faces correspond to D$5$-branes cut-out by NS$5$-branes on $\mT^2$.

\end{itemize}

The method we have described is a more convenient way of generating the zig-zag paths than listing and comparing perfect matchings.
An important simplification was already achieved in~\cite{GarciaEtxebarria:2006aq, Franco:2007ii}.
Using the description of dimers in terms of permutation triples in $S_d$, we have found a further simplification of the zig-zag generating algorithm in terms of a permutation in $S_{2d}$.
Relations between permutations triples in $S_{d}$ and those in $S_{2d}$ play a role in Belyi theory, where they reflect the operation of cleaning a Belyi map.
Cleaning is used to construct Galois invariants such as the cartographic group \cite{schnepsbook}.

\begin{figure}[!h]
\centering
\includegraphics[scale=1]{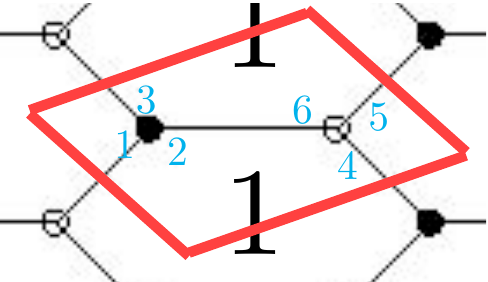}
\caption{$S_6$ description for  $\mathbb{C}^3$ dimer.}
\label{constellation}
\end{figure}

{\em Cleaning} amounts to associating to any dessin a related dessin, called a cleaned version
where the black and white vertices are all turned into black vertices,
and white vertices are introduced in the middle of the edges \cite{Joubert}.
Then we get, by the construction described in Section \ref{sec:dimerperms},
 data associated to the dessin as permutation triples  in $S_{2d} $.
Equivalently, we can stick with the original dessin and rather than attaching labels
to the edges, we attach them to half-edges, {\em i.e.}, there are two numbers
on each edge, one near the black vertex and one near the white (see for example \cite{lando-zvonkin}
where this is described under the heading of constellations).
This is illustrated for the $ \IC^3$ dimer  Figure~\ref{constellation}.
From here we read off the permutation  triple in $S_6$ as
\begin{equation}
\label{constellationC3}
\alpha=(15)\, (26)\, (34)\qquad \sigma =(132)\,(465)\qquad \varphi=(163524)
\end{equation}
The permutation $ \cZ ( \s_B , \s_W^{-1} ) $ we discussed earlier corresponds to  $ \varphi$
here. The procedure we described for constructing $ \cZ$ earlier
is seen as implementing the multiplication of a permutation
in $S_{2d} $,  constructed by collecting the cycles of $\s_B $
and $\s_W$, with another permutation in $S_{2d} $ which is
of cycle structure $[2^d]$.

The permutation descriptions can also  be connected to other constructions such
as triangulations and rhombi that have appeared in the literature.
Given a dessin we can construct a triangulation in the covering space  $ \mT^2$ or the untwisted $ \Sigma $
 by choosing a point inside each polygon and joining it with the neighboring black and white dots (see, {\em e.g.},~\cite{Kenyon1}
in the statistical physics context and \cite{Joubert} in the  Galois theory context).
This defines a triangulation such that one of the edges of each triangle is an edge of the dimer.
Since each such dimer edge has a number between $1$ and $d$ and each edge has two adjacent triangles, it is natural to label each triangle by the edge number.
Thus, the dessins determine a triangulation, with faces labelled $ 1^+ , 1^- , 2^+ , 2^- , \ldots , d^+ , d^- $.
The zig-zag paths can be described as a sequence of integers of alternating signs.
Note that, had we be considering the isoradial embedding,
 these triangles would form the rhombic lattice of \cite{hananyvegh}, where
 each rhombus is labelled by the number of the dimer edge it has as its diagonal.

\subsection{Homology and $(p,q)$ webs from zig-zag permutations}
\label{hompq}

We have described the zig-zags as permutations in the group  $S_{2d}$  of $\{ 1^+ , \ldots, d^+ , 1^- , \ldots , d^- \} $.
We wish to obtain a basis for the homology $H^1 ( \mT^2 , \IZ ) $ and $(p,q)$ webs from an analysis of the zig-zag permutations.
The cycles in this permutation each describe a single zig-zag path.
Let us label these cycles $z_i$, $i=1,\ldots, M$.
We can express the $z_i$ in homology  of the $ \mT^2$,
 as linear combinations of the $a$-cycle and the $b$-cycle:
$z_i = p_i\, a + q_i\, b$, where $p_i, q_i\in \IZ$ define the $(p,q)$ web that yields the theory
(see for example the review \cite{yam08}). The combinatoric description
of zig-zags we have given  as cycles in $S_{2d} $ gives a natural way to construct
the intersection matrix of these homology cycles, which we now describe.
 If $\alpha^-$ appears in the cycle $z_i$ and $\alpha^+$ appears in the cycle $z_j$, this contributes $+1$ to the intersection $z_i\cap z_j$.
When $\beta^+$ appears in $z_i$ while $\beta^-$ appears in $z_j$ this contributes $-1$ to the intersection.
Given a set of zig-zag paths, we can construct the intersection matrix $I_{ij}$, which counts the total intersection of $z_i$ with $z_j$.
Clearly, $I_{ij} = -I_{ji} := \la z_i , z_j \ra$.

The above construction alows us to  prove an interesting property of  the intersection matrix
$I_{ij}$ which is expected from physical considerations. We will show that
the elements in any  given row or a given column of the intersection matrix must sum to zero.
Suppose ${\cal Z} = z_{1} \ldots z_{M} = (e_{1,1}^- \ldots e_{1,i_1}^+)(e_{2,1}^- \ldots e_{2,i_2}^+)\ldots(e_{M,1}^- \ldots e_{M,i_M})$.
The integers $e_{i,m}$ within the cycle $z_i$ are unique, otherwise there would be a self-intersection.
There are an equal number of minus and plus signs within each cycle.
Each element $e_{i,m}^-$ in $z_{i}$ contributes $+1$ to the intersection with some other cycle $z_j$.
Likewise, $e_{i,n}^+$ contributes $-1$ to the intersection with some other cycle $z_k$.
These contributions sum to zero.
Thus, fixing $i$, we see that $\sum_j I_{ij} = 0$.

Fixing conventions for the generators of $H^1 ( \mT^2 , \IZ ) $
so that $\la a, b \ra = 1$, we have $\la a, z_i \ra = q_i$ and $\la b, z_i \ra = -p_i$.
Noting that $I_{ij} = p_i q_j - q_i p_j$, summing on $j$ implies that
\be
p_i \sum_{j=1}^M q_j = q_i \sum_{j=1}^M p_j ~.
\ee
Because this is true for any $i$, if the matrix $I$ is itself non-zero, it follows that
\be
\sum_{i=1}^M (p_i, q_i) = 0 ~.
\ee
This is expected from toric geometry as well as from the brane constructions \cite{yam08}.

Another general property of the matrix $I_{ij}$ is that it always has rank two.
This follows because the $z_i$ are linear combinations of the $a$- and $b$-cycles of the torus on which the dimer sits.
When we diagonalize the intersection matrix, which is skew-symmetric, we find that there are two non-zero eigenvalues $\lambda^\pm = \pm i \lambda \ne 0$ and $M-2$ zero eigenvalues. We denote  the null eigenvectors
as  $ N^{(a)}_i$ and the non-zero eigenvectors $E^{\pm }_{ i}$.
The physical meaning of the null vectors is that they express the linear combinations of the zig-zag paths that vanish in homology.
That is to say, $N^{(a)}_i z_i = 0$.
Na\"{\i}vely, one might expect that $E^{(\pm)}_i z_i = \xi (a \pm b\, i)$, where $\xi$ is some constant of proportionality fixed by the requirement that $\la a, b \ra = 1$.
This is not the case.
Proceeding in this manner does not guarantee that $p_i$ and $q_i$ are integers.

Rather, the simplest way to construct a homology basis given the $z_i$ and $I_{ij}$
obtained from the permutations $ ( \s_B , \s_W)  $, is to choose  $j,k$ such that the element $I_{jk}$ is a minimal entry  of the intersection matrix greater than zero.
We then assign $z_j = \alpha\, I_{jk}\, a$ and $z_k = \alpha^{-1}\, b$.
Consider any $z_i = p_i\, a + q_i\, b$.
We have
\bea
I_{ij} = \la z_i, z_j \ra = \la p_i\, a + q_i\, b, \alpha\, I_{jk}\, a \ra = -\alpha\, I_{jk}\, q_i &\Longrightarrow& q_i = -\alpha^{-1}\, \frac{I_{ij}}{I_{jk}} ~, \\
I_{ik} = \la z_i , z_k \ra = \la p_i\, a + q_i\, b, \alpha^{-1}\, b \ra = \alpha^{-1}\, p_i &\Longrightarrow& p_i = \alpha\, I_{ik} ~.
\eea
We can choose $\alpha$, which divides $I_{jk}$, so that all the $p_i$ and $q_i$ are integers.
The constraint from the null vectors is that $N^{(a)}_i I_{ij} = N^{(a)}_i I_{ik} = 0$.
We find in all the examples we studied explicitly that this assignment is $SL(2,\IZ)$
 equivalent to the $(p,q)$-web discussed in~\cite{hananyvegh}.

Once we have expressed $z_i = p_i a + q_i b $, we know the outward pointing normals of the toric diagram.
The toric diagram is expressed as the dual cone by a standard algorithm~\cite{aharony}.
Knowing the vertices of the dual cone, we may construct the Newton polynomial~\cite{horivafa}.

\subsubsection{Examples}

Let us illustrate this by working out the examples of the $\cN=4$, conifold, and SPP theories.

\paragraph{$\cN=4$:}
The $\cN=4$ theory has zig-zag paths, constructed according to the description in
Section \ref{zzint}, given by
 $z_1 = (1^-\, 2^+)$, $z_2 = (2^-\, 3^+)$, $z_3 = (3^-\, 1^+)$.
The intersection matrix is
\be
I_{\IC^3} = \left(\ba{ccc} 0 & -1 & 1 \cr 1 & 0 & -1 \cr -1 & 1 & 0 \ea \right) ~.
\ee
The eigenvector corresponding to the zero eigenvalue is $(1,1,1)^T$.
This implies that $z_1 + z_2 + z_3 = 0$.
We notice that $I_{13} = I_{21} = I_{32} = 1$ are the minimum positive elements of the intersection matrix.
We may therefore select the first option and put $z_1 = a$ and $z_3 = b$.
We then compute
\be
z_2 = p\, a + q\, b = - \la z_3, z_2 \ra a + \la z_1, z_2 \ra b = I_{23}\, a - \frac{I_{21}}{I_{13}}\, b = -a - b ~.
\ee
The result satisfies the constraint from the null vector.

Suppose we had made some other choice.
If, for example, we put $z_2 = a$ and $z_1 = b$, we will find that $z_3 = -a - b$.
We have
\be
\left(\ba{cc} 0 & 1 \cr -1 & -1 \ea \right) \left(\ba{c} a \cr b \ea \right) = \left(\ba{c} b \cr -a-b \ea \right) ~.
\ee
The matrix on the left hand side is in $SL(2,\IZ)$.
It maps $z_1 = a$, $z_3 = b$ corresponding to using $I_{13}$ to define the basis in homology to $z_1 = b$, $z_3 = -a-b$ corresponding to using $I_{21}$ to define the basis.
Hanany and Vegh~\cite{hananyvegh} choose $z_1 = -a$, $z_2 = -b$, $z_3 = a+b$.
This is as well $SL(2,\IZ)$ equivalent to the identifications we have listed.

\paragraph{Conifold:}
The conifold theory has the zig-zag paths, $z_1 = (1^-\, 2^+)$, $z_2 = (2^-\, 3^+)$, $z_3 = (3^-\, 4^+)$, $z_4 = (4^-, 1^+)$.
The intersection matrix is
\be
I_{{\rm conifold}} = \left(\ba{cccc} 0 & -1 & 0 & 1 \cr 1 & 0 & -1 & 0 \cr 0 & 1 & 0 & -1 \cr -1 & 0 & 1 & 0 \ea \right) ~.
\ee
For the conifold, the null vectors are $(1,0,1,0)^T$ and $(0,1,0,1)^T$.
Thus, $z_1 + z_3 = z_2 + z_4 = 0$.
Our algorithm presents four choices:
$(z_1, z_4) = (a,b)$, $(z_2, z_1) = (a,b)$, $(z_3, z_2) = (a,b)$, or $(z_4, z_3) = (a,b)$.
The first option yields the result of~\cite{hananyvegh}:
\be
z_2 = I_{24}\, a - \frac{I_{21}}{I_{14}}\, b = -b \qquad
z_3 = I_{34}\, a - \frac{I_{31}}{I_{14}}\, b = -a ~.
\ee
The null-vector constraints are directly checked to be satisfied.
The other choices are $SL(2,\IZ)$ equivalent to this one.

\paragraph{SPP:}
The suspended pinch point has the zig-zag paths, $z_1 = (1^-\, 3^+\, 7^-\, 6^+)$, $z_2 = (2^-\, 1^+\, 5^-\, 4^+)$, $z_3 = (3^-\, 2^+)$, $z_4 = (4^-\, 7^+)$, and $z_5 = (6^-\, 5^+)$.
The intersection matrix is
\be
I_{{\rm SPP}} = \left(\ba{ccccc} 0 & 1 & -1 & 1 & -1 \cr -1 & 0 & 1 & -1 & 1 \cr 1 & -1 & 0 & 0 & 0 \cr -1 & 1 & 0 & 0 & 0 \cr 1 & -1 & 0 & 0 & 0 \ea \right) ~.
\ee
For SPP, the null vectors are $(1,1,0,0,1)^T$, $(-1,-1,0,1,0)^T$, $(1,1,1,0,0)^T$.
Hence, $z_1 + z_2 + z_5 = -z_1 - z_2 + z_4 = z_1 + z_2 + z_3 = 0$.
Noting that $I_{12} = 1$, we have
\be
z_1 = a ~, \quad z_2 = b ~, \quad z_3 = -a-b ~, \quad z_4 = a+b ~, \quad z_5 = -a-b ~.
\ee
In~\cite{hananyvegh}, the zig-zags are expressed as
\be
z_1 = a+b ~, \quad z_2 = -a ~, \quad z_3 = -b ~, \quad z_4 = b ~, \quad z_5 = -b ~.
\ee
We see that
\be
\left(\ba{cc} 1 & 1 \cr -1 & 0 \ea \right) \left(\ba{c} a \cr b \ea \right) = \left(\ba{c} a+b \cr -a \ea \right) ~.
\ee
The two choices of homology basis are therefore $SL(2,\IZ)$ equivalent.

\paragraph{Orbifolds:}
The examples we have given are somewhat trivial since the intersection matrix contains only $0, \pm1$.
This is no longer true for the orbifold theories, which we discuss in Appendix
 \ref{orbexamps}.
The same recipe allows us to write the zig-zag paths in terms of the $a$- and the $b$-cycles of the torus.

\section{A conjecture about Belyi complex structure and $R$-charges }\label{tauBtauR}

We have so far introduced a very convenient set of tools which allow for an efficient combinatorial/analytic codification of the low energy theory on D$3$-branes probing a toric CY$_3$ cone.
The analytic description is under the form of a Belyi pair, in which we should specify a torus as well as a
holomorphic  map from this torus into a marked $\mathbb{P}^1$.
This torus contains, in turn, the dimer, which can be then thought of as a dessin d'enfant.
It is then natural to ask for the physical significance of this construction, which in particular requires a specific choice of the torus where the dimer lives.
Various brane or geometrical  interpretations for the dimer have been provided in the literature.
However, none of them affords a first principles understanding of the isoradial dimer.
Let us first discuss these isoradial dimers.

\begin{figure}[h]
\begin{center}
\includegraphics[scale=0.5]{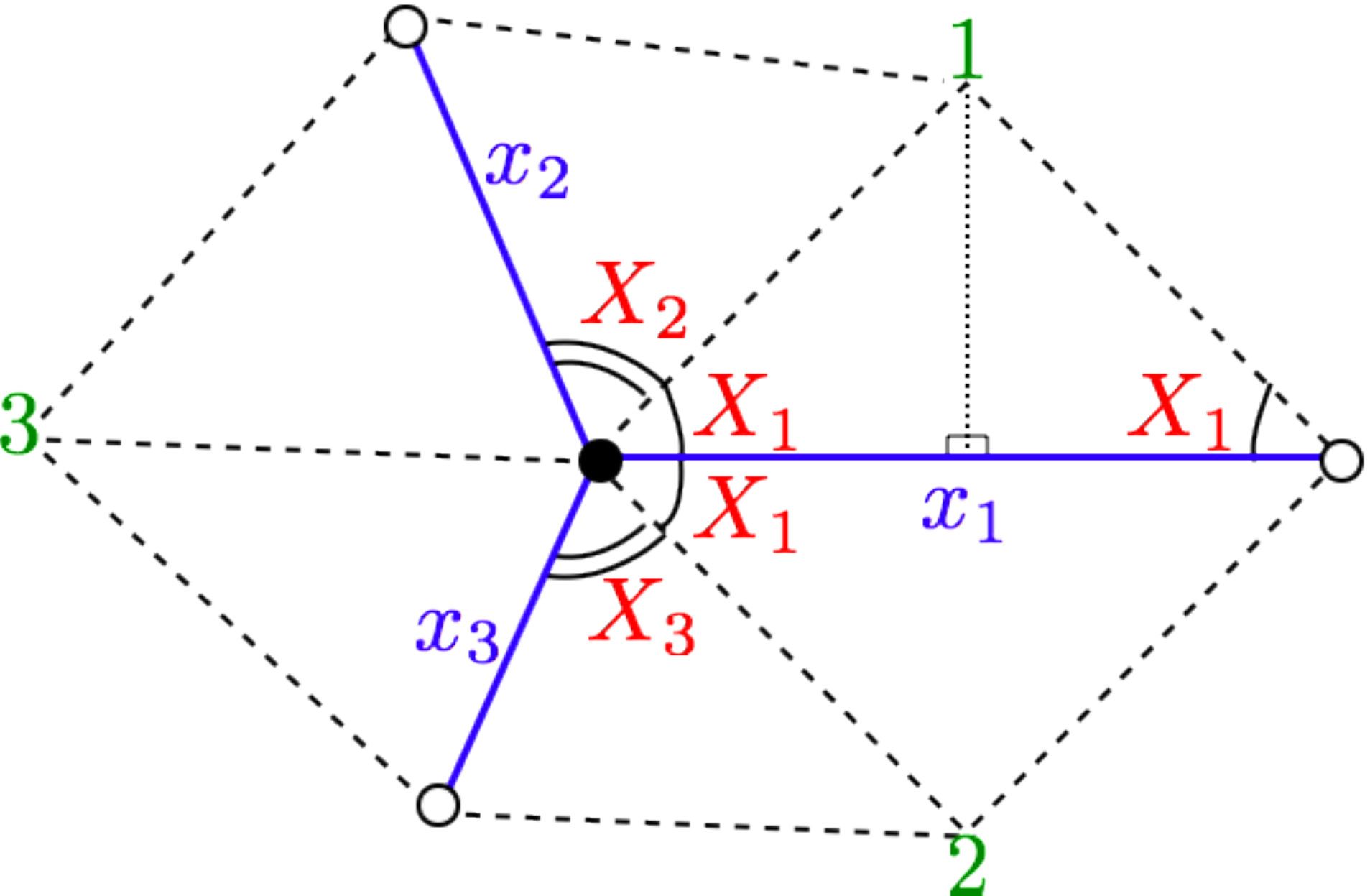}
\end{center}
\caption{The isoradial embedding; the dashed lines are of equal length.}
\label{fig:angles}
\end{figure}

The edges of the bipartite graph are associated to fields in the gauge theory.
Each field has an $R$-charge in the IR.
The $R$-charge is determined by $a$-maximization~\cite{Intriligator:2003jj}.
Not all bipartite graphs give consistent CFTs.
For some bipartite graphs, the $a$-maximization can result in vanishing $R$-charges.
A consistency condition is that the bipartite graphs must admit isoradial embeddings \cite{hananyvegh}.
The isoradial embedding demands that the lines from the center of a face to a vertex in the graph be of equal length.
These are the dashed lines in Figure~\ref{fig:angles}.
We normalize this distance to unity.
This condition constrains  the lengths of the edges in the graph.

We first restrict to these isoradial bipartite graphs.
Then, we implement the $a$-maximization procedure (reviewed briefly in Appendix \ref{nutshell})
 to determine the charges.
It was observed that, if we associate the $R$-charges to angles in the bipartite graph in the isoradial setup, then the vanishing of the beta function guarantees that the bipartite graph lives on a torus, or equivalently can be drawn as a periodic structure on the plane.

To get $\tau_R$, we draw what may be called the $R$-dimer, which is simply the bipartite graph, drawn according to the rule that each edge has a length determined by the $R$-charge of the corresponding field.
To compute the complex structure parameter of the torus, one computes the lengths of the edges
and relative angles  that prescribe the unit cell.

The angle $X$ that the line from the centre of the face makes to a vertex associated with the edge $x$ is fixed by the $R$-charge of the associated quantum field:
\be
X = \frac{\pi R_x}{2} .
\label{eq:angle}
\ee
The angle between two edges $x_1$ and $x_2$ that meet at a vertex is $X_1+X_2$.
The sum of the angles at a vertex is $2\pi$.
This is simply the condition that the combinations of fields that appear in the superpotential have a total $R$-charge two.
Dropping a perpendicular from the center of the face to the midpoint of an edge, it is clear that the length associated to $x$ is $2\cos X$.

We can start with any chosen edge to be horizontal and draw the remaining edges according to the above to generate a periodic lattice.
A unit cell can be chosen to have its corners at one of the vertices of the dimer.
Because of the double periodicity, there are two lattice displacements which can take the original vertex to nearest copies of the same vertex.
Each of these lattice displacements is a sum of displacements along edges of the dimer, which are drawn as straight lines joining a black vertex to a white vertex.
Each of these edge displacements can be represented as a complex number.
Adding the edge displacements we get the lattice displacements.
The ratio of the two lattice displacements gives a $\tau$ parameter, which is the complex structure of the base torus as determined by the $R$-charges.
This is what we call $\tau_R$.

For $\IC^3$, $\tau_R = {1 \over 2} + {i \sqrt {3} \over 2}$.
Very interestingly the same value is obtained for the complex structure of the Belyi curve, using the explicit construction of the Belyi pair in Section \ref{sec:dimerperms}, {\em i.e.},
$\tau_B = {1 \over 2} + {i \sqrt {3} \over 2}$.
This is the value such that $j (\tau) = 0$.
A similar story applies to the conifold.
We find that $ \tau_R = \tau_B = i $.
This is the value such that $j (\tau) = 1728$.
As explained in Section \ref{sec:orbs},
for any orbifold, there is an associated unbranched cover $\widehat {\mT}^2 $
of the original $ \mT^2$.   The unbranched cover is specified combinatorially
by a pair of commuting permutations up to
 conjugation equivalence, or alternatively by three  integers as shown
 in Figure \ref{toruscover}. The complex structure $ \tau $ of the
cover  $ \widehat{\mathbb{T}}^2 $ is determined by this
 combinatoric data of the cover,  along with the  original complex structure,
 using (\ref{tauofcover}).  The same relation  gives
 $ \tau_B ( { \widehat \mT} ^2  ) $ in terms of  $ \tau_B ( \mT^2 ) $
 and $ \tau_R (  { \widehat \mT} ^2  )  $ in terms of $ \tau_R ( \mT^2 ) $.
 This means that, once we know $ \tau_B = \tau_R $ for a given toric Calabi-Yau,
 the agreement will continue for their orbifolds. So for branes at  $\IC^3$
 and the conifold, as well as their orbifolds,  we have $ \tau_B = \tau_R $.

Rather than being more  explicit about the simpler examples, let us illustrate the calculation with the SPP theory.
In Figure~\ref{SPPdimer}, we choose the origin to be the vertex where the edges $2$, $3$, and $7$ meet.
The endpoints of the torus that describe the unit cell are at the black vertices where the same edges meet.
Thus, to determine the complex structure, one must compute the vectors:
\be
P = -\vec{x}_6 + \vec{x}_5 ~, \quad
U = \vec{x}_3 + \vec{x}_1 + \vec{x}_6 + \vec{x}_7 ~.
\ee
For the case of SPP, consistent with $a$-maximization~\cite{Intriligator:2003jj}, there are three angles associated to the fields:
\be
\alpha = X_1 = \left(1 - \frac{1}{\sqrt3} \right) \pi ~, \quad
\beta = X_2 = X_3 = X_5 = X_6 = \frac{1}{2\sqrt3} \pi ~, \quad
\gamma = X_4 = X_7 = \left(\frac12 - \frac{1}{2\sqrt3} \right) \pi ~.
\ee
The subscripts denote the edges.
The edges that meet at the vertices in the graph enable us to deduce that
\be
\pi = \alpha + 2 \beta = 2 \beta + 2 \gamma ~.
\ee
In this way, all the angles can be expressed in terms of $\gamma$.
After a bit of trigonometry, we may succinctly express
\be
P = \chi^{-1}(\chi^4 - 1) ~, \quad
U = \chi(3 - \chi^2) ~,
\ee
where $\chi = e^{i \beta}$. 
The modular parameter is then the ratio
\be
\tau_{\rm SPP} = \frac{P}{U} = \frac{-1+e^{\frac{2i\pi}{\sqrt3}}}{3e^{\frac{i\pi}{\sqrt3}}-e^{\frac{2i\pi}{\sqrt3}}} \approx -0.16465 + 0.549718\, i ~.
\ee

\begin{figure}[h!]
\begin{center}
\includegraphics[scale=0.6]{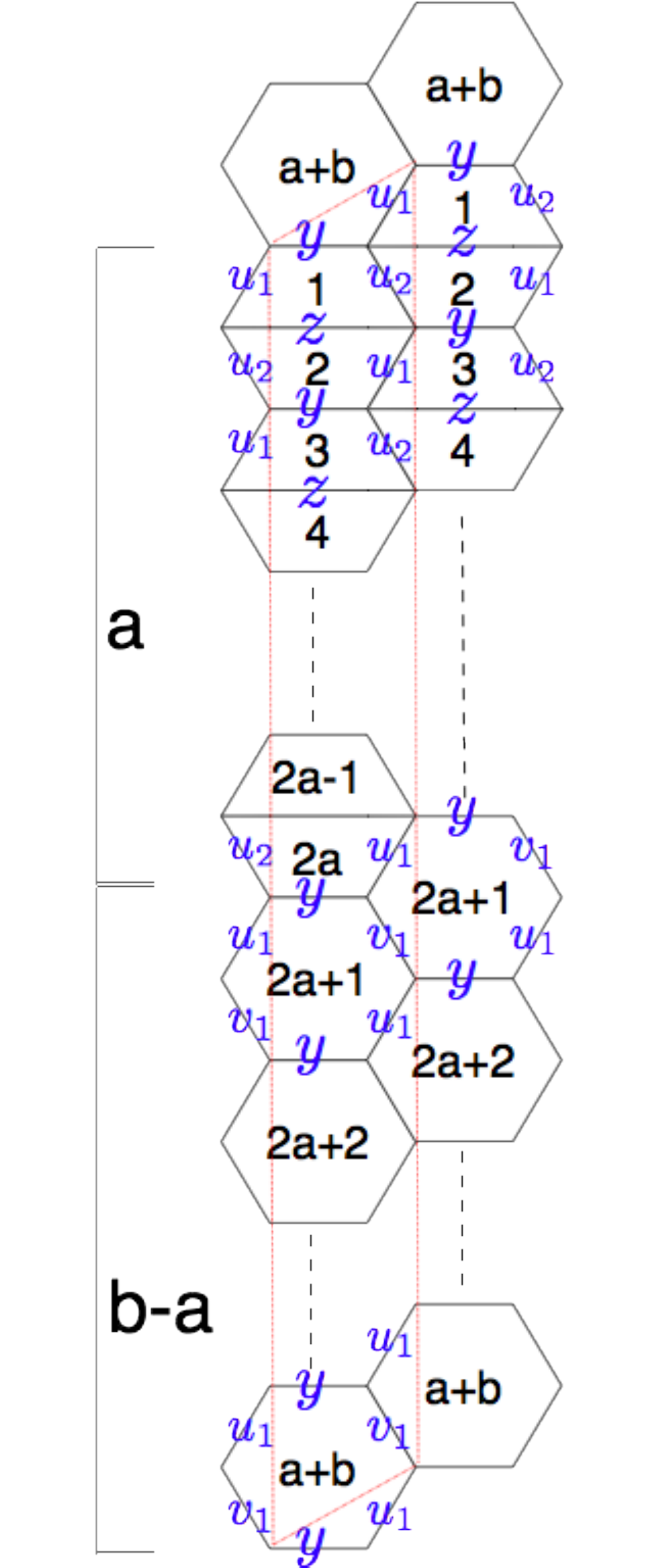}
\end{center}
\caption{Dimer for the $L^{aba}$ theory.}
\label{fig:laba}
\end{figure}

We may, of course, perform a similar calculation for a more complicated dimer.
Consider the $L^{aba}$ theories~\cite{FHMSVW05}, whose bipartite graph is shown in Figure~\ref{fig:laba}.

Here, the $R$-charges consistent with $a$-maximization are
\be
R(u_1) = R(y) = \frac13 \frac{b-2a+w}{b-a} ~, \quad
R(u_2) = R(z) = \frac12 R(v_1) = \frac13 \frac{2b-a-w}{b-a} ~,
\ee
where $w = \sqrt{a^2 + b^2 - ab}$.
The angles associated to the edges may therefore be expressed in terms of $U_1 = \frac\pi2 R(u_1)$ and $U_2 = \frac\pi2 R(u_2)$, etc. with relations $\pi = U_1 + U_2 + Y + Z = U_1 + V_1 + Y$.
Placing the origin at the top left corner of the fundamental cell, we must therefore compute
\be
P = \vec{y} - \vec{u}_1 ~, \quad
U = a (\vec{u}_1 + \vec{u}_2) + (b-a) (\vec{u}_1 + \vec{v}_1) ~.
\ee
After a bit of geometry, we find that
\be
P = (\chi_1 + \chi_1^{-1})(1 + \chi_2^{-2}) ~, \quad
U = b (\chi_1 + \chi_1^{-1}) \chi_2^{-2} - a (\chi_2 + \chi_2^{-1}) \chi_1 \chi_2 - (b-a) (\chi_2^2 + \chi_2^{-2}) \chi_1 ~,
\ee
where $\chi_j = e^{i U_j}$.
The modular parameter $\tau_{L^{aba}} = \frac{P}{U}$.

The theory $L^{121}$ is the same as SPP.
\begin{figure}[h]
\begin{center}
\includegraphics[scale=0.5]{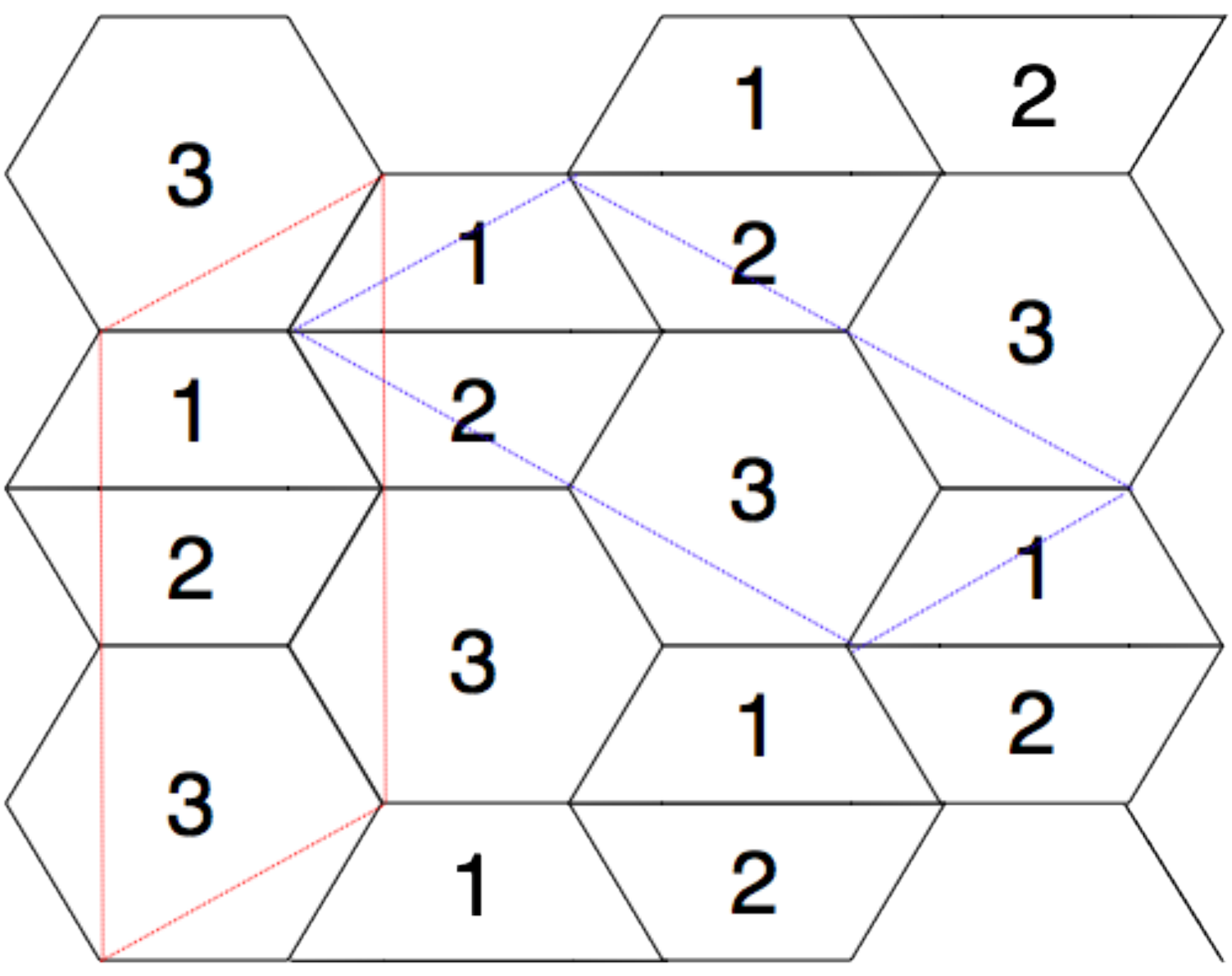}
\end{center}
\caption{Dimer for the $L^{121}$ theory with equivalent fundamental domains sketched.}
\label{fig:L121}
\end{figure}
We find
\be
\tau_{L^{121}} = \frac{-1 + e^{\frac{2i\pi}{\sqrt3}}}{-2 + 3 e^{\frac{i\pi}{\sqrt3}} + e^{\frac{2i\pi}{\sqrt3}}} \approx 0.297811+ 0.331435\, i ~.
\ee
One can check computationally that $j(\tau_{\rm SPP}) = j(\tau_{L^{121}})$.
Thus, the two values of $\tau$ are related by a modular transformation.
We derive the relation:
\be
\tau_{\rm SPP} = \frac{\tau_{L^{121}}}{-2 \tau_{L^{121}} + 1} ~.
\ee
The two fundamental cells we have used (the red and the blue domains in Figure~\ref{fig:L121}) are $SL(2,\IZ)$ equivalent.

\subsection{A number theoretic consistency check of $\tau_R=\tau_B$}

While our conjecture that $ \tau_{B}= \tau_R$ works in the simple examples checked, we would like to verify it in the more involved case of SPP, already involving irrational $R$-charges and thus exhibiting much of the generic properties of the $L^{a\,b\,c}$ theories. That would require explicit knowledge of the SPP Belyi pair, which unfortunately is currently not available.
It is possible however to develop  arguments to rule out $ \tau_B = \tau_R $
in any specific case where $ \tau_R $ is known by investigating some number theoretic
properties of $ \tau_R$. We will explain this type argument for SPP and will show that
the argument fails to rule out the equality thanks to the transcendentality of $ \tau_R$

Let us first note that, upon writing the torus for the dimer as
\be
y^2 = x(x-1)(x-\lambda) ~,
\ee
the modular invariant $j(\tau)$ for the elliptic curve is expressed in terms of $\lambda$ by algebraic operations
(see equation (\ref{eq:jlambda})).
For a Belyi curve, the parameter $ \lambda $ is in $ \bmQ $.
Since the field $ \bmQ$ is closed under the algebraic operations used to determine $j$ from $ \lambda$,
this means that, for a Belyi curve, $ j $ itself is algebraic. %
Furthermore, it is known that if $j (\tau) $ is algebraic and $ \tau $ is also algebraic, then $ \tau $ has to satisfy a
{\em quadratic}  equation with rational coefficients \cite{mazur}.
Thus, one way to prove the proposed equality wrong would be to show that $ \tau_R $ is algebraic but not a solution of a quadratic equation. Happily, we show, assuming a conjecture in the theory of transcendental numbers, that $ \tau_R$ is actually transcendental, thus disposing of this possible counterargument against $ \tau_B = \tau_R$ for SPP.
To prove that, we first show  that $ x = e^{i \pi \over 2\sqrt{3}} $ is transcendental.
This proceeds by assuming a widely accepted conjecture in the mathematical literature \cite{Lang, Ramachandra, Waldschmidt} which goes under the name of {\em four exponentials conjecture}, for which there is no counterexample known.

It is known that if $(x_1, \ldots, x_m)$ is a set of $m$ complex numbers that are linearly independent over the rational numbers and $(y_1,\, \ldots, y_n)$ is a set of $n$ complex numbers that are linearly independent over the rational numbers and $m\,n > m+n$, then at least one of $\exp(x_i y_j)$ must be transcendental.
The minimal case, $m=3$, $n=2$, is known as the {\em six exponentials theorem} \cite{Lang, Ramachandra}.
A conjecture replaces the strict inequality in the hypothesis by $m\,n \ge m+n$, which permits $m=n=2$ as a minimal case.
Thus, the four exponentials conjecture states that if $(x_1,\, x_2)$ and $(y_1,\, y_2)$ are two pairs of complex numbers, with each pair linearly independent over the rational numbers, then at least one of $\exp(x_i\, y_j)$ must be transcendental.

Writing $ \chi = e^{i \pi \over 2\sqrt{3}} $ we have
\bea
\tau_R ({\rm SPP}) = {\chi^3 - \chi^{-1} \over 3\chi - \chi^3 } ~.
\eea
Putting $x_1 = 1$, $x_2 = {1\over2\sqrt{3}}$, $y_1 = i\pi$, $y_2 = {i\pi\over2\sqrt{3}}$, the four exponentials conjecture tells us that one of $e^{i\pi}$, $e^{i\pi\over2\sqrt{3}}$, $e^{i\pi\over12}$ is transcendental.
The first is $-1$ and the last solves $x^{12}+ 1=0$, so this implies that $\chi$ is transcendental.
Similarly, we may conclude that $\chi^3$ and $\chi^{-1}$ are also transcendental.
If the four exponentials conjecture is true, then $\tau$ is a ratio of combinations of transcendental numbers.

The equation for $\tau_R$ can be recast as
\bea
(\tau_R + 1) \chi^{4} - 3 \tau_R \chi^2 -1 = 0 ~.
\eea
If $\tau_R  $ were algebraic, since $ \bmQ$ is algebraically closed, it would follow that $\chi$ is algebraic.
But we have shown, assuming the four exponentials conjecture, that $\chi$ is transcendental, so we conclude that $ \tau_R $ is transcendental. We have thus ruled out a possible argument against $ \tau_B = \tau_R$.
A stronger argument in favor would be to show that $ j ( \tau_R ) $  in this case is algebraic
for this transcendental $ \tau_R$.

\subsection{Discussion of the conjecture}

To summarize, we have found that for the $\cN=4$ and the conifold, as well as their orbifolds, there is the equality $\tau_R=\tau_B$.
We have shown, using the four exponentials conjecture,
 that the complex structure parameter $\tau_R$ is transcendental for the SPP theory, but we cannot compare this directly with $\tau_B$, since we do not have the explicit Belyi pair.

A word of caution is in order at this point.
The identification of complex structures is not intended to imply a straightforward identification of the $R$-dimer with the dessin.
Recall that the dessin is defined as the inverse image of the interval $[0,1]$.
Close to a vertex, the Belyi map will go like $z^n$, where $n$ is the number of edges incident on the vertex.
This implies that the angle between these lines is $\frac{2\pi}{n}$.
In the $R$-dimer construction, the angle between two edges is determined by the $R$-charges of the corresponding fields (see \eref{eq:angle}).
These $R$-charges are themselves solved by $a$-maximization, which leads to a quadratic system of equations.
There is, in general, no reason for the angles at a vertex to be equal.
For the case of $\IC^3$ and the conifold, it so happens that the angles are equal.
So one might suspect that the equality $\tau_R=\tau_B$ is a consequence of this rather special circumstance and should be easy to rule out for any case where the angles are not equal in the $R$-dimer.
This makes the SPP case discussed above is a particularly interesting laboratory for testing our conjecture.

The calculation of $ \tau_R $ from the $R$-dimer, and indeed the definition of the $R$-dimer, relies on the geometry of geodesics on the torus, which are straight lines when the torus is realized as the unit cell of a periodic tiling of the plane.
The calculation of $ \tau_B$ arises from the holomorphic geometry of Belyi pairs.
More precisely, when a topological torus is realized as a branched cover of $\mP^1$, there is a natural complex structure induced on it which is the pull-back of the standard complex structure on the $\mP^1$.

The string theory construction of the toric SCFT offers a number of different approaches.
Among the different regions of string theory moduli space, there are D$3$-branes at the Calabi--Yau singularity.
There is the picture of D$5$-branes and NS$5$-branes in type IIB string theory.
There is the T-dual type IIA system in terms of D$6$-branes.
At weak coupling type IIB provides a picture featuring a $T^2$ in spacetime and holomorphic surfaces.
At strong coupling type IIB provides a picture where NS$5$-branes form geodesic zig-zags on $\mathbb{T}^2$, which enclose the vertices of the dimer.
It may be expected that the different regimes of string theory moduli space will provide realizations of the holomorphic geometry of Belyi pair and the geodesic geometry of $R$-dimer.
In this way, the question of equality of $ \tau_B$ and $ \tau_R $ can be translated into a physical question about the moduli of string theory which allow different realizations of the same superconformal gauge theory.

A reasonable guess is that the holomorphic type IIA or type IIB constructions would provide natural setups to explore a derivation of the Belyi pair in string theory.
The strong coupling limit of type IIB, which incorporates geodesic zig-zags on $\mathbb{T}^2$, is a natural setting for exploring a derivation of the geodesic geometry of the $R$-dimer.
There are important challenges along the way:
the physics of systems containing both D$5$- and NS$5$-branes is not well-understood from a worldsheet string point of view.
Furthermore, the $R$-dimer is based on $R$-charges which only become well-defined in the deep infrared.

These ideas could be explored, in the first instance, in the more symmetric cases of $\IC^3$ and its orbifolds which preserve $ \cN=2$ supersymmetry, notably $ \IC^2/\mathbb{Z}_n \times \IC $.
The structure of $\mathcal{N}=2$ theories is much more constrained, since the K\"ahler potential is also under control.
Thus, one might hope for a deeper understanding of the Belyi pair, which might be related to the Seiberg--Witten curve of the theory under consideration, perhaps along the lines of~\cite{ACD06}.

\section{Summary and outlook}

In this paper we have initiated the study of dimer models for D$3$-branes at toric CY$_3$ singularities in terms of Belyi theory.
This involves an elegant description of the combinatoric data for the dimer in terms a triple of permutations, as well as algebraic equations for the Belyi curve and Belyi map, which are defined over algebraic numbers.
The permutation description allows for a neat description of zig-zag paths, and should provide a new approach to the computational classification of consistent dimers, a problem recently considered in~\cite{dhp}.

The explicit construction of the Belyi curve and Belyi map provides a geometric realization of physical data, such as the
permutation  symmetries of superpotential terms.
Here, we have considered symmetries which leave the positive terms and negative terms of the superpotential separately invariant as well as symmetries which exchange these terms.
Automorphisms of the elliptic curve supporting the Belyi map play an interesting role in both cases.
Braid group actions on covers, which can be described combinatorially and geometrically, are key to the latter problem.
These themes were explored in some basic examples of $ \IC^3 $, conifold, and SPP, as well as orbifolds.
We have made general remarks about superpotentials of orbifold theories, relating them to automorphisms of unbranched covers of a torus by a torus.

The counting of torus covers appears in the context of two-dimensional SYM~\cite{GT}.
The zero area limit, which has particular interest as a topological point~\cite{CMR}, involves these unbranched covers.
Another connection to matrix models is that permutation triples appear in the computation of correlators of observables in Hermitian or complex matrix models~\cite{dMR,tom}.
An interesting problem is to express the consistency condition of dimers, such as non-self-intersection of zig-zags, in the matrix model language.
This would allow a matrix model counting for toric CFTs.

The analytic Belyi pair constructions will also be very interesting to explore in the context of orbifolds.
We would hope to find an algorithm that would generate the Belyi pair for any orbifold, using as data an analytic description of the unbranched cover along with the original Belyi pair. This could be very helpful in connection with
 the  inverse algorithm, given that one can always embed the threefold singularity of interest in a sufficiently large orbifold. In that respect, studying the realization of RG flows whereby the singularity is partially resolved would also be a very interesting and important aspect.
Multiple quiver diagrams can map to a single toric diagram~\cite{td}.
The different toric phases of a given theory are believed to manifest Seiberg duality~\cite{sd}.
It would be enlightening to make these ideas explicit in the language of permutation triples and Belyi pairs, particularly in connection with our conjectured equality of $\tau_R$ and $\tau_B$.

Given a dessin d'enfant on a genus $g$ Riemann surface, finding the associated Belyi map is an open problem in mathematics.
In the physical examples of which we are aware, whereas the cycle structures of $\sigma_B$ and $\sigma_W$ may differ, the number of cycles in the permutations is the same.
This is to say that there are an equal number of black and white nodes in the dimer.
Such bipartite graphs are called {\em balanced} in the mathematics literature.
It may be that the special properties of balanced graphs are essential to understanding the physics of supersymmetric gauge theories.

From a mathematical point of view, much of the interest
in Belyi theory stems from the role of the absolute Galois
group. It is known that this group acts faithfully
on the set of all dessins. It is also known that it acts
faithfully on subsets such as genus zero dessins, or genus one dessins,
or tree-like dessins on genus zero \cite{lando-zvonkin}. It is natural to ask if
the set of ``consistent dimers'' which arise in AdS/CFT,
characterized in terms of properties of zig-zag paths,
 form closed Galois orbits, and, if so,
 whether they provide a faithful action.

We have explored an intriguing equality between the complex structure of the Belyi curve and a complex structure determined by the $R$-charges, which was found to hold true for the conifold and $\IC^3$ as well as their orbifolds.
It is tempting to conjecture that this equality holds for all toric varieties.
Since explicit constructions of Belyi pairs are not easy in general, such an equality would give non-trivial data about the Belyi complex structure for Belyi curves corresponding to all consistent dimers.
If true, the equality would also give a large class of values of (generically transcendental) $\tau$ for which the Klein $j$-invariant $j(\tau)$ is algebraic.
Since it might be argued that the agreement of the two complex structures is an accident due to symmetry reasons, we made some consistency checks for the case of SPP.
Remarkably, these consistency checks relate our problem to the well-known mathematical problem of proving certain numbers to be algebraic.
Using the four exponentials conjecture, we have been able to argue that both conjectures are compatible.
This deep relation with number theory is intriguing, and deserves further study.

Perhaps the most important open problem is to find a derivation of the Belyi pair from string theory constructions associated with branes or geometries associated with the SCFT for D$3$-branes at toric Calabi--Yau singularities.
Given the relations we have uncovered between the complex structure of the Belyi curve and that of the curve which supports the $R$-dimer, this question is closely related to the outstanding question of constructing the $R$-dimer, or as a first step identifying the corresponding torus complex structure, in the physics of strings and branes related to the toric Calabi--Yaus.

Finally, while our investigations here have been entirely in the context of four dimensional SCFTs,
extensions to other dimensions can be considered. In, {\em e.g.},~\cite{Ueda:2008hx, Imamura:2008qs, Hanany:2008cd, Hanany:2008fj} it has been noted that three-dimensional SCFTs dual to M$2$-branes probing CY$_4$ singularities might also be encoded in bipartite graphs drawn on a torus. While a number of issues regarding these proposals remain to be clarified\footnote{A partial list would include in particular the actual SCFT character of the theories given that no efficient analog of $a$-maximization is available, and the role of the collapsing eleven dimensional circle potentially giving rise to non-isolated singularities which might suggest the inclusion of fundamental fields (see \cite{Benini:2009qs, Jafferis:2009th}).}, it seems that the tools we have introduced could be applied to these theories. In particular, a combinatorial description and the application of the Belyi theorem should be possible. However, it remains to be understood how to encode the Chern--Simons levels, and we point out that many such theories would involve self-intersecting zig-zags. 

We leave these very interesting questions for future investigations.

\section*{Acknowledgements}

We thank Yang-Hui He, Jurgis Pasukonis, and David Turton for discussions. DR-G is indebted to Sergio Benvenuti and
Sebastian  Franco for many insights and explanations over the years. VJ and SR are supported by an STFC grant ST/G000565/1.
DR-G is supported by the Israel Science Foundation under grant no.\ 392/09.
He also acknowledges support from the Spanish Ministry of Science through research grant FPA2009-07122 and Spanish Consolider-Ingenio 2010 Programme CPAN (CSD2007-00042) as well as to the European Commission, under whose financial support (MOIF-CT-2006-38381) this work was started.

\begin{appendix}

\section{Dimers and SCFTs in a nutshell}\label{nutshell}

For the sake of completeness, let us collect some relevant facts about dimers and their relation with brane physics.
A thorough explanation can be found in the reviews, {\em e.g.},~\cite{Kennaway:2007tq, yam08} and references therein.

Dimers provide an economic way of encoding the SCFT dual to D$3$-branes probing a given toric CY$_3$.
As emphasized in the text, dimers are bipartite graphs drawn on a torus.
They can be obtained as the dual graph to the periodic quiver.
Because of this, faces represent gauge groups while edges represent chiral multiplets.
More specifically, an edge separating face $i$ from face $j$ corresponds to a bifundamental chiral multiplet which transforms in the $(N_i,\, \overline{N}_j)$ representation.
The superpotential terms are encoded in the vertices of the dimer.
Choosing an orientation on the torus, one can go around the black nodes according to the ordering thus induced.
Each node gives rise to a cyclic sequence of fields which corresponds to a gauge invariant trace of the fields.
One goes around the white nodes according to the orientation opposite to that induced by the torus and similarly constructs another list of gauge invariant operators.
The terms coming from the black nodes are added together to form $W_+ $.
The terms from the white nodes are added  to form $W_-$.
The superpotential is $ W = W_+ - W_- $, {\em i.e.}, the black nodes contribute positive terms while the white nodes contribute negative terms.

\begin{figure}[h!]
\begin{center}
\includegraphics[scale=.25]{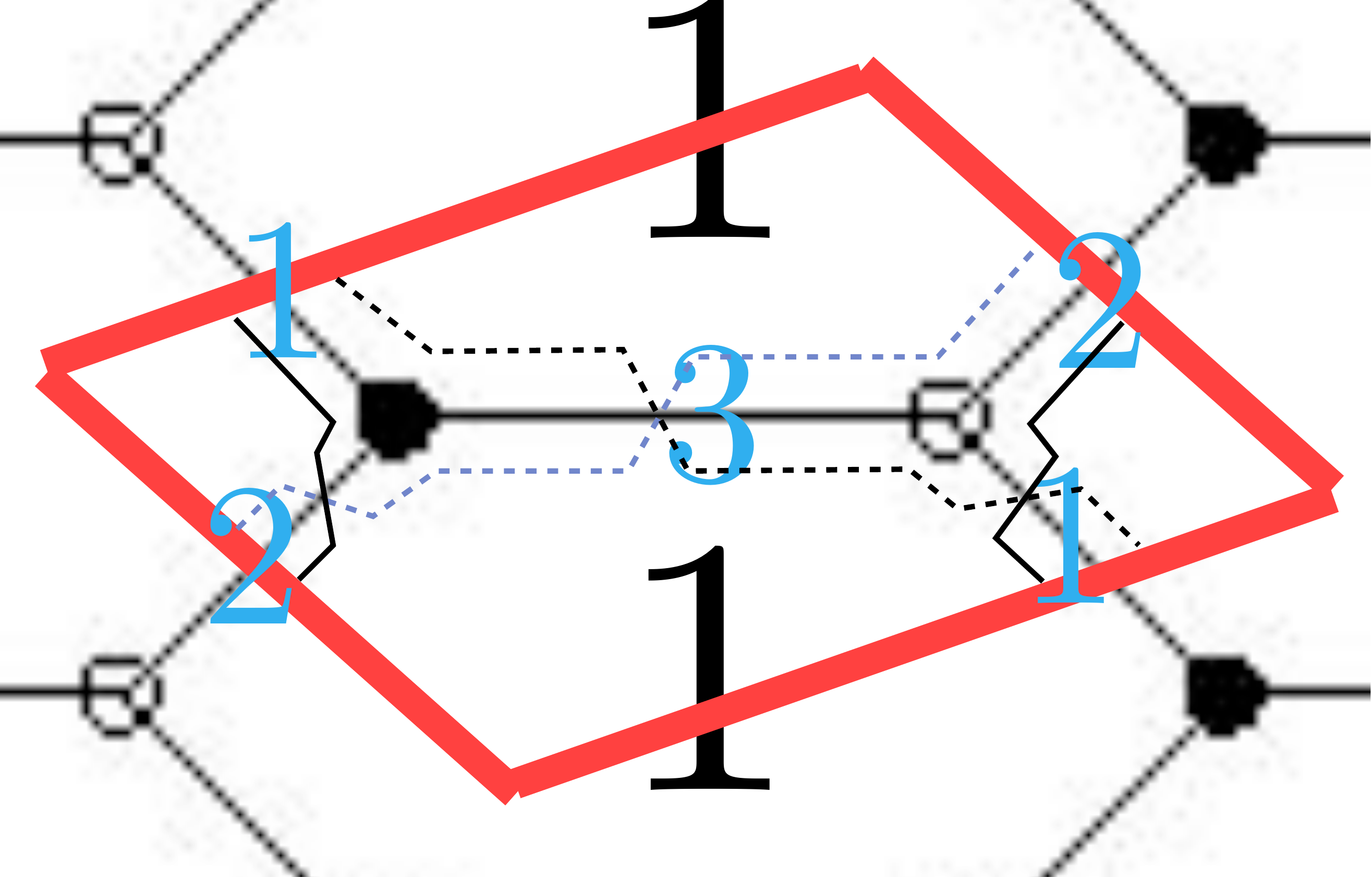}
\end{center}
\caption{Dimer for the $\mathcal{N}=4$ SYM theory with its three zig-zag paths.}
\label{C3BitZZ}
\end{figure}

As an example consider the $\mathcal{N}=4$ case given in Figure~\ref{C3BitZZ}.
There are three edges and a single face.
Thus, there are three adjoint chiral multiplets of a single gauge group.
Since there are two vertices, the superpotential contains two terms, one positive and one negative.
Choosing to encircle the black node anticlockwise and the white node clockwise, and following the above rules we obtain the standard $\mathcal{N}=4$ superpotential
\begin{equation}
W= {\rm Tr} (\, X_1 X_2 X_3 -  X_1 X_3 X_2  \,) ~.
\end{equation}
Sometimes this is written as ${\rm Tr} ( X^1_{11} X^2_{11} X^3_{11} -  X^1_{11} X^3_{11} X^2_{11} ) $, where the lower indices label the gauge groups; in this case there is only one.

In connecting the dimer to dessin d'enfant, we label the edges $ 1, 2, 3 $ as usual, and construct permutations $ \s_B$ and $ \s_W $, where both permutations are read according to the orientation of the surface.
The connection between dessin d'enfants and branched covers also motivates the definition of $ \s_{ \infty} = ( \s_B \s_W)^{-1} $, which in this case is $ (123)$.
With these data, there is a simple relation between the genus of the surface and the cycle structure of the permutations.
Since we are considering genus one here, this relation (given in general form in (\ref{genonecond})) becomes
\bea
0 = 3 - 1 -1 -1 ~.
\eea
The two permutations $ \s_B , \s_W $ contain all the information about the faces, which can be described using a permutation in $S_{2d} $ constructed from $ \sigma_B , \sigma_W $ using a construction we describe in section \ref{consdimzig}.
A closely related construction gives an elegant combinatoric description of zig-zag paths without referring to the picture.
These arise in characterizing {\em consistent dimers}, which form the subset of toroidal dimers which give rise to consistent SCFTs.
For the case of $ \cN=4$ there are three zig-zag paths, which are drawn in Figure~\ref{C3BitZZ}.

Due to the high degree of symmetry of this example, the $R$-charges of all fields under the relevant $U(1)_R$ are $\frac{2}{3}$ as can be deduced from the superpotential.
In particular, this means that the isoradially drawn dimer has edges meeting at each node with an equal $\frac{2\pi}{3}$ angle.
Indeed, as computed in the main text, it follows that
\begin{equation}
\tau_R=\frac{1}{2}+i\, \frac{\sqrt{3}}{2} ~.
\end{equation}

More generically, the $R$-charges at the superconformal fixed point can be found by {\em $a$-maximization}~\cite{Intriligator:2003jj}. This proceeds by assigning   trial $R$-charges to all  the fields and enforcing  the vanishing of the $\beta$-functions for all couplings. This amounts to setting to zero the NSVZ exact $\beta$ function\footnote{As discussed in, {\em e.g.},~\cite{Benini:2009mz}, it is convenient to match supergravity conventions to keep the vector fields holomorphically normalized while imposing canonic normalization for the chiral multiplets.} for the gauge coupling $x_I=\frac{8\pi^2}{g_I^2}$ of the $I$-th gauge group
\begin{equation}
\beta(x_I) = 3\,N_c-\sum_i\, T_2(\mathbf{r}_i)\, (1-\gamma_i) ~,
\end{equation}
where the sum runs to all fields charged under the $I$-th gauge group in the representation $\mathbf{r}_i$, where $T_2(\mathbf{r}_i)$ is its quadratic Casimir. At the superconformal fixed point the scaling dimensions are related to $R$-charges as $\Delta=\frac{3}{2}\, R$. It then follows that the anomalous dimensions of the fields can be expressed in terms of the $R$-charges at the superconformal fixed point as
\begin{equation}
\gamma=3\, R-2 ~.
\end{equation}
Thus, the vanishing of the NSVZ gauge fields beta function imposes the condition
\begin{equation}
N_c-\sum_i\, T_2(\mathbf{r}_i)\, (1-R_i)=0\quad \forall\ {\rm gauge\ groups} ~.
\end{equation}
Furthermore, the vanishing of the $\beta$ function superpotential couplings requires (see, {\em e.g.},~\cite{Kennaway:2007tq})
\begin{equation}
\sum_i\, R_i=2\quad \forall \ {\rm monomials\ in}\ W ~,
\end{equation}
where the sum runs to all fields on each superpotential monomial. This set of equations, together with the vanishing of the NSVZ $\beta$-functions, generically admits a continuous set of solutions. The $a$-maximization procedure allows to uniquely fix the exact superconformal $R$-charge among all these possibilities by selecting that which maximizes the central charge $a$. More explicitly, the central charge $a$ reads, in terms of the trial $R$-charges
\begin{equation}
\label{acentralcharge}
a=\frac{3}{32}\, \Big(3\, \sum_i\, (R_i-1)^3-\sum_i\, (R_i-1)\Big) ~.
\end{equation}
where the minus one stands for the fact that only the fermions (which have one less unit of $R$-charge than the bosons) contribute to the central charge. Upon inserting in (\ref{acentralcharge}) the solutions from the vanishing of the $\beta$ functions, the unique solution which maximizes $a$ is guaranteed to be the exact $R$-charge assignation. Indeed, it is straightforward to check that the assignation for $\mathcal{N}=4$ above coincides with the one arising from $a$-maximization.
Note also that the construction of the faces, starting from $ ( \s_B , \s_W ) $ in terms of a permutation in $S_{2d} $
in Section \ref{zzint}, 
gives the information about the fields entering the summation for each gauge group.

\section{Some remarks on elliptic curves}\label{ellipticcurves}

Tori are elliptic curves defined over $\mathbb{C}$. For most of the purposes, the relevant form of an elliptic curve is given by the Weierstrass equation
\begin{equation}
y^2=x^3+Ax+B ~, \qquad 4A^3+27B^2\ne 0 ~,
\end{equation}
where $y,\,x$ and $A,\,B$ are elements in $\mathbb{C}$. One can show that this curve describes a Riemann surface of genus one, that is, a torus.

The above expression can be embedded in $\mathbb{P}^2$:
\begin{equation}
Y^2\,Z=X^3+A\,X\,Z^2+B\, Z^3
\end{equation}
by using the relation between projective and affine coordinates
\begin{equation}
x=\frac{X}{Z} ~, \qquad y=\frac{Y}{Z} ~.
\end{equation}
Embedding the affine curve into projective space thus includes the closure, which is the point at infinity $Z=0$. That point is usually denoted as $O$.

Given two points $P_1(x_1,\,y_1)$ and $P_2(x_2,\,y_2)$ in the curve, there is  an  operation which gives a
third point $P_3(x_3,\,y_3)$. This is  denoted as $P_1+P_2=P_3$. It turns out that the set of points on the curve together with this ``$+$'' operation, upon the addition of the point at infinity, forms a group. Indeed, the point at infinity has the same properties as the ``zero.'' One can then see that in the Weierstrass form $-P(x,\,y)=(x,\,-y)$.

Another standard form (Legendre form) for the elliptic curve is
\be
y^2 = x (x -1) (x - \lambda) ~,\qquad \lambda\ne0,\, 1 ~.
\ee

We will be interested in morphisms of the curve. Let us begin in a generic way, and consider two elliptic curves $E_1$, $E_2$, between which we can consider a generic morphism
\begin{equation}
F:\, E_1\rightarrow E_2 ~.
\end{equation}
A special subclass of such morphisms are those, denoted as $\phi$, which preserve $O$
\begin{equation}
\phi:\, E_1\rightarrow E_2 ~,\qquad \phi(O)=O ~.
\end{equation}
This type of morphisms is called an {\em isogeny}. An important result is that for isogenies we have
\begin{equation}
\phi:\, E_1\rightarrow E_2\,/\,\phi(O)=O\,\Longrightarrow \phi(P+Q)=\phi(P)+\phi(Q) ~.
\end{equation}

We will be particularly interested in the case where $E_1=E_2=E$. When $\phi$ is invertible, it defines an isomorphism of the curve to itself.
Isogenies of this type are
called {\em automorphisms} of $E$, which
we will denote as ${\rm Aut_g}(E)$, where the subscript indicates that they preserve the
group structure of the elliptic curve.

Another particularly important type of morphism between a curve and itself is the {\em translation-by}-$P$ map $\tau_P$ defined as
\bea
\tau_P: && E \rightarrow E ~, \\
&& Q \mapsto \tau_P(Q)=P+Q ~. \nn
\eea

Making use of the translation-by-$P$ map, we can turn any isomorphism of the curve into an isogeny as
\begin{equation}
\phi=\tau_{-F(O)}\circ F ~.
\end{equation}
Since the inverse of the translation-by-$P$ is the translation-by$(-P)$, we can therefore write any isomorphism $F$ of a curve as
\begin{equation}
F=\tau_{F(O)}\circ\phi ~.
\end{equation}
That is, we have that any isomorphism %
of a curve is a composition of an isogeny and a translation. More precisely, ${\rm Aut}(E)=
E\ltimes {\rm Aut_g}(E)$. Since for every point in the curve we have a translation, the set of translations is naturally identified with $E$. Thus, we have a natural bijection
\begin{equation}
E\ltimes {\rm Aut_g}(E)\rightarrow {\rm Aut}(E) ~.
\end{equation}
The composition of two elements gives the group law in ${\rm Aut_g}(E)$ as
\begin{equation}
(P_1,\,\phi_1)\cdot(P_2,\,\phi_2)=(P_1+\phi_1\,P_2,\, \phi_1\circ\phi_2) ~.
\end{equation}
This can be understood as follows.
A generic isomorphism is the composition of an isogeny and a translation. Thus, setting $F_1(P)=\phi_1(P)+Q_1$, $F_2(P)=\phi_2(P)+Q_2$, the composition is
\begin{equation}
F_2\circ F_1(P)=F_2(\phi_1(P)+Q_1)=\phi_2(\phi_1(p)+Q_1)+Q_2=\phi_2\circ\phi_1(P)+\phi_2(Q_1)+Q_2 ~,
\end{equation}
which reproduces the result above.\footnote{We have used that for an isogeny $\phi(P_1+P_2)=\phi(P_1)+\Phi(P_2)$.}

We have used  the letter $E$ here for the elliptic curve, as conventional in 
discussions of the group structure of the torus. We use $\mT^2$ in the bulk of the paper. 
The group ${\rm Aut} (E) = {\rm Aut} (\mT^2) $ contains as a subgroup the automorphism
of any covering map which realizes $\mT^2$ as a cover. These subgroups
can be computed from the combinatorial data of permutations
 as in Section \ref{sec:dimerperms}.

\subsection{$j$-function}\label{appsec:jfunction}

Klein's $j$-invariant is the primitive modular invariant function on the upper half plane.
Defining $q=e^{2\pi i \tau}$, we have the expressions
\be
j(q) := 256\, \frac{(1-\lambda(q)+\lambda(q)^2)^3}{\lambda(q)^2(1-\lambda(q))^2} = 1728\, J(\sqrt{q}) ~,
\label{eq:jlambda}
\ee
where
\be
\lambda(q) = \left(\frac{\vartheta_2(q)}{\vartheta_3(q)} \right)^4 ~.
\ee
Note that, because of the definition of the function $j$ in terms of $\lambda$ as a rational function, if $\lambda\in\overline{\mathbb{Q}}$ we will automatically have that $j$ is also an algebraic number.

This function is invariant under
\be
\tau\mapsto \tau' = \frac{a\tau + b}{c\tau + d} ~, \qquad
ad-bc = 1 ~, \qquad
a,b,c,d\in \IZ ~.
\ee
Thus $j(\tau)$ defines a map from the fundamental domain ${\cal H}/SL(2,\IZ)$ to the complex numbers $\IC$.
Any meromorphic function invariant under $SL(2,\IZ)$ is a function of $j(\tau)$.

More generally, instead of considering $j$ as function on the fundamental domain of the upper half plane, it can be defined purely algebraically.
Given an elliptic curve
\be
y^2 + a_1 xy + a_3 y = x^3 + a_2 x^2 + a_4 x + a_6
\ee
over any field, define
\bea
&& b_2 = a_1^2 + 4 a_2 ~, \quad
b_4 = a_1 a_3 + 2 a_4 ~, \quad
b_6 = a_3^2 + 4 a_6 ~, \quad
b_8 = a_1^2 a_6 - a_1 a_3 a_4 + a_2 a_3^2 + 4 a_2 a_6 - a_4^2 ~, \nn \\
&& c_4 = b_2^2 - 24 b_4 ~, \quad
c_6 = -b_2^3 + 36 b_2 b_4 - 216 b_6 ~, \quad
\Delta = -b_2^2 b_8 + 9 b_2 b_4 b_6 - 8 b_4^3 - 27 b_6^2 ~.
\eea
The $j$ function in terms of the coefficients in the elliptic curve is then
\be
j = \frac{c_4^3}{\Delta} = 1728 \frac{c_4^3}{c_4^3-c_6^2} ~,
\label{eq:jalg}
\ee
where the last equality applies for fields with characteristic $k\ne 2,3$.
Two elliptic curves are isomorphic if and only if their $j$-invariants are the same.
When the elliptic curve is written as
\be
y^2 = x(x-1)(x-\lambda)
\label{eq:ellambda}
\ee
the $j$-invariant of the curve may be found simply by substituting this $\lambda$ into the expression~\eref{eq:jlambda}.

Any elliptic curve of the form~\eref{eq:ellambda} enjoys an invariance under $(x,y)\mapsto (x,-y)$.
Thus there is a $\IZ_2$ symmetry of the curve.
The values $j=0$ and $j=1728$ are special because these correspond to the enhancement of symmetries.
\bi
\item In the former case, $j=0$, the elliptic curve $\Sigma$ can be written as
\be
y^2 = x^3 + A ~.
\ee
This is invariant separately under $(x,y)\mapsto (x,-y)$ and $(x,y)\mapsto (\omega_3 x, y)$, where $\omega_3$ is a cubed root of unity.
Thus, $j=0$ implies that $\Sigma$ enjoys a $\IZ_2\times \IZ_3$ symmetry.
\item In the latter case, $j=1728$, the elliptic curve $\Sigma$ can be written as
\be
y^2 = x^3 + A x ~.
\ee
This is invariant under $(x,y)\mapsto (-x, i y)$.
Thus, $j=1728$ implies that $\Sigma$ enjoys a $\IZ_4$ symmetry.
\ei

The function $j(\tau)$ can be inverted.
We have
\be
j^{-1}(z) = i\, \frac{r(\frac{z}{1728})-s(\frac{z}{1728})}{r(\frac{z}{1728})+s(\frac{z}{1728})} ~,
\ee
where $r$ and $s$ are expressed in terms of the standard ${}_2F_1$ hypergeometric functions:
\bea
r(z) = \Gamma(\frac{5}{12})^2 \, {}_2F_1(\frac{1}{12}, \frac{1}{12}; \frac12; 1-z) ~, \quad
s(z) = 2(\sqrt{3}-2) \, \Gamma(\frac{11}{12})^2 \sqrt{z-1} \, {}_2F_1(\frac{7}{12}, \frac{7}{12}; \frac32; 1-z) ~.
\eea
In particular, $j^{-1}(0) = \frac12 + i \frac{\sqrt{3}}{2} = e^{\pi i/3}$ and $j^{-1}(1728) = 1 + i \simeq i = e^{\pi i/2}$.

In the examples we consider in Section \ref{basexamp},
 $j=0$ arises for the Belyi curve associated with $ \cN=4$ SYM for branes
transverse to $\IC^3$, and $j = 1728$ arises for the case of the conifold.

\section{Orbifolds of the basic examples}
\label{orbexamps}

\subsection{$\frac{\mathbb{C}^2}{\mathbb{Z}_n}\times\mathbb{C}$ orbifolds of $\mathbb{C}^3$}

The  $\frac{\mathbb{C}^2}{\mathbb{Z}_n}\times\mathbb{C}$ orbifolds of $\mathbb{C}^3$
are defined by
\begin{equation}
\frac{\mathbb{C}^2}{\mathbb{Z}_n}\times\mathbb{C}=\{\,(x,\,y\,z)\in\mathbb{C}^3\,/\,(x,\,y\,z)\sim(\omega\, x,\, \omega^{-1}\,y,\,z),\quad \omega^n=1\,\} ~.
\end{equation}
These orbifolds preserve the natural holomorphic three-form in $\mathbb{C}^3$, and thus are at least $\mathcal{N}=1$ supersymmetric.
(In fact, they are $\mathcal{N}=2$.)
There is also a natural brane construction for them, which makes them especially interesting.

The corresponding dimer is obtained by adjoining $n$ copies of the fundamental cell of the $\mathbb{C}^3$ dimer (along a given direction).
We may write the structure of the permutations defining the dimer as follows:
\begin{equation}
\sigma_B=(123)\,(456)\,\ldots ((3n-2)(3n-1)(3n)) ~,\qquad
\sigma_W=(1\, M_{C_2}\, 3)\, (4\, M_{C_3}\, 6)\,\ldots\, ((3n-2)\, M_{C_1}\, (3n)) ~,
\end{equation}
where $M_{C_i}$ is the ``middle element'' in each permutation. That is, if we consider the permutation $((i)\, (i+1)\, (i+2))$, $M_{\sigma_i}=(i+1)$. Put in words, the procedure amounts to writing $n$ copies of the fundamental three-cycle cell in the canonical order\footnote{We will refer to a permutation with the structure $(123)(456)\ldots((n-2)\,(n-1)\,n)$ as ``having the canonical order.''} for $\sigma_W$, and constructing from there $\sigma_W$ by shuffling in a cyclic way the middle element. From here we see that the combinatorial data predicts a Belyi map with $n$ preimages of $0$, all with ramification two, $n$ preimages of $1$, all with ramification two, and $n$ preimages of $\infty$, all with ramification two. The total ramification of such a map is then $B=6\, n$, while the degree is $d=3\, n$, so this indeed corresponds to a graph drawn on a torus.

 The automorphism group of the pair  is  given by
\begin{equation}
\gamma_A=(1\,4\,\ldots\,(3n-2))\, (2\,5\,\ldots\,(3n-1))\,(3\,6\,\ldots\, (3n)) ~.
\end{equation}
One can check that $\gamma^n=1$. Furthermore, from the structure of $\gamma$ one can see that none of the cycles composing the $\sigma_B$, $\sigma_W$ permutations is left fixed (indeed, these fundamental cycles are just shuffled).

Importantly, note that, since the ramification structure is the same over all the three points in the target $\mathbb{P}^1$, we should expect a ``symmetry'' which exchanges $0\leftrightarrow 1$ and $0\leftrightarrow \infty$.

\subsubsection{$\frac{\mathbb{C}^2}{\mathbb{Z}_3}\times \mathbb{C}$}

\begin{figure}
\centering
\includegraphics{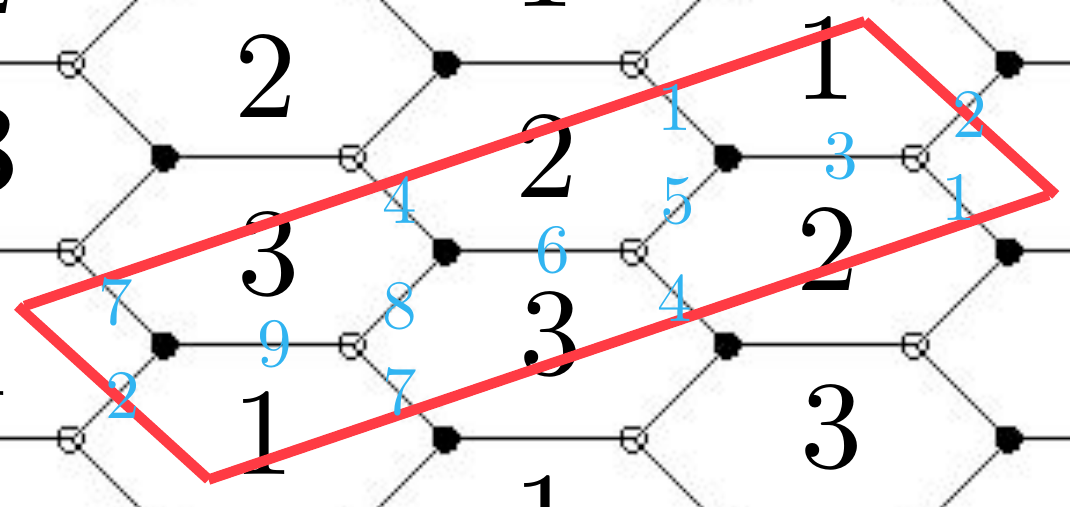}
\caption{$\frac{\mathbb{C}^2}{\mathbb{Z}_3}\times \mathbb{C}$ dimer.}
\label{fig:c2modz3}
\end{figure}

The dimer for the $\frac{\mathbb{C}^2}{\mathbb{Z}_3}\times \mathbb{C}$ theory is shown in Figure~\ref{fig:c2modz3}. 
The permutations are
\begin{equation}
 \sigma_B=(153)\,(486)\,(729) ~, \qquad \sigma_W=(123)\,(456)\,(789) ~, \qquad \sigma_{\infty}=(165)\,(273)\,(498) ~,
\end{equation}
which coincides with the general formula above. Furthermore, the automorphism group is generated by
\begin{equation}
\gamma_A=(147)\,(258)\,(369) ~,
\end{equation}
and it is such that $\gamma_A^3=1$.
These statements have been checked with {\tt SAGE}.

\subsubsection{$\frac{\mathbb{C}^2}{\mathbb{Z}_5}\times \mathbb{C}$}

\begin{figure}
\centering
\includegraphics{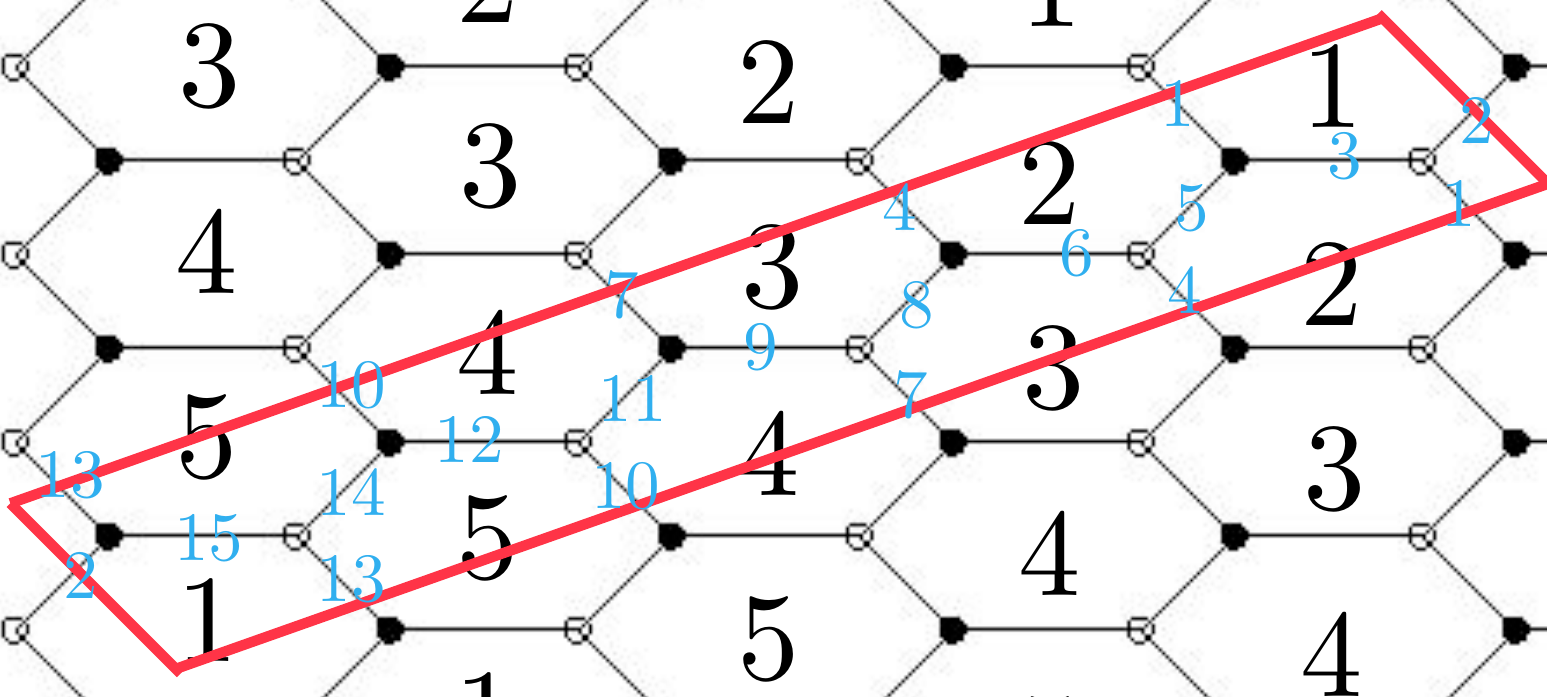}
\caption{$\frac{\mathbb{C}^2}{\mathbb{Z}_5}\times \mathbb{C}$ dimer.}
\label{fig:c2modz5}
\end{figure}

The dimer for the $\frac{\mathbb{C}^2}{\mathbb{Z}_3}\times \mathbb{C}$ theory  is shown in Figure~\ref{fig:c2modz5}.
The permutations are
\begin{equation}
\sigma_B=(153)\,(486)(7\,11\,9)\,(10\,14\,12)\,(13\,2\,15) ~,
\qquad \sigma_W=(123)\,(456)\,(789)\,(10\,11\,12)\,(13\,14\,15) ~,
\end{equation}
and the $\gamma$ generating the automorphism group
\begin{equation}
\gamma_A=(1\,4\,7\,10\,13)\,(2\,5\,8\,11\,14)\,(3\,6\,9\,12\,15) ~,
\end{equation}
which satisfies $\gamma_A^5=1$.

\subsubsection{Zig-zags in orbifolds $\IC^2/\IZ_n \times \IC$ of $\cN=4$}

For the orbifold of $\cN=4$, we have seen that
\be
\sigma_B = (123)(456)\ldots((3n-2)(3n-1)(3n)) ~, \quad
\sigma_W = (1(M_{C_2})3)(4(M_{C_3})6)\ldots((3n-2)(M_{C_1})(3n)) ~.
\ee
The zig-zag paths are then given by the cycles
\bea
&& z_1 = (1^-2^+(3n-2)^-(3n-1)^+(3n-5)^-(3n-4)^+\ldots4^-5^+) ~, \nn \\
&& z_2 = (2^-3^+5^-6^+\ldots(3n-1)^-(3n)^+) ~, \nn \\
&& z_3 = (3^-1^+) ~, \quad z_4 = (6^-4^+) ~, \quad \ldots ~, \quad z_{2+n} = ((3n)^-(3n-2)^+) ~.
\eea
We therefore have two long zig-zags along the skeleton of the dimer and $n$ zig-zags of length two.
The long zig-zags intersect each other $n$ times.
The short zig-zags intersect each of the long ones once with opposite signs and the short ones do not intersect each other at all.
The intersection matrix may be readily computed from these observations.

For $\IC^2/\IZ_3\times \IC$, for example, the intersection matrix is
\be
I_{\IC^2/\IZ_3\times \IC} = \left(\ba{ccccc} 0 & -3 & 1 & 1 & 1 \cr 3 & 0 & -1 & -1 & -1 \cr -1 & 1 & 0 & 0 & 0 \cr -1 & 1 & 0 & 0 & 0 \cr -1 & 1 & 0 & 0 & 0 \ea \right) ~.
\ee
The constraints from the null vectors are $z_1 + z_2 + 3 z_i = 0$, $i=3,4,5$.
A solution is $z_1 = a$, $z_2 = -a-3b$, $z_3 = b$, $z_4 = b$, and $z_5 = b$.

\subsection{Orbifolds of the conifold}

\subsubsection{Non-chiral $\mathbb{Z}_n$ orbifolds of the conifold}

 The conifold is given by
\begin{equation}
\mathcal{C}(T^{11})=\{x,\,y,\, u,\,v\,/\, x\,y=u\,v\,\} ~.
\end{equation}
The non-chiral $Z_n$  orbifolds are quotients under
\begin{equation}
(x\,y)\sim(\omega\, x,\, \omega^{-1}\, y) ~,\qquad \omega^n=1 ~.
\end{equation}

\begin{figure}
\centering
\includegraphics[scale=.7]{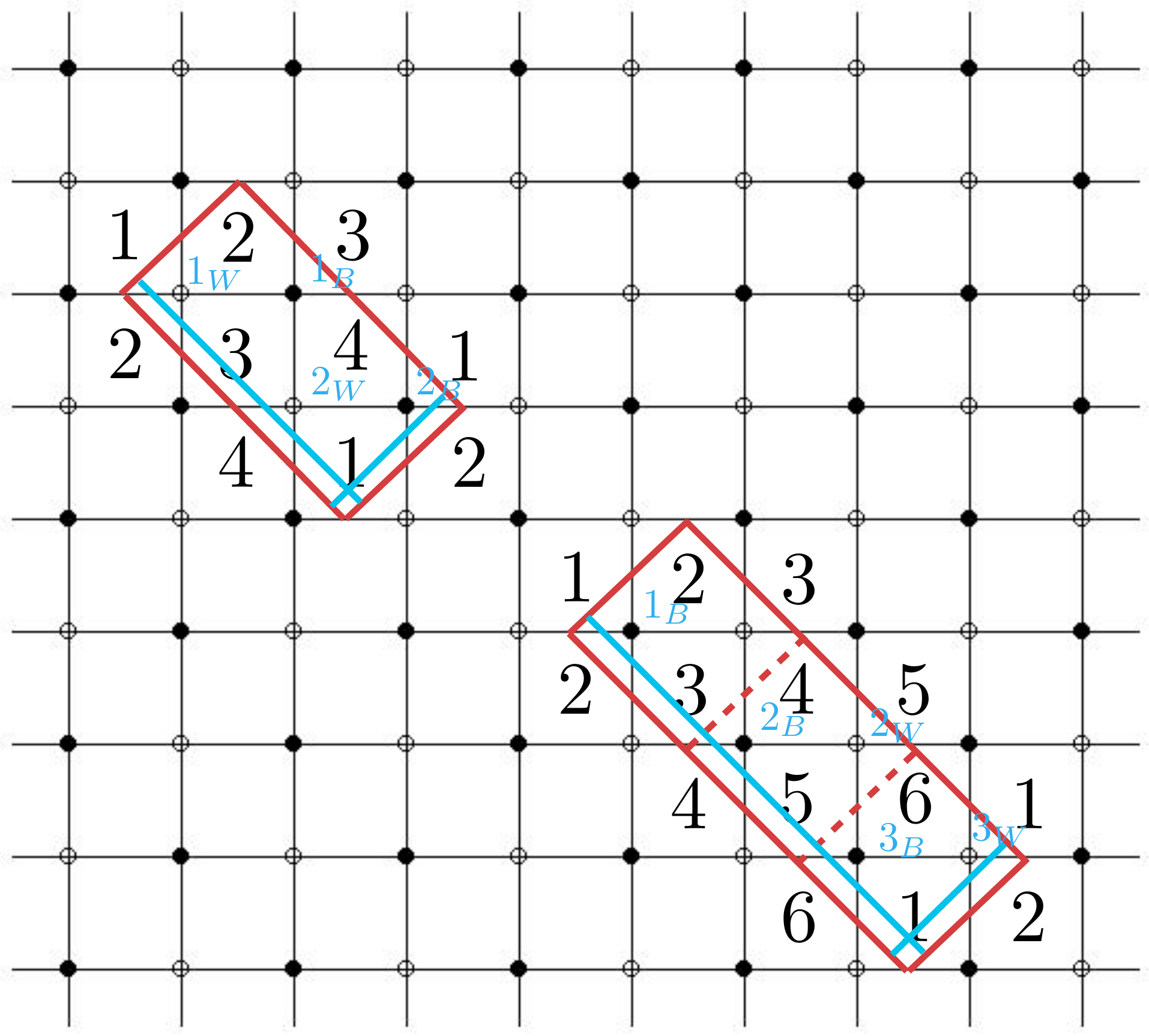}
\caption{$\mathbb{Z}_n$ orbifolds of the conifold for $n=2,\, 3$.}
\label{fig:kast}
\end{figure}

In order to make the discussion concrete, let us explicitly consider the $n=2,\, 3$ cases. Their dimers (with the cycles and  of the nodes in the dimer suitable prepared to compute the Kasteleyn matrix) appears in Figure~\ref{fig:kast}. One can explicitly check from those dimers that the determinant of the corresponding Kasteleyn matrix obtained in each case does indeed reproduce the expected Newton polynomial of each toric variety.

\begin{figure}
\centering
\includegraphics[scale=.7]{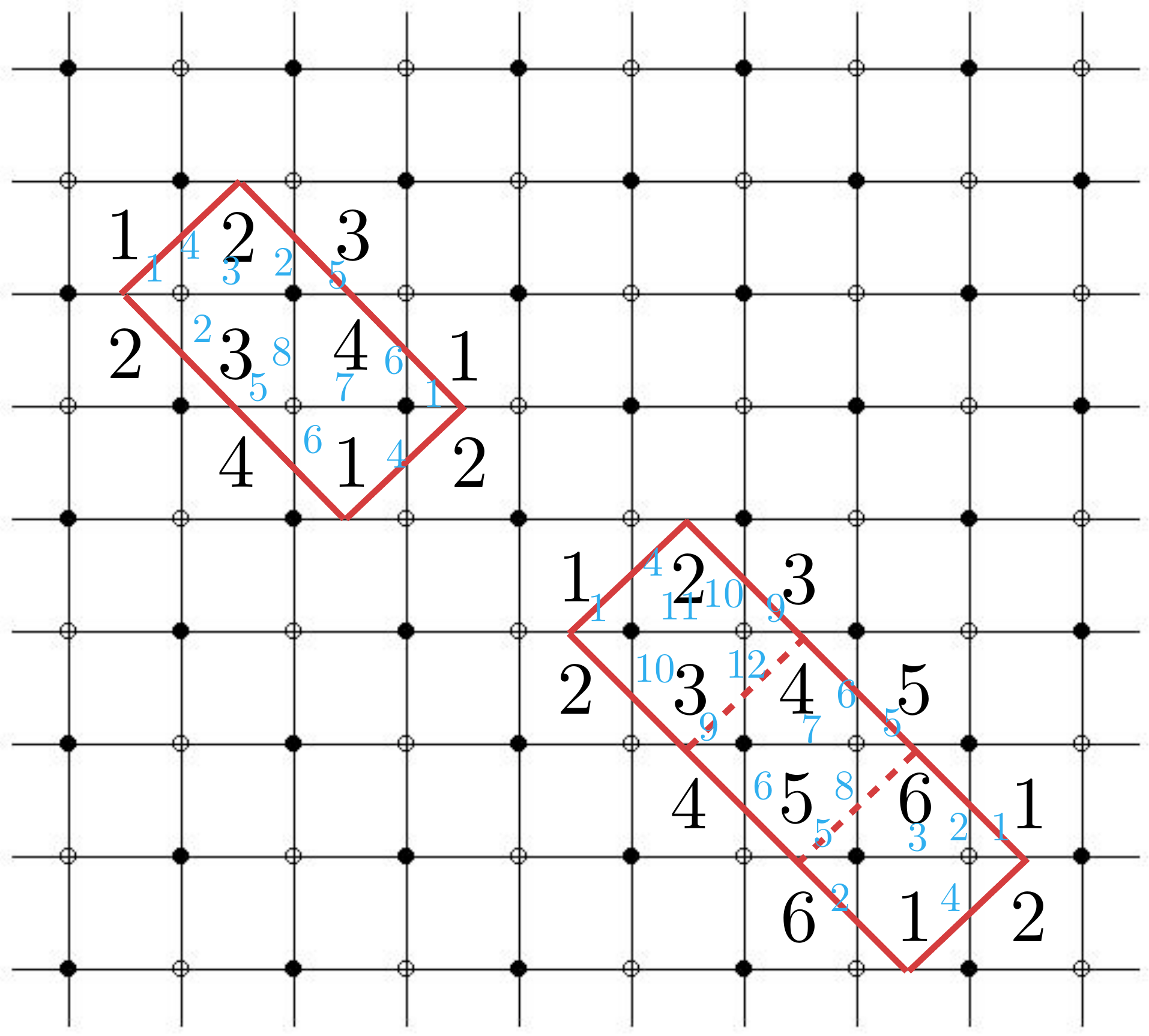}
\caption{$\mathbb{Z}_n$ orbifolds of the conifold for $n=2,\, 3$.}
\label{fig:orbofcon}
\end{figure}

Let us now redraw the dimer in Figure~\ref{fig:orbofcon}, labelling the edges in a way to read the permutation structure describing this graph as a dessin.

In the $\mathbb{Z}_2$ case we have
\begin{equation}
\sigma_B=(1674)\,(5238) ~, \qquad \sigma_W=(1234)\,(5678) ~.
\end{equation}
In the $\mathbb{Z}_3$ case we have
\begin{equation}
\sigma_B=(1\,10\,11\,4)\,(5\,2\,3\,8)\,(9\,6\,7\,12) ~, \qquad
\sigma_W=(1234)\,(5678)\,(9\,10\,11\,12) ~.
\end{equation}
We don't write the $\sigma_{\infty}$ permutation explicitly: as usual, it is $(\sigma_B \sigma_W)^{-1}$.

From these two examples we  read off the pattern for the $\mathbb{Z}_n$ orbifold:
\begin{equation}
\sigma_W=(1234)\ldots((4n-4)\,(4n-3)\,(4n-2)\,(4n-1)\,4n) ~, \quad \sigma_B=(1\,M_{C_n}\,4)\ldots((4n-3)\,M_{C_1}\,4n) ~.
\end{equation}

We can now turn to the endomorphisms. The matrix $\gamma$ generating the automorphism group is
\begin{equation}
\gamma_A=(1\,5\,\ldots\,(4n-3))\, (2\,6\,\ldots\,(4n-2))\,(3\,7\,\ldots\, (4n-1))\,(4\,8\,\ldots\,4n)) ~.
\end{equation}
One can check that $\gamma^n=1$, and that again no fundamental cycle is left fixed.

\subsubsection{Zig-zag paths}

Recalling the permutations  for the orbifold of the conifold, we have
\bea
&& \sigma_B = (1234)(5678)\ldots((4n-3)(4n-2)(4n-1)(4n)) ~, \\
&& \sigma_W = (1(M_{C_2})4)(5(M_{C_3})8)\ldots((4n-3)(M_{C_1})(4n)) \nn ~.
\eea
We use Section \ref{zzint} to   write the zig-zags:
\bea
&& z_1 = (1^-2^+(4n-3)^-(4n-2)^+(4n-7)^-(4n-6)^+\ldots5^-6^+) ~, \nn \\
&& z_2 = (3^-4^+7^-8^+\ldots(4n-1)^-(4n)^+) ~, \\
&& z_3 = (2^-3^+) ~, \quad z_4 = (6^-7^+) ~,\quad \ldots ~, \quad z_{2+n} = ((4n-2)^-(4n-1)^+) \nn \\
&& z_{3+n} = (4^-1^+) ~, \quad z_{4+n} = (8^-5^+) ~, \quad \ldots ~, \quad z_{2+2n} = ((4n)^-(4n-3)^+) \nn ~.
\eea
The only intersections are short zig-zags with long ones; this happens once for each pair $(z_i, z_j)$, $i=1,2$, $j=3,\ldots,2+2n$.
Thus the intersection matrix always contains elements $0, \pm1$.
We can then define $a$- and $b$-cycles $z_j = a$, $z_k = b$ by choosing any element $I_{jk} = 1$.

\subsubsection{Chiral orbifold: $\mathbb{F}_0$}

Let us rewrite the conifold in terms of
\begin{equation}
x=z_2+i\, z_1 ~,\quad y=z_2+i\,z_2 ~,\quad u=z_4+i\,z_3 ~,\quad v=-z_3-i\, z_4 ~,
\end{equation}
such that the defining equation becomes
\begin{equation}
z_1^2+z_2^2+z_3^2+z_4^2=0 ~.
\end{equation}
In these coordinates, the holomorphic three-form is
\begin{equation}
\Omega=\frac{dz_2\wedge dz_3\wedge dz_4}{z_1} ~.
\end{equation}

In view of this, $z_i\rightarrow \omega\, z_i$ will be a supersymmetric orbifold only if $\omega^2=1$, that is, only for a $\mathbb{Z}_2$ orbifold.
Let us now consider precisely this $\mathbb{Z}_2$ orbifold, in other words the chiral orbifold of the conifold leading to the cone over the zeroth Hirzebruch surface $\mathbb{F}_0$
\begin{equation}
\mathbb{F}_0=\{(x,\,y,\,u,\,v)\in \mathbb{C}\,/\,x\,y=u\,v,\quad (x,\,y,\,u,\,v)\sim(-x,\,-y,\,-u,\,-v) ~.
\end{equation}

The dual theory is encoded in the dimer in Figure~\ref{fig:f01} below.
\begin{figure}[!h]
\centering
\includegraphics[scale=.7]{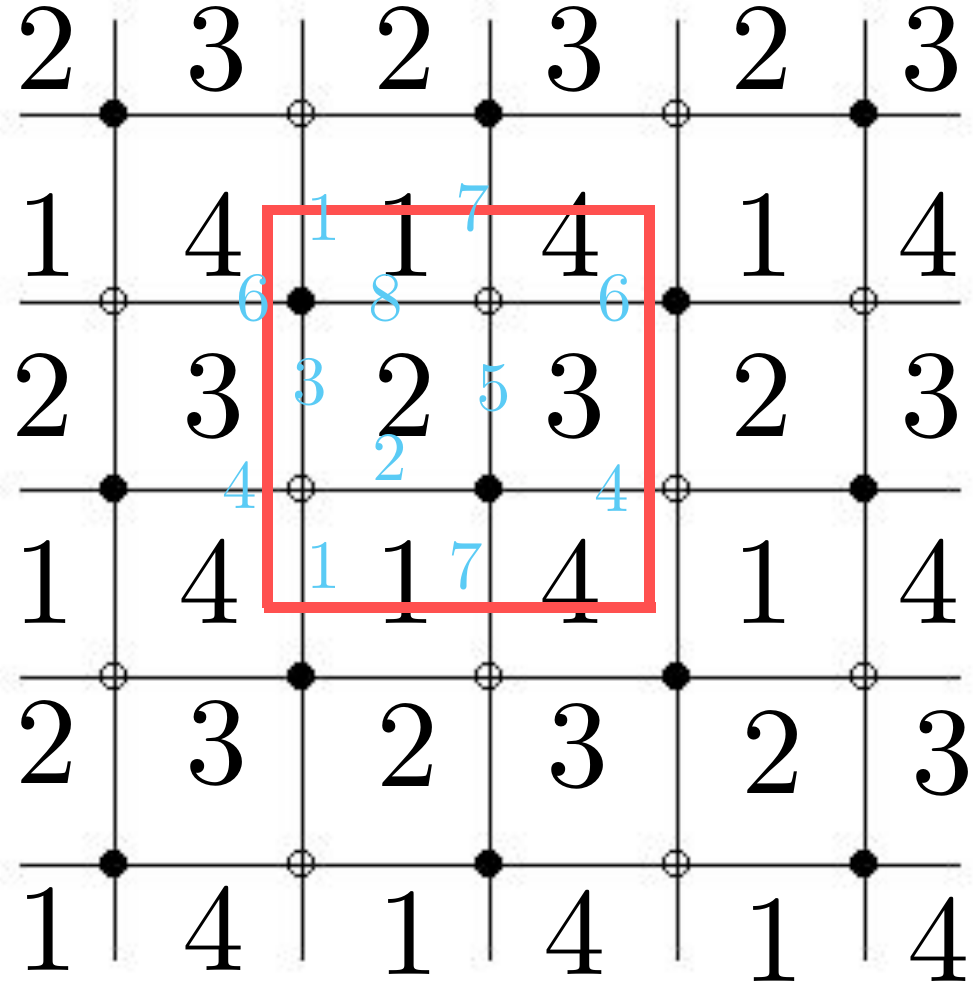}
\caption{Dimer for the phase 1 of $\mathbb{F}_0$.}
\label{fig:f01}
\end{figure}
From the dessin, we read the permutations at each of the nodes:
\begin{equation}
 \sigma_B=(1638)(5274)~, \quad\sigma_W=(1234)\,(5678) ~, \quad \sigma_{\infty}=(17)\,(28)\,(35)\,(46) ~.
\end{equation}
The ramification is $B=3+3+3+3+1+1+1+1=16$, while the degree is eight, thus corresponding to a dessin on a genus one surface.
We can explicitly construct the Belyi pair. We find
\begin{equation}
\beta=\frac{i\, (i+x)^4}{8\,x\, (1-x^2)} ~.
\end{equation}
The curve is the same as in the unorbifolded case, namely $y^2=x^3-x$.
The generators of the automorphism group leaving the pair invariant are
\begin{equation}
\phi_{\pm}(x,\, y)=(\frac{x+1}{1-x},\, \pm \frac{2\, i\, y}{(1-x)^2}) ~.
\end{equation}
Since $\phi_{\pm}^4=1$ and $\phi_-\circ\phi_+=\phi_+\circ\phi_-$, we have that ${\rm Aut}(\mathbb{T}^2,\, \beta)=\mathbb{Z}_4\times \mathbb{Z}_4$.

From a combinatorial perspective we find that there are two permutations (modulo conjugatios)
which leave the pair invariant
\begin{equation}
\gamma_{A_1}=(1234)\,(5678) ~,\qquad \gamma_{A_2}=(1638)\,(5274) ~.
\end{equation}
Clearly $\gamma_{A_i}^4=1$ while $[\gamma_{A_1},\, \gamma_{A_2}]=0$, and thus we recover the expected $\mathbb{Z}_4\times \mathbb{Z}_4$ automorphism group of the pair.

We can now turn to the exchange of black and white nodes. One can check that automorphisms $b_i$
in $ {\rm Aut} ( \mT^2 ) $ obey $ 1- \beta = \beta \circ b_i $  corresponding to the $B_1$ braiding operation
\begin{equation}
\begin{array}{l c l}
b_1:\,(x,\,y)\rightarrow(-x,\,\pm i\,y) ~, & & b_2:\,(x,\,y)\rightarrow(\frac{1}{x},\, \pm\frac{i\,y}{x^2}) ~, \\ & & \\
b_3:\,(x,\,y)\rightarrow(\frac{1-x}{1+x},\,\pm \frac{2\,y}{(1+x)^2}) ~, & & b_4:\,(x,\,y)\rightarrow (-\frac{1+x}{1-x},\,\pm\frac{2\,y}{(1-x)^2}) ~.
\end{array}
\end{equation}
However, after taking into account the automorphisms of the pair, it follows that we can just keep
\begin{equation}
b_1(x,\,y)=(-x,\, i\, y) ~.
\end{equation}
This transformation squares to $b_1^2(x,\,y)=(x,\,-y)$, which is, up to the action of an automorphism of the pair (essentially this is due to $\phi_+\circ\phi_-(x,\, y)=(x,\,-y)$), equivalent to $(x,\,y)$, thus recovering the expected $\mathbb{Z}_2$.

From a combinatorial perspective, modulo conjugation, we find, like in the unorbifolded case, two permutations implementing the expected twisted action of $B_1$:
\begin{equation}
\gamma_+=(13)\,(26)\,(48)\,(75) ~,\qquad \gamma_-=(28)\,(46)\,(1)\,(3)\,(5)\,(7) ~.
\end{equation}
As in the unorbifolded case, we can obtain $\gamma_1$ starting with $\gamma_2$ and acting with an automorphism, so that we can keep
\begin{equation}
\gamma_{B_1}=(13)\,(26)\,(48)\,(75) ~,
\end{equation}
which squares to one. In this case, as in the unorbifolded case,  the $B_2$ transformation is not implemented
by any $\gamma_{B_2}$ since $ \s_W $ and $ \s_{\infty}  $  have different cycle structures.
These statements have as well been verified with {\tt SAGE} mathematical software.

\subsection{Nonchiral $\mathbb{Z}_n$ orbifolds of SPP}

Let us now consider the SPP non-chiral orbifolds:
\begin{equation}
(u,\, v)\sim (\omega\, u,\, \omega^{-1}\, v) ~, \qquad \omega^n=1 ~.
\end{equation}
The dimer for $n=2$ is shown in Figure~\ref{fig:spporb} below.
\begin{figure}[!h]
\centering
\includegraphics[scale=.7]{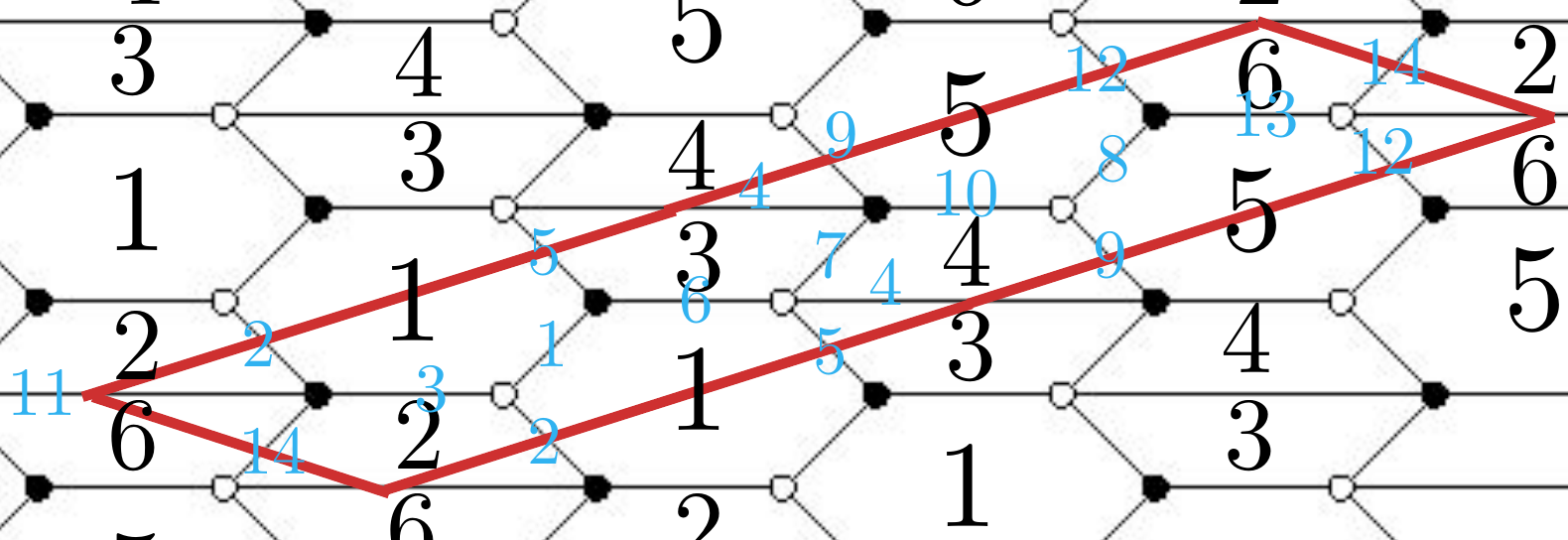}
\caption{$\mathbb{Z}_2$ orbifold of SPP}
\label{fig:spporb}
\end{figure}
The permutations are now
\begin{equation}
\sigma_B=(1\,5\,6)\,(4\,9\,10\,7)\,(8\,12\,13)\,(11\,2\,3\,14) ~,
\qquad \sigma_W=(123)\,(4567)\,(8\,9\,10)\,(11\,12\,13\,14) ~.
\end{equation}
Thus, we again see the common pattern of the orbifolds: we repeat the original (unorbifolded) $\sigma_W$ $n$ times ordering it in a canonical way. As for $\sigma_B$, we take another copy of this orbifolded $\sigma_W$, take the middle elements of the longest cycle and cyclically permute them.

The $\gamma_A$ generating the automorphisms are as well computed according to the usual pattern:
we group together in $\sigma_W$ the cycles of the same size and take first element of each permutation for each length of cycles.
In the particular $n=2$ case, this is
\begin{equation}
\gamma_A=(1\,8)\,(2\,9)\,(3\,10)\,(4\,11)\,(5\,12)\,(6\,13)\,(7\,14) ~.
\end{equation}
Once again, we can check that $\gamma_A^n=1$ and that no elementary cycle is left fixed.

\end{appendix}

\end{document}